\documentclass[a4paper]{amsart}
\usepackage[english]{babel}
\usepackage[utf8]{inputenc}
  
\usepackage[cm]{fullpage}
\usepackage{amsmath, amssymb, amsthm}
\usepackage{graphicx}
\usepackage{floatrow}
\usepackage{physics}

\usepackage{tikz}
\usetikzlibrary{decorations.pathmorphing}
\usetikzlibrary{patterns}
\usepackage{todonotes}
\usepackage{mathrsfs}
\usepackage{comment}
\usepackage{verbatim}
\usepackage{sidecap}
\usepackage{soul}
\usepackage{enumerate}

\usepackage{caption}
\usepackage{subcaption}

\usepackage[hidelinks]{hyperref}

\newcommand{\R}{\mathbb{R}}

\newcommand{\T}{\mathbb{T}}
\newcommand{\C}{\mathbb{C}}
\newcommand{\N}{\mathbb{N}}

\newcommand{\Z}{\mathbb{Z}}
\newcommand{\Li}{\mathcal{L}}
\newcommand{\Ft}{\mathcal{F}}
\newcommand{\Ei}{\mathcal{E}}
\newcommand{\Hi}{\mathcal{H}}

\newcommand{\id}{1\! \!1}
\newcommand{\n}[1]{ \left \vert \left \vert #1 \right \vert  \right \vert}

\DeclareMathOperator*{\vectorize}{vec}
\DeclareMathOperator*{\lcu}{lcu}
\DeclareMathOperator*{\diag}{diag}
\DeclareMathOperator*{\per}{per}
\DeclareMathOperator*{\supp}{supp}
\DeclareMathOperator*{\res}{res}
\DeclareMathOperator*{\ess}{ess}
\DeclareMathOperator*{\eff}{eff}
\DeclareMathOperator*{\appt}{appt}
\DeclareMathOperator*{\HS}{HS}
\DeclareMathOperator*{\TC}{TC}
\DeclareMathOperator*{\esssup}{esssup}

\DeclareMathOperator*{\conv}{conv}

\definecolor{mulbb}{RGB}{246,76,25}

\newcommand{\Ad}{\operatorname{Ad}}

\theoremstyle{plain}
\newtheorem{theorem}{Theorem}[section]

\newtheorem{lemma}[theorem]{Lemma}
\newtheorem{claim}[theorem]{Claim}
\newtheorem{remark}[theorem]{Remark}

\newtheorem{example}[theorem]{Example}
\newtheorem{corollary}[theorem]{Corollary}
\newtheorem{conjecture}[theorem]{Conjecture}

\newtheorem{question}[theorem]{Question}
\newtheorem{proposition}[theorem]{Proposition}

\newtheorem{convention}[theorem]{Convention}

\usepackage{cleveref}

\crefname{lemma}{lemma}{lemmas}
\Crefname{lemma}{Lemma}{Lemmas}

\crefname{definition}{definition}{definitions}
\Crefname{definition}{Definition}{Definitions}
\crefname{claim}{claim}{claims}
\Crefname{claim}{Claim}{Claims}

\crefname{remark}{remark}{remarks}
\Crefname{remark}{Remark}{Remarks}

\crefname{notation}{notation}{notations}
\Crefname{notation}{Notation}{Notations}

\crefname{example}{example}{examples}
\Crefname{example}{Example}{Examples}

\crefname{corollary}{corollary}{corollaries}
\Crefname{corollary}{Corollary}{Corollaries}

\crefname{conjecture}{conjecture}{conjectures}
\Crefname{conjecture}{Conjecture}{Conjectures}

\renewcommand\bra[1]{{\langle{#1}|}}
\renewcommand\ket[1]{{|{#1}\rangle}}
\newcommand{\ResultTitle}[2]{\textbf{\hyperref[#1]{Thm.\ref*{#1}} (#2).}}

\newcommand{\ResultTitleCor}[2]{\textbf{\hyperref[#1]{Cor.\ref*{#1}} (#2).}}

\usepackage{caption}

\usetikzlibrary{arrows.meta,positioning}

\tikzset{
  >=Latex,
  imply/.style={->,thick},
  box3/.style={draw,rounded corners,fill=blue!10,align=left,inner sep=2.5pt,text width=6.8cm,font=\footnotesize},
  box4/.style={draw,rounded corners,fill=green!12,align=left,inner sep=2.5pt,text width=6.8cm,font=\footnotesize},
}

\title{\small{Spectra of generators of Markovian evolution in the thermodynamic limit:\\ From non-Hermitian to full evolution via tridiagonal Laurent matrices}}
\author{Frederik Ravn Klausen}
\address{Frederik Ravn Klausen \\ Princeton University, Department of Mathematics, Fine Hall}
\email{fk3206@princeton.edu}

\usepackage{enumitem}

\begin{document}

\begin{abstract}
\noindent 
It is shown that generators of single-particle, translation-invariant Lindblad operators on the infinite line are unitarily equivalent to direct integrals of 
finite-range bi-infinite Laurent operator with finite-range perturbations. 
This representation enables rigorous calculation of spectra for local dissipation such as dephasing and incoherent hopping, and yields proofs of gaplessness, absence of residual spectrum and a condition for convergence of finite volume spectra to their infinite volume counterparts.
The analysis relies on new results on the spectra of direct integrals of non-normal operators which may be of independent interest. 
\end{abstract}

\vspace{-4cm}
\maketitle
\vspace{-1cm}

\section{Introduction}

Spectral properties of the Hamiltonian encode many important physical properties of quantum systems, such as the speed of propagation as described by the RAGE theorem \cite{ruelle1969remark,amrein1973characterization,enss1978asymptotic}.  Beyond the closed-system paradigm described by unitary dynamics, a natural setting is that of Markovian time-evolution generated by a completely positive dynamical semigroup with generator $\Li$.  
Here, gaps in the spectrum around the origin in the complex plane provide information about relaxation times towards the non-equilibrium steady state of the system (as shown in Appendix \ref{sec:spec_dyn} where it is also discussed why the relation between spectra and dynamics is weaker in infinite volume). 


This paper concerns the spectrum $\sigma(\Li)$ for translation-invariant, finite-range, one-dimensional single-particle systems. The results are outlined in Figure 1.
In \Cref{sec:spec_direct_integral}, general formulas of independent interest for the spectra of direct-integral operators are proven. In \Cref{sec:translation_invariance}, using translation invariance, $\Li$ is decomposed as a direct integral of a Laurent operator (corresponding to the effective non-Hermitian Hamiltonian) and a finite-rank operator (corresponding to the quantum jumps).
\Cref{sec:general_app} applies the decomposition to prove gaplessness of $\Li$, describe its approximate point spectrum, and establish a condition for convergence of finite-volume spectra to their infinite-volume counterparts.
In \Cref{sec:examples}, spectra of open quantum systems studied in the literature \cite{Esposito2005EmergenceOD, Znidaric2015RelaxationTO, flynn2021topology} are computed in infinite volume.
These formulas can be viewed as approximations to large finite-volume systems (see \Cref{thm:convergence_finite} for convergence conditions).
Understanding the interplay between disorder and dissipation is an emerging problem \cite{plenio2008dephasing, clark2011diffusive, xu2018interplay, frohlich2016quantum}; \Cref{sec:Lindblad_random} provides a spectral perspective.

\begin{figure}
\begin{tikzpicture}[x=1cm,y=1cm,
  >=Latex, imply/.style={->,thick}, box2/.style={draw,rounded corners,fill=blue!30,align=left,inner sep=2.5pt,text width=5cm,font=\footnotesize},
  box3/.style={draw,rounded corners,fill=blue!10,align=left,inner sep=2.5pt,text width=5cm,font=\footnotesize},
  box4/.style={draw,rounded corners,fill=green!12,align=left,inner sep=2.5pt,text width=4.8cm,font=\footnotesize},
  box5/.style={draw,rounded corners,fill=yellow!12,align=left,inner sep=2.5pt,text width=4.8cm,font=\footnotesize},
  box6/.style={draw,rounded corners,fill=orange!12,align=left,inner sep=2.5pt,text width=4cm,font=\footnotesize}
]

\node[box2] (t35) at (0,2) {%
  \ResultTitle{directint}{pseudospectra} \\
  Measurable family of operators $A(q)$,\\
  \begin{center} $\displaystyle \sigma\left(\int^{\oplus}_{I}\hspace{-8pt}A(q)dq \right) =  \bigcap_{\varepsilon > 0} \bigcup^{\ess}_{q \in I} \sigma_\varepsilon( A(q)).$ \end{center}};

\node[box2] (t37) at (6,2) {%
  \ResultTitle{direct_continuity}{union of fibers}\\
  Norm-continuous operators $A(q)$,\\
  $\displaystyle \sigma \left(\int^{\oplus}_{I}\hspace{-8pt}A(q)dq \right) =\bigcup_{q\in I}\sigma\!\big(A(q)\big).$};

\node[box3] (t33) at (0,0) {%
  \ResultTitle{main}{direct integral form}
  \begin{center}$\Li \simeq \int_{[0,2\pi]}^\oplus \!\big(T(q)+F(q)\big)\,dq$,\end{center} 
  with $T(q)$ Laurent ($r$-diag.),\\ $F(q)$ finite rank.};

\node[box3] (c38) at (6,0) {%
  \ResultTitleCor{cor:general_rank}{spectrum of $\Li$}\\
  $\displaystyle 
\sigma(\Li)=\bigcup_{q\in \lbrack 0, 2 \pi \rbrack}\hspace{-0.7em} \sigma( T(q) + F(q)).
$};

\node[box3] (c39) at (6,-1.7) {%
  \ResultTitleCor{cor:rank_one}{rank-one update} \\
  $F(q)=|\mathtt{v}(q)\rangle\langle \mathtt{u}(q)|$. \\ If $z\in\sigma(\Li)$ then $z\in \sigma(T(q))$ or \\
  $ \bra{\mathtt{u}(q)}(T(q)-z)^{-1} \ket{\mathtt{v}(q)}= -1.$};

\node[box4] (f41) at (12.5,-0.4) {%
  \ResultTitle{thm:no_gap}{$\Li$ is gapless} 
  };

  \newcommand{\veryshortarrow}[1][3pt]{\mathrel{%
   \hbox{\rule[\dimexpr\fontdimen22\textfont2-.2pt\relax]{#1}{.4pt}}%
   \mkern-4mu\hbox{\usefont{U}{lasy}{m}{n}\symbol{41}}}}
   
\node[box4] (f43) at (12.5,-0.9) {  \ResultTitle{thm:convergence_finite}{Cond. \hspace{-3pt}$\sigma(\Li_n^{\per})\veryshortarrow \sigma(\Li)$}};

\node[box4] (f42) at (12.5,-1.45) {%
  \ResultTitle{thm:approx_point}{$\sigma_{\mathrm{appt}}(\Li)=\sigma(\Li)$}};
\node[box5] (f44) at (12.5,-2) {  \ResultTitle{thm:incoherent_hopping}{$\sigma(\Li)$ hopping}};

\node[box5] (f45) at (12.5,-2.75) {  \ResultTitle{theorem:dephasing_spectrum}{$\sigma(\Li)$ local dephasing}\\$\left( - G +  i  \lbrack -4,  4 \rbrack \right) \cup   \lbrack -G + \sqrt{G^2-16} , 0\rbrack$};


\node(a) at (10.2,-0.4) { };
\node(b) at (10.2,-0.8) { };
\draw[imply] (t35) -- (t37);
\draw[imply] (t33) -- (c38);
\draw[imply] (t37) -- (c38);
\draw[imply] (c38) -- (c39);
\draw[imply] (c39) -- (a);
\draw[imply] (c39) -- (f42);
\draw[imply] (c39) -- (b);
\draw[imply] (c39) -- (f44);
\draw[imply] (c39) -- (f45);

\node[anchor=north west, text width=4cm, align=left, font=\small]
      at (10,2.9){\stepcounter{figure}\begin{minipage}{4.6cm}
         Figure \thefigure: Overview of the main results. The Lindbladian $\Li$ satisfies finite range and translation invariance and from Corollary \ref{cor:rank_one} also $\text{rank}(L_k)=1$.
         \label{fig:overview}
      \end{minipage}};
\end{tikzpicture}
\end{figure}

With some exceptions, \cite{avron2012adiabatic,holevo,tamura2016dynamical,clark2011diffusive,frohlich2016quantum},  the study of Lindbladian evolutions has, in recent decades, focused on finite spin systems.
Our rigorous analytic results also complement the large body of recent work on random Lindblad operators studied from a random matrix theory point of view \cite{PhysRevLett.123.234103,CanRMT,tarnowski2021random,lange2021random}.




  \begin{figure}
  \centering
  \includegraphics[scale =0.5]{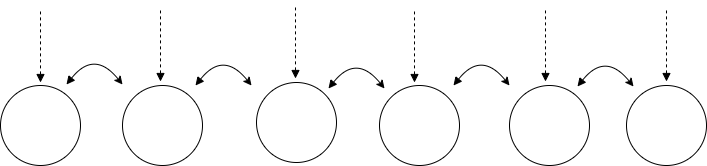}
    \caption{A single particle on a lattice with hopping between lattice sites and local dissipation. In this example, the hopping and dissipation have the shortest possible range, but the methods presented apply as long as these properties stay local and translation-invariant.  \label{fig:first_sketch}}
\end{figure}

\subsection{Lindblad systems on the infinite lattice\label{inf_tech}}
Let $ \mathcal{H} = \ell^2(\Z)$ with distinguished (position) basis $\{ \ket{\delta_k} \}_{k \in \Z}$ and $\HS(\Hi)$ the Hilbert-Schmidt operators.
Markovian, open-system dynamics of a single quantum particle on $\mathcal{H}$ is generated by a \emph{Lindbladian} \cite{Gorini:1975nb,lindblad1976generators}, 
$\mathcal{L}: \HS( \mathcal{H}) \to  \HS( \mathcal{H}) $ of the form 
\begin{align} \label{Lindblad_form}
\mathcal{L}(\rho) = - i \lbrack H, \rho \rbrack + G \sum_{k} L_k \rho L_k^* - \frac{1}{2}(L_k^* L_k \rho  + \rho L_k^* L_k),
\end{align}
where $H$ is the Hamiltonian of the system (a self-adjoint bounded operator), coupling constant $G > 0$, and $*$ denotes the adjoint. 
The operators $L_k \in \mathcal{B}( \mathcal{H})$ in the dissipative part are referred to as Lindblad operators.
Time-evolution is given by the completely positive semigroup $e^{t \Li}$. 
$\Li$ decomposes into a non-Hermitian evolution $\mathcal{T}$  and a quantum jump term $\mathcal{F}$,
\begin{align}\label{eq:Lindblad_decomposition}
\mathcal{L}(\rho)  =  - i(H_{\eff} \rho - \rho H_{\eff}^*) + G \sum_{k} L_k^* \rho L_k = \mathcal{T}(\rho) + \mathcal{F}(\rho),
\end{align}
where $\mathcal{F}(\rho) = G \sum_{k} L_k^* \rho L_k$ and the effective non-Hermitian Hamiltonian is defined by
$
H_{\eff}= H  - \frac{iG}{2}\sum_{k} L_k^* L_k.
$



\subsection{Assumptions}
Define the operator shift $S_n$ on $\Hi = \ell^2(\Z)$ by 
\begin{align}\label{eq:shift_operator}
(S_n \psi)(x) = \psi(x-n)
\end{align}
and let $S_1 = S$. Since $ \Hi = \ell^2( \Z)$ the operator $S_n$ is unitary and $S_n^{-1} = S_n^* = S_{-n}$. Say that $\Li$ of form \eqref{Lindblad_form} is
\begin{align}
   \text{Finite range } r < \infty : \text{ if }   \bra{\delta_y}, L_k \ket{\delta_x} = 0 =\bra{\delta_y}, H \ket{\delta_x},   \text{ whenever }  \max \{ \abs{x-k},  \abs{y-k} \}> r  \tag{Finite Range}\label{eq:finite-range}. \\
\text{Translation-invariant}: \text{ if } S_1 L_k S_{-1} = L_{k+1}  \text{ for each } k \in \Z \text{ and } H = S_1 H S_1^*\tag{TransInv}\label{eq:translation-invariance}. 
\end{align}
 Assumption \eqref{eq:translation-invariance} implies that $\Li$ is translation-\emph{covariant},  i.e. 
\begin{align}  \label{eq:translation_covariant}
\Li(S_1 \rho S_1^*) = S_1 \Li(\rho)S_1^*.
\end{align}
In addition, many of our motivating examples are rank 1.
\begin{align*}
\text{Rank one: $\text{Rank}(L_k) = 1$ for all $k \in \Z$} \tag{Rank 1}\label{eq:Rank1}. 
 \end{align*}

By \eqref{eq:translation-invariance}, $\sum_k L_k L_{k}^* $ is constant on diagonals. Therefore, \eqref{eq:finite-range} and \eqref{eq:translation-invariance}  imply the assumption in \Cref{bounded_op} in \Cref{sec:bound_op} and so $\Li$ is bounded (in fact on all Schatten spaces):
\begin{proposition}
  If $\Li$ of Lindblad form \eqref{Lindblad_form} is \eqref{eq:translation-invariance} and \eqref{eq:finite-range}, then $\Li \in \mathcal{B}(\HS(\Hi))$. 
\end{proposition}
Throughout $\Li \in \mathcal{B}(\HS(\Hi))$ and the exact formulas for spectra in \Cref{sec:examples} are only obtained there because spectral independence is not known. 
  The Lindblad form ensures that every point in the spectrum has non-positive real part (see e.g. \cite{473322,baumgartner2008analysis}). For finite-dimensional systems, $\Li(X^*) = \Li(X)^*$  which implies  invariance under complex conjugation $\sigma( \mathcal{L}) = \overline{ \sigma(\mathcal{L})}$. 
In finite dimensions, there is always a steady state $\rho_\infty$, meaning $\Li(\rho_\infty) = 0$ (see e.g. \cite[Prop. 5]{baumgartner2008analysis}). The symmetries of $L_k$ are inherited by $\mathcal{L}$ and the steady state \cite{buvca2012note, Nigro, symmetries, Georgios}. In particular, assuming \eqref{eq:translation-invariance} the steady state subspace is translation-invariant. Although translation-invariant infinite volume Lindbladians have not been studied extensively some results exist \cite{Holevo1993ANO, clark2011diffusive}.

\section{Spectrum of direct integral of operators}\label{sec:spec_direct_integral}
In this section, we prove formulas for the spectrum of a direct integral of non-normal operators in the general case and in the norm-continuous case that is relevant in this paper. First we introduce the direct integral, for more information, for example on the details of measurability, see \cite[XII.16]{reed1978iv}.
\subsection{The direct integral of Hilbert spaces}\label{sec:direct_integral}
Suppose that $\{\Hi_q \}_{q \in I}$ is a family of Hilbert spaces indexed by some index set $I \subset \R$ and let $dq$ denote the Lebesgue measure on $I$. 
The direct integral 
$\int_{I}^\oplus \Hi_q dq$ consists (of equivalence classes up to sets of measure 0) of vectors $v$ such that $v_q \in \Hi_q$ for each $q\in I$. It is a Hilbert space with inner product
\begin{align*}
\langle v, w \rangle_{\int_{I}^\oplus \Hi_q dq} = \int_I \langle v_q, w_q \rangle_{\Hi_q}dq.
\end{align*}
If $q \mapsto A(q)\in \mathcal{B}(\Hi_q)$ is measurable define $ \int_{I}^{\oplus} A(q) dq$ through $\int_{I}^{\oplus}A(q) dq \cdot v = \int_{I}^{\oplus}A(q) v(q) dq $.
If for an operator $\mathcal{A}$ on  $\int_{I}^\oplus \Hi_q dq$ the converse is true, i.e. there exists a measurable family of operators $\{A(q) \}_{q\in I}$ such that $\mathcal{A} = \int_{I}^{\oplus}A(q) dq$ then $\mathcal{A}$ is \emph{decomposable}.
The norm of a decomposable operator is given as 
\begin{align}\label{eq:esssup_norm}
\norm{\mathcal{A}}  = \esssup_{q\in I}\norm{A(q)}. 
\end{align}

\subsection{Pseudospectra and essential unions} For self-adjoint, translation-invariant operators $A$ the spectrum of $A$ coincides with the union of the spectra of the operators contained in the direct integral representation of $A$ after the Fourier-transform \cite[XIII.85]{reed1978iv}. However, in the non-normal case the point-wise spectrum may not be sufficient.
In fact, already for the direct sum (direct integral with respect to the counting measure), the spectrum is not the union of the spectra of the fibers as may be seen from computing the spectra of an infinite direct sum of Jordan blocks of size $2,3, \dots $ with $0$ on the diagonal \cite[Problem 98]{halmos2012hilbert}. 

The pseudospectrum can be used to recover a connection between the spectra of operators and their direct integral decompositions. So, for a bounded operator $B \in \mathcal{B}(X) $ on a Banach space $X$, for every $\varepsilon > 0$ define the \emph{$\varepsilon$-pseudospectrum}  $ \sigma_\varepsilon(B)\subset\C$ as the $\lambda$ for which $ \norm{ (B- \lambda )^{-1} } > \frac{1}{ \varepsilon}$, where we set $\norm{B^{-1}}= \infty$ whenever $B$ is not boundedly invertible. 
Then 
$\label{eq:pspectrum}
\sigma(B) = \bigcap_{\varepsilon > 0} \sigma_{\varepsilon}(B).
$

For a family of measurable sets $M(q)$ with respect to the Lebesgue measure on some interval\footnote{The following also holds for other measures, but we consider the Lebesgue measure for clarity}  say that $x$ is in the \emph{essential union},  $x\in  \bigcup^{\ess}_{q \in I} M(q) $ if and only if there exists a set $M$ of positive measure such that $M \subset \{q \mid x \in M(q)\}$. 

For more information on direct integrals and their spectral theory see \cite{azoff1974spectrum,chowspectrum}.
A related theorem is proven in the self-adjoint case in \cite[XIII.85]{reed1978iv} and spectra and direct integrals were also studied in \cite{ismailovsome}. In addition the direct sum case was studied in \cite[Theorem 5]{ismailov2020spectra}.  

\begin{theorem} \label{directint}
Let $I \subset \R$ be an interval and $\Hi = \int_{I}^{\oplus} \Hi_q dq$ for some family of separable Hilbert spaces  $\{\Hi_q \}_{q\in I}$. For any measurable family $\{A(q) \}_{q \in I} $ of bounded operators  on $\Hi$ with direct integral $\int^{\oplus}_{I} A(q) dq \in \mathcal{B}(\Hi)$, 

\begin{align}
\sigma \left(\int^{\oplus}_{I}\hspace{-8pt}A(q)dq \right) =  \bigcap_{\varepsilon > 0} \bigcup^{\ess}_{q \in I} \sigma_\varepsilon( A(q)).
\end{align}
\end{theorem}
\begin{proof}
Set $\mathcal{A} = \int^{\oplus}_{I}\hspace{-4pt}A(q)dq$.  Suppose that $\lambda \not \in    \bigcup^{\ess}_{q \in I} \sigma_\varepsilon( A(q))$ for some $\varepsilon >0$.
Then  $
\norm{ (A(q) - \lambda)^{-1} }_{\Hi_q}  \leq \frac{1}{\varepsilon}
$ for almost all $q \in I$. Thus, $$ \esssup_{q \in I} \norm{ (A(q) - \lambda)^{-1}}_{\Hi_q}  \leq \frac{1}{\varepsilon}.$$ By \eqref{eq:esssup_norm}, $ \int_I^{\oplus}(A(q) - \lambda)^{-1} dq$ is a well-defined bounded operator, which is an inverse of $\mathcal{A}-\lambda$ from both sides and so $\mathcal{A}- \lambda$ is invertible,  $ \lambda \not \in \sigma(\mathcal{A})$.

Conversely, let $\lambda \in \bigcap_{\varepsilon > 0}  \bigcup^{\ess }_{q \in I} \sigma_\varepsilon( A(q))$. 
By \cite[Lemma 1.3]{chowspectrum} if $\mathcal{A} - \lambda$ is boundedly invertible then the inverse is decomposable $(\mathcal{A} - \lambda)^{-1} =  \int_I^{\oplus} E(q)  dq$, where $E(q) = (A(q) - \lambda)^{-1}$ for almost all $q$. Since $\lambda \in \bigcap_{\varepsilon > 0}  \bigcup^{\ess }_{q \in I} \sigma_\varepsilon( A(q))$, then either $\lambda \in \sigma(A(q))$ for a set of $q$s with positive measure (in which case $\mathcal{A} - \lambda$ is not boundedly invertible) or for each $n \in \N$ there exists a set $I_n$ such that $\abs{I_n} > 0$ with $ n \leq \norm{(A(q) - \lambda)^{-1}}_{\Hi_q} < \infty$ for all $q \in I_n$. 
In the latter case, for each $q \in I_n$ and $n \in \N$ there exists a $v_{q,n}\in \Hi_q $ with $\norm{v_{q,n}}_{\Hi_q}  = 1$ and such that $ \norm{ (A(q) - \lambda)^{-1}v_{q,n} }_{\Hi_q}  \geq \frac{n}{2}$. Define\footnote{One may worry whether there is a measurable choice of $q \mapsto v_{q,n} $. This concern we address in Appendix \ref{sec:measurability}.} the normalized $w_n = \abs{I_n}^{-1/2}\int_{\lbrack 0, 2\pi  \rbrack}^{\oplus}v_{q,n}\id_{q \in I_n} dq.$ 
To see that $\mathcal{A} -\lambda$ is not boundedly invertible, combine \eqref{eq:esssup_norm} with 
 \begin{equation*}
\norm{ \int_{\lbrack 0, 2\pi \rbrack}^{\oplus} (A(q) - \lambda)^{-1}dq\, w_n }^2  
  = \frac{1}{\abs{I_n}}\int_{I_n}\norm{(A(q) - \lambda)^{-1} v_{q,n}}_{\Hi_q}^2 dq \geq \frac{1}{\abs{I_n}}\int_{I_n}  \left( \frac{n}{2} \right)^2 dq = \left( \frac{n}{2} \right)^2. \hfill \qedhere
 \end{equation*}
\end{proof}

To determine the essential union of the pseudo-spectra it suffices to have continuity in $q$ in the pseudospectra of $T(q)$.  The continuity is reminiscent of a theorem for self-adjoint operators which also finds the spectrum of the direct integral in terms of its fibers \cite{fialovatwo}. 
  A  recent related result that, in some sense, is in between the generality of Theorem \ref{directint} and Theorem \ref{direct_continuity} appeared in \cite{ng2020direct}.

A family of measurable bounded operators $\{A(q)\}_{q\in I}$ is \emph{norm continuous} if the function $ q \mapsto \norm{A(q)}$ is continuous. 
\begin{lemma}\label{lemma:double_essential_spectrum}
Let $I \subset \R$ be compact and $\{A(q)\}_{q\in I}$  norm-continuous. 
For any $q_0\in I$ and $\varepsilon > 0$,
\begin{align} \label{lemma:psuedo_inclusion}
\sigma_{\varepsilon}(A(q_0)) \subset   \bigcup_{q\in I}^{\ess}  \hspace{-0.1 em} \sigma_{2\varepsilon}(A(q)).
\end{align}
\end{lemma}
\begin{proof}
Let $\lambda \in \sigma_{\varepsilon}(A(q_0))$ and find a vector $v$ such that $\norm{(A(q_0)-\lambda)v} \leq \varepsilon$. 
By norm continuity, there exists a $\delta >0$ such that for all $q \in (q_0 - \delta, q_0 + \delta)$, $\norm{A(q)-A(q_0)} \leq \varepsilon$. For all such $q$, 
$$
\norm{(A(q)-\lambda)v} \leq \norm{A(q)-A(q_0)} + \norm{(A(q_0)-\lambda)v} \leq 2\varepsilon. 
$$
So $\lambda \in \sigma_{2\varepsilon}(A(q))$ for all $q \in (q_0 - \delta, q_0 + \delta)$ and therefore also in their essential union. 
\end{proof}
 Whenever $\lambda \not\in \sigma(A(q)) $ define the resolvent $R(q) = (A(q)- \lambda)^{-1}$.
If both $R(p), R(q)$ exist, by the resolvent equation, 
\begin{align}\label{eq:resolvent_bound}
 \abs{\norm{R(p)} - \norm{ R(q)} } \leq \norm{R(p)}\norm{A(p) - A(q)}\norm{R(q)}.
\end{align}

\begin{theorem} \label{direct_continuity}
Let $I \subset \R$ be compact and $\{A(q)\}_{q\in I}$ norm-continuous family of bounded operators. Then
\begin{align*}
\bigcup_{q \in I } \sigma(A(q)) = \sigma \left( \int^{\oplus}_{I}A(q) dq \right).
\end{align*}
\end{theorem}
\begin{proof}
"$ \subset $": 
Combining Lemma \ref{lemma:double_essential_spectrum} with Theorem \ref{directint} yields for any $q_0\in I$, 
\begin{align*}
\sigma(A(q_0)) =\bigcap_{\varepsilon > 0}\sigma_{\varepsilon}(A(q_0)) \subset \bigcap_{\varepsilon > 0}   \bigcup_{q\in I}^{\ess} \sigma_{2\varepsilon}(A(q)) = \sigma \left( \int^{\oplus}_{I} \hspace{-0.6 em} A(q) dq \right).
\end{align*}
"$ \supset $": Let $\lambda\in \bigcap_{\varepsilon > 0}  \bigcup^{\ess}_{q\in I}  \sigma_\varepsilon( A(q))$.
Assume for contradiction that $ \lambda \not\in \cup_{q\in I}\sigma( A(q)) $.
By  \eqref{eq:resolvent_bound} the function $N: I \to \R$ defined by $N(q)= \norm{(A(q) - \lambda)^{-1} } $ is continuous.
Set
$
S_n =   \left \{q\in I \mid  \norm{A(q) - \lambda)^{-1} } \geq n  \right \}=  N^{-1}\left( \lbrack n, \infty ) \right).
$
Since $N$ is continuous and $ \lbrack n, \infty ) $ is closed  $S_n$ is closed.
Furthermore, $S_n \subset S_{n-1}$ and $S_n$ is non-empty for each $n\in \N$. By the finite intersection property (since $I$ is compact) $\bigcap_{n\in \N} S_n $ is non-empty. But if $q_0\in \cap_{n\in \N}S_n$ then $\norm{(A(q_0) - \lambda)^{-1} } \geq n$ for all $n\in \N$ and thus $\lambda\in \sigma( A(q_0)) $ which is a contradiction. 
\end{proof}

\section{From translation-invariance to a direct integral decomposition}\label{sec:translation_invariance}

\noindent In this section, translation-covariance is used to obtain a direct integral decomposition for $\Li \in \mathcal{B}( \HS( \ell^2(\Z)))$. 

First recall that a map between separable Hilbert spaces $U:\Hi_1 \to \Hi_2$ is an \emph{isometric isomorphism} if it is linear, bijective, and preserves the inner product 
$
\langle U(x), U(y) \rangle_{ \Hi_2} = \langle x, y \rangle_{ \Hi_1}. 
$
Every isometric isomorphism of separable Hilbert spaces is a unitary operator \cite[Theorem 5.21]{durhuus2014mathematical}, which gives rise to an isomorphism of $C^*$-algebras  $\Ad_U:\mathcal{B}(\Hi_1) \to \mathcal{B}(\Hi_2)$ acting by 
conjugation.

\subsection{Explicit vectorization of the Lindbladian} 
Define $ \ell^2(\Z) \otimes  \ell^2(\Z)$ to be the set of all formal symbols $v \otimes w$ where $v,w \in \ell^2(\Z)$ with inner product $ \langle v_1 \otimes w_1, v_2 \otimes w_2 \rangle=  \langle v_1 , v_2 \rangle  \langle  w_1, w_2 \rangle $ and taking closure with respect to that inner product.  
Following (\cite[(4.88)]{vectorization, lecturenotes}), define  $\vectorize: \HS( \ell^2(\Z)) \to  \ell^2(\Z) \otimes \ell^2(\Z) $ by extending the map
$
\vectorize(\ket{\delta_i}\bra{\delta_j}) = \ket{\delta_i} \otimes \ket{\delta_j}   
$ linearly, which by definition of the Hilbert-Schmidt norm is an isometric isomorphism. 
The vectorization of a product is given by
\begin{align}
\vectorize(ABC) =  A \otimes C^T \vectorize(B)
\end{align}
  analogous to \cite[(4.84)]{lecturenotes}\footnote{We make a slight change of notation.  The vectorization map in \cite[(4.84)]{lecturenotes} is defined as $\ket{i}  \bra{j} \to \ket{\delta_j} \otimes  \ket{i}$.}, where $C^T$ is the transpose of $C$. Thus, 
\begin{align}\label{eq:dissipVec}
\vectorize(\Li(\rho)) =  \left(- i(H \otimes \id) + i (\id  \otimes H^T)  +  G \sum_k L_k \otimes (L_k^*)^T   - \frac{1}{2} L_k^* L_k \otimes \id -  \frac{1}{2}  \id \otimes (L_k^* L_k)^T \right) \vectorize(\rho). 
\end{align}

Recall the definition of $H_{\eff}= H  - \frac{iG}{2}\sum_{k} L_k^* L_k$, by \eqref{eq:dissipVec},
\begin{align}\label{eq:vectorization} 
\Ad_{\vectorize}(\Li) =  -i H_{\eff}\otimes \id + i\id \otimes \overline{H_{\eff}} + G \sum_k L_k \otimes \overline{L_k}.
\end{align}

\subsection{Isometric isomorphisms and direct integral decomposition}
Assuming \eqref{eq:translation-invariance},  $\Li$ is covariant under the joint translations $(x,y) \mapsto (x+1, y+1)$ cf. \eqref{eq:translation_covariant}, which means that Fourier transformation in that direction will simplify $\Li$. 
To do it, define an operator $C: \ell^2(\Z) \otimes\ell^2(\Z) \to\ell^2(\Z) \otimes\ell^2(\Z)$ by 
\begin{align}\label{eq:Cdef}
C \ket{\delta_{j,k}}= \ket{\delta_{j,k-j}}. 
\end{align}
Since $f:\Z^2\to\Z^2$ given by $f(j,k) = (j,k-j)$ is a bijection, $C$ maps an orthonormal basis to an orthonormal basis so it is unitary (and hence an isometric isomorphism) with inverse
$ C^*\ket{\delta_{j,k}} = C^{-1}\ket{\delta_{j,k}} = \ket{\delta_{j,k+j}}.$
Conjugation by  $C$  changes joint translation invariance to translation invariance in the first tensor factor\footnote{We were made aware of this trick in \cite{355246}.}. 
\begin{lemma} \label{lemma:C_relations}
Let $S$ be the shift operator defined in \eqref{eq:shift_operator}. The unitary operator $C$ satisfies the following relations:
$$\textit{ (i) } C(\id \otimes S)C^* = \id\otimes S,
\textit{ (ii) } C(S \otimes \id)C^* = S\otimes S^*, 
\textit{ (iii) } C(S \otimes S)C^* = S\otimes \id. $$ 
\end{lemma}
\begin{proof}
First (i) and then (ii) follows by computing, for any $k,j \in \Z$: 
\begin{align*}
&C(\id \otimes S)C^* \ket{\delta_{j,k}} = C(\id \otimes S) \ket{\delta_{j,k+j}} = C \ket{\delta_{j,k+j+1}} =  \ket{\delta_{j,k+1}}  = (\id \otimes S)\ket{\delta_{j,k}}\\
&C(S \otimes \id)C^* \ket{\delta_{j,k}} = C(S \otimes \id) \ket{\delta_{j,k+j}} = C \ket{\delta_{j+1,k+j}} =  \ket{\delta_{j+1,k-1}}  = S\otimes S^* \ket{\delta_{j,k}}. 
\end{align*}
The third relation follows from multiplying the previous two. 
\end{proof}
Define $\Ft_1 = \Ft \otimes \id$, the Fourier transform in the first coordinate. 
Then by Lemma \ref{lemma:transform_Laurent}  
\begin{align} \label{eq:ft1}
\Ft_1 \left(  \sum_{k\in \Z} S_k \ket{\delta_{a}}\bra{\delta_{b}}S_k^*  \otimes \ket{\delta_{a'}}\bra{\delta_{b'}} \right) \Ft_1^* = \mathcal{F}S_{a-b} \mathcal{F}^* \otimes  \ket{\delta_{a'}}\bra{\delta_{b'}} = e^{-iq(a-b)} \ket{\delta_{a'}}\bra{\delta_{b'}}. 
\end{align}

The map $I:  L^2(\lbrack 0, 2 \pi \rbrack) \otimes \ell^2(\Z)  \to  L^2\left( \lbrack 0, 2 \pi \rbrack,  \ell^2(\Z) \right) $ defined\footnote{Here $L^2 \left( \lbrack 0, 2 \pi \rbrack, \ell^2(\Z) \right)$ is the space of (equivalence classes of) functions $f: \lbrack 0, 2 \pi \rbrack \to  \ell^2(\Z)$,  $\int_{\lbrack 0, 2\pi \rbrack} \norm{f(q)}_{2}^2 dq < \infty $, with inner product  
$
\langle f, g \rangle
= \int_{\lbrack 0,2\pi \rbrack} \langle f(q), g(q) \rangle_{ \ell^2(\Z)} dq. 
$} by $I(g \otimes \psi) = \psi_g$ where $\psi_g(q) = g(q) \psi$ for any $q \in \lbrack 0,2\pi\rbrack, g\in L^2 \left( \lbrack 0, 2 \pi\rbrack \right), \psi\in \ell^2(\Z)$ is an isometric isomorphism. In total, we have the isometric isomorphisms, 
\begin{align}\label{eq:lemma:isomorphisms}
\HS(\ell^2(\Z)) \overset{\vectorize}{\cong} \ell^2(\Z) \otimes \ell^2(\Z)  \overset{C}{\cong} \ell^2(\Z) \otimes \ell^2(\Z)  \overset{ \Ft_1}{\cong}  L^2( \lbrack 0, 2 \pi \rbrack) \otimes \ell^2(\Z) \overset{I}{\cong} L^2 \left( \lbrack 0, 2 \pi \rbrack,  \ell^2(\Z) \right) =  \int_{\lbrack 0, 2\pi \rbrack }^\oplus \ell^2(\Z)_q dq. 
\end{align}
The composition $\mathcal{J}: \HS(\ell^2(\Z))  \to   \int_{\lbrack 0, 2\pi \rbrack }^\oplus \ell^2(\Z)_q dq $ defined by 
$\mathcal{J}= I \circ \Ft_1 \circ C \circ \vectorize,
$
 is an isometric isomorphism. Conjugation by $\mathcal{J}$, $\Ad_{\mathcal{J}}$, simplifies $\Li$.  A result, which is central to the rest of this paper. 

\begin{theorem} \label{main}
Suppose that  $\Li = \mathcal{T} + \mathcal{F}$ is of Lindblad form \eqref{Lindblad_form}, satisfies \eqref{eq:finite-range}, \eqref{eq:translation-invariance} and is decomposed into non-hermitian evolution and quantum jump part as in \eqref{eq:Lindblad_decomposition}. 
Using $\mathcal{J}$ defined by \eqref{eq:lemma:isomorphisms},  $\Li$ is unitarily equivalent to
$$
\Ad_{\mathcal{J}}(\Li ) = \int_{\lbrack 0, 2 \pi \rbrack}^{\oplus} T(q) + F(q) dq,
$$
with $T(q)$  a bi-infinite $r$-diagonal Laurent operator and $F(q)$ a finite-rank operator with finite-range for each $q \in \lbrack 0, 2 \pi \rbrack$. Moreover, the non-hermitian evolution $\mathcal{T}$ is unitarily equivalent to $
\Ad_{\mathcal{J}}(\mathcal{T}) = \int_{\lbrack 0, 2 \pi \rbrack}^{\oplus} T(q) dq.
$
If $h_l \in \C$ is defined through the relation $H_{\eff} =   H  - \frac{iG}{2}\sum_{k} L_k^* L_k =  \sum_{l=-r}^r h_l S_l $, where $S_l$ is the shift operator,  then
\begin{align}\label{eq:T(q)_in_main}
 T(q)  =  -i \sum_{l=-r}^r h_l   e^{iql} S_l^* + i\sum_{l=-r}^r \overline{h_l}  S_l. 
\end{align}
If $L_0 =  \ket{\phi}\bra{\psi} $ is rank-one (with  $\alpha_l, \beta_l \in \C$, such that $\ket{\phi}= \sum_{l}\alpha_l \ket{\delta_{l}}$ and $ \ket{\psi}= \sum_{l} \beta_l \ket{\delta_{l}}$), then so $F(q)$ and 
\begin{align}\label{eq:main_theorem_rank_one_case}
F(q) = G  \left( \sum_{l_1,l_2}\alpha_{l_1} e^{iql_1}\overline{\alpha_{l_2}}\ket{\delta_{l_2- l_1}}  \right)  \left( \sum_{l_1',l_2'}\beta_{l_1'} e^{-iql_1'}\overline{\beta_{l_2'}}\bra{\delta_{l_2'- l_1'}} \right) =\ket{\mathtt{v}(q)} \bra{\mathtt{u}(q)}. 
\end{align}
\end{theorem}
\begin{proof}
Consider the three terms in \eqref{eq:vectorization} separately.  
For the first, by Lemma \ref{lemma:C_relations} ii), Lemma \ref{lemma:transform_Laurent}, and slight notation abuse $q \mapsto e^{-iql}$ by $e^{-iql}$ 
\begin{align*}
 \Ad_{\Ft_1 \circ C}(H_{\eff} \otimes \id ) =   \sum_{l=-r}^r h_l \Ad_{\Ft_1} \left(  \Ad_{C} (S_l \otimes \id) \right) = \sum_{l=-r}^r h_l \Ad_{\Ft \otimes \id}(S_l \otimes S_l^*)= \sum_{l=-r}^r h_l  e^{-iql} \otimes S_l^*.
\end{align*}
Thereby,
$
\Ad_{I \circ \mathcal{F}_1\circ C}( H_{\eff} \otimes \id ) = \sum_{l=-r}^r h_l   e^{-iql} S_l^*. 
$
 Similarly, for the second term by Lemma \ref{lemma:C_relations} i)  and Lemma \ref{lemma:transform_Laurent}, 
\begin{align*}
 \Ad_{\Ft_1 \circ C}(\id \otimes \overline{H_{\eff}}  ) = \sum_{l=-r}^r \overline{h_l} \Ad_{\Ft_1} \left(  \Ad_{C} (\id \otimes S_l) \right)
= \sum_{l=-r}^r \overline{h_l} \Ad_{\Ft_1} \left(  \id \otimes S_l \right) =  \sum_{l=-r}^r \overline{h_l} 1 \otimes S_l, 
\end{align*}
where $1 \in L^2( \lbrack 0, 2\pi \rbrack)$ is the constant function $1$ 
and so  
$
\Ad_{ I \circ \mathcal{F}_1\circ C}(\id \otimes \overline{H_{\eff}}) =\sum_{l=-r}^r \overline{h_l} S_l. 
$

\noindent For the quantum jump terms $\mathcal{F}$,  use that $L_k = S_k L_0 S_k^*$ combined with Lemma \ref{lemma:C_relations} iii):
\begin{align*}
 \Ad_{C} \left( \sum_{k \in \Z} L_k \otimes \overline{L_k} \right)  
 = \sum_{k \in \Z} (S_k \otimes \id)  C (  L_0 \otimes  \overline{L_0})C^* (S_k^* \otimes \id).
\end{align*}
The operator $  C (  L_0 \otimes  \overline{L_0})C^*$ is local, so conjugation by $ \mathcal{F}S_k \otimes \id $ is a local operator ($q$-dependent matrix) by \eqref{eq:ft1}.

\noindent For the explicit form \eqref{eq:main_theorem_rank_one_case} in the rank-one case,  combine the following calculation with \eqref{eq:ft1}

 $ \displaystyle
 \Ad_{C}(L_0 \otimes  \overline{L_0}) = \Ad_{C}\left( \sum_{l_1, l_2}\alpha_{l_1}\overline{\alpha_{l_2}} \ket{\delta_{l_1, l_2}}  \sum_{l_1', l_2'}\beta_{l_1'}\overline{\beta_{l_2}'} \bra{\delta_{l_1', l_2'}} \right) = \sum_{l_1, l_2}\alpha_{l_1}\overline{\alpha_{l_2}} \ket{\delta_{l_1, l_2-l_1}} \sum_{l_1', l_2'}\beta_{l_1'}\overline{\beta_{l_2}'} \bra{\delta_{l_1', l_2'-l_1'}}. 
$ \hfill \qedhere
\end{proof}

\subsection{Spectrum of the full Lindbladian}
With the results at hand, a formula for $\sigma(\Li)$ is obtained. 
\begin{corollary}\label{cor:general_rank}
Let $\mathcal{L} = \mathcal{T}+ \mathcal{F} \in \mathcal{B}(\HS(\Hi))$ be a Lindbladian of the form \eqref{Lindblad_form} satisfying  \eqref{eq:finite-range} and \eqref{eq:translation-invariance}. Let $T(q)$ and $F(q)$ be as in Theorem \ref{main}. Then $
\sigma(\mathcal{T}) = \bigcup_{q\in \lbrack 0, 2 \pi \rbrack}\sigma(T(q)), \text{ }\sigma(\mathcal{T}) \subset \sigma(\Li)
$ and 
\begin{align}
\sigma(\Li)=\bigcup_{q\in \lbrack 0, 2 \pi \rbrack}\hspace{-0.7em} \sigma( T(q) + F(q)).
\end{align}
\end{corollary}
\begin{proof}
The two identities follow from Theorem \ref{main} and Theorem \ref{direct_continuity} since \eqref{eq:translation-invariance} and \eqref{eq:finite-range} imply norm continuity since the operators $T(q)$ and $F(q)$ are respectively $2r$-diagonal and finite range with coefficients that are polynomials in $\{ e^{iq}, e^{-iq} \}$ (see Lemma \ref{lemma:trivial_norm_continuity} for completeness). 
To prove the inclusion, note that by Theorem \ref{main}, $F(q)$ is finite rank for each $q$ and the essential spectrum is invariant under finite rank perturbations \cite[IV.5.35]{kato2013perturbation}, 
\begin{align*}
\sigma(T(q)) = \sigma_{\ess}(T(q))  =  \sigma_{\ess}(T(q)+F(q)) \subset \sigma(T(q)+F(q))  \subset \sigma(\Li), 
\end{align*}
since translation invariance of $T(q)$ implies its spectrum is essential (cf. Corollary \ref{range_of_symbol}). 
\end{proof}
In the applications in \Cref{sec:examples},  each $L_k$ is rank-one, so by Theorem \ref{main}, $F(q) =   \ket{\mathtt{v}(q)}\bra{\mathtt{u}(q)}$ defined in \eqref{eq:main_theorem_rank_one_case}
is also rank-one. In that case, the description $\sigma(\Li)$ is even more explicit.
\begin{corollary}\label{cor:rank_one}
Let $\mathcal{L} \in \mathcal{B}(\HS(\Hi))$ be a Lindbladian of the form \eqref{Lindblad_form} satisfying \eqref{eq:finite-range}, \eqref{eq:translation-invariance} and \eqref{eq:Rank1}. Let $T(q)$ and $F(q)= \ket{\mathtt{v}(q)}\bra{\mathtt{u}(q)}$ be as in Theorem \ref{main}. 
Then
\begin{align}
\sigma(\Li) = \bigcup_{q\in \lbrack 0, 2 \pi \rbrack}\hspace{-0.7em}\sigma( T(q))   \cup  \left\{\lambda \in \C \mid   \bra{\mathtt{u}(q)},(T(q)-\lambda)^{-1} \ket{\mathtt{v}(q)}   = -1  \right\}.
\end{align}
\end{corollary}
\begin{proof}
By the decomposition in Theorem \ref{direct_continuity} it suffices to prove for each $q \in \lbrack 0, 2\pi\rbrack$ that
\begin{align*}
\sigma( T(q) + F(q)) = \sigma( T(q))   \cup  \left\{\lambda \in \C \mid   \bra{\mathtt{u}(q)},(T(q)-\lambda)^{-1} \ket{\mathtt{v}(q)}= -1  \right\}.
\end{align*}
This is known as rank-one update. For completeness, we give a proof in Appendix \ref{sec:rank_one_pert}.
\end{proof}

\section{General applications of the direct integral decomposition}\label{sec:general_app}
Before continuing with the concrete applications in the next section, we discuss some more abstract consequences of the results presented in the previous section.

\subsection{Gaplessness of translation-covariant Lindblad generators}

Say that $\Li$ has a gap if $0$ is an isolated point in the spectrum, i.e. there exist a ball $B_r(0)$ of radius $r > 0$ such that $\sigma(\Li)\cap B_r(0) = \{0\}$. Otherwise, $\Li$ is \emph{gapless}. 

For some $r \in \N$ let $T(q)$ be an $r$-diagonal Laurent operator with  smooth  diagonals $a_i: S^1 \to \C$. Let $\varrho(A) = \C \setminus \sigma(A)$ denote the resolvent set of $A$. On $\varrho(T(q))$, resolvents and their matrix elements 
\begin{align}\label{eq:resolvent_definition}
R^k_q(z) = \bra{\delta_{0}},(T(q) -z)^{-1} \ket{\delta_{k}}
\end{align}
are holomorphic (proven in \cite[Theorem III-6.7]{kato2013perturbation}).
The following continuity argument is relegated to Appendix \ref{sec:resolvent_norm_estimates}. 
\begin{lemma}\label{lemma:assumptions_satisfied}
Let $I \subset \R$ be a compact set, and $\{T(q)\}_{q\in I}$ be a norm-continuous family of bounded operators. Suppose that $V \subset \cap_{q \in I} \rho(T(q))$ is compact. Then $\sup_{z \in V, q \in I}\norm{(T(q)-z)^{-1}} \leq C < \infty$. 
\end{lemma}

The next lemma proving local uniform convergence follows from the resolvent equation,  continuity and $r$-diagonality.  
\begin{lemma} \label{lemma:balls}
Let $q_n \to q_0$ be a sequence in $S^1$ and  suppose $V \subset \varrho(T(q_n))\cap\varrho(T(q_0)) $ for all $n\in\N$ and is open. Assume that $\sup_{z \in V, n \in \N}\norm{(T(q_n)-z)^{-1}} \leq C < \infty$ and $\sup_{z \in V}\norm{(T(q_0)-z)^{-1}} \leq C < \infty$ for some $C > 0$. 
Then on $V$,
$$
R^k_{q_n}( \cdot) \overset{(\lcu)}{\longrightarrow} R^k_{q_0}(\cdot).
$$

\end{lemma}
\begin{proof} 
Take any compact set $K \subset V$. Then by the resolvent equation, 
\begin{align*}
\abs{R^k_{q_n}(z) - R^k_{q_0}(z)} & = \abs{\bra{\delta_{0}}, (T(q_n)-z)^{-1}- (T(q_0)-z)^{-1}\ket{\delta_{k}}}
  \leq \norm{(T(q_n)-z)^{-1} - (T(q_0)-z)^{-1}} \\
 &\leq  \norm{(T(q_n)-z)^{-1}}\norm{T(q_n) - T(q_0)}\norm{(T(q_0)-z)^{-1}}
   \leq  C^2 \norm{T(q_n) - T(q_0)} \to 0, 
\end{align*}
since $a_i$ are all smooth functions $\norm{T(q_n) - T(q_0)} \to 0$ whenever $q_n \to q_0$ (cf. Lemma \ref{lemma:trivial_norm_continuity}). 
\end{proof}
\begin{theorem} \label{thm:no_gap}
If $\Li$ is \eqref{eq:finite-range}, \eqref{eq:translation-invariance},  \eqref{eq:Rank1} and $0 \in \sigma(\Li)$.
Then $\Li$ is gapless or $\dim(\ker(\Li))=\infty.$
\end{theorem}
\begin{proof}
 If (as is the case in Section \ref{non_normal_dissipators}) the non-Hermitian evolution is gapless we are done by Corollary \ref{cor:general_rank}. So suppose that the non-Hermitian evolution is gapped around 0. 
Since $R^k_{q}(\cdot)$ is holomorphic on $\varrho(T(q))$ and so on the complement of $\bigcup_{q\in \lbrack 0, 2\pi \rbrack} \sigma(T(q)) = \sigma( \mathcal{T}),$  $R^k_q$ are holomorphic for each $q\in\lbrack0, 2\pi \rbrack$. Thus, if $0 \not \in \sigma( \mathcal{T})$  then $\varrho(\mathcal{T})$ is an open set containing $0$ and thus $B_{2r}(0) \subset \varrho(\mathcal{T})$ for some $r>0$. Now, let $V = \overline{B_r(0)}$ be the ball around $0$ with radius $r$.

So for any sequence $q_n \to q_0$ if we define
$
f, f_n: V \to \C
$
by 
$$
f_n(z) = \bra{\mathtt{u}(q_n)},(T(q_n)-z)^{-1} \ket{\mathtt{v}(q_n)}+1
\text{, }
f(z) = \bra{\mathtt{u}(q_0)},(T(q_0)-z)^{-1} \ket{\mathtt{v}(q_0)}+1, 
$$
then since the vectors $\bra{\mathtt{u}(q_0)}$ and $ \ket{\mathtt{v}(q_0)}$ have only finitely many (smooth) non-zero entries (and local uniform converge is preserved under multiplying by smooth functions) 
conclude that 
$
\{f_n \}_{n\in\N}\overset{(\lcu)}{\to}  f 
$
on $V$. 

Since $0 \in \sigma(\Li) \setminus \sigma(\mathcal{T})$ by Corollary \ref{cor:rank_one} there exists a $q_0 \in \lbrack 0,2\pi \rbrack$ such that $f(q_0) = 0$. As $f$ is analytic and $f$ is not identically zero, every zero of $f$ is isolated. 
Since $\{f_n \}_{n\in\N}\overset{(\lcu)}{\to}  f$ by Hurwitz's theorem for every $a > 0$ there is a sufficiently large $n$ such that $f_n$ has a zero $z_n$ with $\abs{z_n}  < a$. 

Let $I$ be the set of $q\in [0,2\pi]$ such that $0 \not \in \sigma(T(q) + F(q))$:  
$$
I = \left \{ q \in  \lbrack 0, 2\pi \rbrack  \mid  \bra{\mathtt{u}(q)},T(q)^{-1} \ket{\mathtt{v}(q)} + 1\neq 0 \right \} \subset \lbrack 0, 2\pi \rbrack. 
$$
Either there is a sequence $\{q_n\}_{n \in \N} \subset I$ such that $q_n \downarrow q_0 $ or there exists an $\varepsilon > 0$ such that $(q_0 - \varepsilon, q_0+ \varepsilon) \cap   \lbrack 0, 2\pi \rbrack  \subset I^c$.

In the first case,  $f_n(z_n) =0$ is equivalent to $ \bra{\mathtt{u}(q_n)},(T(q_n)-z_n)^{-1} \ket{\mathtt{v}(q_n)}= -1$, which again means that $z_n \in \sigma(\Li)$.  
As  $\{q_n\}_{n \in \N} \subset I$ then $z_n$ converges to $0$ without being equal to zero, $\Li$ is gapless. 

In the second case, for every $q\in (q_0 - \varepsilon, q_0+ \varepsilon) \cap   \lbrack 0, 2\pi \rbrack  \subset I$ there is a normalized vector $\ket{v(q)} \in \ell^2( \Z) $ such that 
\begin{align}\label{eq:is0}
(T(q) + F(q)) \ket{v(q)} = 0. 
\end{align}

Now, split $(q_0 - \varepsilon, q_0+ \varepsilon) \cap   \lbrack 0, 2\pi \rbrack  \subset I$  up into $N$ disjoint intervals $I_1, \dots I_N$ and define for $1 \leq i \leq N$, 
$$
w_i =  \frac{1}{ \sqrt{\abs{I_i}}} \int_{ \lbrack 0, 2\pi \rbrack }^{\oplus} v(q) \id_{q \in I_i}dq. 
$$
Then 
$
\langle w_i,  w_j \rangle  = \delta_{i,j} 
$
and combining \Cref{main} and \eqref{eq:is0} means that $
\Li w_i =0$ for each $1 \leq i \leq N$. 
Thus, the kernel of $\Li$ is at least $N$ dimensional for any $N \in \N$ and hence infinite-dimensional. 
\end{proof}

\subsection{Approximate point spectrum of Lindblad operators}
For an operator $T$ and $\lambda \in \sigma(T)$, a \emph{Weyl sequence} $\{v_n \}_{n \in \N}$ satisfies $\norm{v_{n}} = 1$ and $ \norm{(T- \lambda)v_{n}}\to 0 $ as  $ n \to \infty$. 
The approximate point spectrum $\sigma_{\appt}(T)$ is the set of $\lambda \in \sigma(T)$, which admits a Weyl sequence. $\sigma(T)\setminus \sigma_{\appt}(T)$ is the residual spectrum.   For normal operators, the residual spectrum is always empty \cite[Lemma 12.11]{einsiedler2017functional} that is also true for the class of models we consider. 
For completeness, note that translation covariance  \eqref{eq:translation_covariant} implies that the entire spectrum is essential.
\begin{theorem} \label{thm:approx_point}
If $\Li \in \mathcal{B}(\HS( \ell^2(\Z)))$ of form \eqref{Lindblad_form}  satisfies \eqref{eq:finite-range}, \eqref{eq:translation-invariance}, \eqref{eq:Rank1}, then 
$
\sigma_{\appt}(\Li) = \sigma( \Li).
$
\end{theorem}
\begin{proof}
As in Theorem \ref{main} write $\Li = \mathcal{T} + \mathcal{F}$, with $\mathcal{T}  =  \int_{\lbrack 0,2\pi \rbrack}^{\oplus}T(q)  dq $ and $\mathcal{F}  =  \int_{\lbrack 0,2\pi \rbrack}^{\oplus}F(q)  dq $. \\
 \emph{Step 1: Direct integrals of Laurent operators only have approximate point spectrum.}
  Let construct a Weyl sequence for $\lambda \in  \sigma(\mathcal{T}) = \bigcup_{q\in \lbrack 0, 2 \pi \rbrack}\sigma(T(q)) $.  
Find $q_0$ such that $ \lambda \in  \sigma(T(q_0))$. Since Laurent operators are normal, there is a Weyl sequence $v_n$ corresponding to $\lambda$ for the operator $T(q_0)$. Now, define the normalized $w_n =\int_{\lbrack 0, 2 \pi \rbrack}^{ \oplus} v_n \sqrt{n} \id_{ \lbrack q_0 - \frac{1}{2n}, q_0 + \frac{1}{2n} \rbrack}dq$. 
Using $\norm{ (T(q) - T(q_0)) } \to 0 $  as $ q \to q_0$  (cf.  Lemma \ref{lemma:trivial_norm_continuity}) as well as $ \norm{ (T(q_0) - \lambda)  v_n } \to 0$ as $n \to \infty$,
\begin{align*}
\norm{\int_{ \lbrack 0, 2 \pi \rbrack}^{ \oplus} (T(q) - \lambda) dq w_n }^2 &= \int_{q_0 - \frac{1}{2n}}^{q_0 + \frac{1}{2n}} \hspace{-0.3cm} n\norm{(T(q) - \lambda) v_n }^2 dq \leq  \int_{q_0 - \frac{1}{2n}}^{q_0 + \frac{1}{2n}} \hspace{-0.3cm} n \left( \norm{ (T(q) - T(q_0))  v_n } + \norm{ (T(q_0) - \lambda)  v_n } \right)^2 dq \to 0. \end{align*}

 \emph{Step 2: Direct integrals of Laurent operators with finite-range perturbations only have approximate point spectrum.}
 Assume that $\lambda \in \sigma(\Li)$. Then we split up into cases according to whether $\lambda \in \sigma(\mathcal{T})$

  \emph{2a) Spectrum of the non-Hermitian evolution is approximate point:} Suppose first that $\lambda \in \sigma(\mathcal{T})$.
 Then translation-invariance of $T(q)$ means that $ S_1 T(q) S_{-1} = T(q)$. Consider a Weyl sequence $v_n$ for $T(q)$. Then $\{ S_a v_n \}_{n \in \N} $ is also a Weyl sequence for $T(q)$ corresponding to $\lambda$ for any $a \in \Z$ since
\begin{align*}
\norm{ T(q) S_a v_n} = \norm{ S_a T(q) S_{-a} S_a v_n} = \norm{ S_a T(q)  v_n} =\norm{T(q)  v_n} \to 0. 
\end{align*}
Since $F(q)$ is finite range and $v_n$ is finite norm for every $\varepsilon > 0$ there exist an $a \in \Z$ such that $ \norm{ F(q) S_a v_n} \leq \varepsilon$. Thus, there exists a sequence $a_n$ such that $ \norm{F(q) S_{a_n} v_n} \leq \frac{1}{n}$.
Now,  $w_n = S_{a_n} v_n$ is a Weyl sequence for $T(q) + F$ since 
\begin{align*}
\norm{ ( T(q) + F) S_n v_n} \leq  \norm{T(q)  v_n} +  \norm{F S_n v_n}  \to 0. 
\end{align*}
By an argument as in Step 1, we can use continuity to conclude that this eigenvector in the fiber gives rise to an approximate eigenvector in the direct integral.

  \emph{2b) Additional spectrum from quantum jump terms is approximate point :}
Suppose that $\lambda \not \in \sigma(\mathcal{T}) $ then using the \eqref{eq:Rank1} assumption by Corollary \ref{cor:rank_one} there is a $q$ such that  $ \bra{\mathtt{u}}(T(q)-\lambda)^{-1} \ket{\mathtt{v}}  = -1$ and so 
 \begin{align*}
 \left( T(q) - \lambda  + \ket{\mathtt{v}}\bra{ \mathtt{u}} \right) (T(q)- \lambda)^{-1}\ket{\mathtt{v}}  = \ket{\mathtt{v}}+ \bra{ \mathtt{u}} (T(q)- \lambda)^{-1}\ket{\mathtt{v}}   \ket{\mathtt{v}}= 0, 
  \end{align*}
meaning that $ (T(q)- \lambda)^{-1}\ket{\mathtt{v}}  $ is an eigenvector of $(T(q) +   \ket{\mathtt{v}}\bra{ \mathtt{u}})$ with eigenvalue $\lambda$. Again, as in Step 1, use continuity to conclude that this eigenvector in the fiber gives rise to an approximate eigenvector in the direct integral. \qedhere
\end{proof}
Say that a Weyl sequence  $\{\rho_n\}_{n \in \N} \subset \HS( \Hi)$ is \emph{approximately classical} if (in the position basis $\{\ket{\delta_{k}}\}_{k\in\Z}$) there exist $C,\mu > 0$ such that for all $n$ sufficiently large and $x,y \in \Z$,
$
\abs{ \bra{x},\rho_n \ket{\delta_y}} \leq C e^{- \mu \abs{x-y}}. 
$

\begin{proposition}
Let $\Li \in \mathcal{B}(\HS( \Hi))$ satisfy \eqref{eq:finite-range}, \eqref{eq:translation-invariance} and \eqref{eq:Rank1} and suppose that $\mathcal{T} =  \int_{\lbrack 0,2\pi \rbrack}^{\oplus}T(q) dq$ where $T(q)$ is tridiagonal for each $q\in \lbrack 0,2\pi \rbrack$. 
Any $\lambda \in \sigma(\Li)\slash \sigma(\mathcal{T})$  has an approximately classical Weyl sequence. 
\end{proposition}
\begin{proof} 
By Corollary \ref{cor:rank_one} there is a $q_0$ such that $ (T(q_0) - \lambda)$ is invertible and 
$
\mel{\mathtt{u}}{(T(q_0) - \lambda)^{-1}} {\mathtt{v}}= -1.
$
This implies that
 $(T(q_0) - \lambda)^{-1} \ket{\mathtt{v}}$ is in the kernel of $( (T(q_0) - \lambda)+ \ketbra{\mathtt{v}}{\mathtt{u}}) $ for fixed $q_0$. Let $N = \norm{(T(q_0) - \lambda)^{-1} \ket{\mathtt{v}}}$.  Then, as before, define for each $n \in \mathbb{N}$ the normalized vector $v_n$ by 
$
v_n =  \frac{\sqrt{n}}{N} \int_{\lbrack q_0 - \frac{1}{2n}, q_0 + \frac{1}{2n} \rbrack }^{\oplus}(T(q_0) - \lambda)^{-1} \ket{\mathtt{v}} dq.
$
$v_n$ is a Weyl sequence for $\int_{ \lbrack 0, 2 \pi \rbrack}^{\oplus} (T(q) + \ket{\mathtt{v}}\bra{\mathtt{u}}) dq $   by norm-continuity of $T(q)$.
Since $(T(q_0) - \lambda)$ is tridiagonal, we can explicitly determine the inverse. Unwinding the Fourier transform.  Back in $\ell^2(\Z) \otimes \ell^2(\Z)$,
 \begin{align*}
v_ n = \frac{\sqrt{n}}{N} \sum_{x,y \in \Z} \int_{q_0 - \frac{1}{2n}}^{q_0 + \frac{1}{2n} }e^{iqx}\bra{\delta_y}(T(q_0) - \lambda)^{-1} \ket{\mathtt{v}}dq  \ket{\delta_x}\ket{\delta_y}.
 \end{align*}
As $\ket{\mathtt{v}}$ is local around 0, consider the matrix elements $\bra{\delta_y}(T(q_0) - \lambda)^{-1} \ket{\delta_j}$, which by Lemma \ref{inverse} are exponentially decaying in $\abs{y}$. 
 Thus,  $v_n$ is an approximately classical Weyl sequence corresponding to $\lambda$ since for some $c>0$,
\begin{equation*}
    \abs{v_ n(x,y)}\leq   \frac{\sqrt{n}}{N} \int_{q_0 - \frac{1}{2n}}^{q_0 + \frac{1}{2n} } \abs{\bra{\delta_y}(T(q_0) - \lambda)^{-1} \ket{\mathtt{v}}}dq \leq e^{-c\abs{y}}. \hfill  \qedhere 
\end{equation*}
\end{proof}
See also Section \ref{improving_upper_bound} for a related result in the case of local dephasing. 
Since $\Li \in \mathcal{B}(\HS(\ell^2(\Z)))$ there are no eigenvectors in the direct integral picture and in particular no steady states! E.g. for dephasing noise if we instead defined $\Li$ on $\mathcal{B} ( \mathcal{B}( \Hi))$ then $\Li( \id ) = 0$ and so the identity would be a steady state. Since $\id $ is neither Hilbert-Schmidt nor trace-class $\Li$ does not have a steady state although $0 \in \sigma(\Li)$.

\subsection{Sufficient conditions for convergence of finite volume spectra to infinite volume spectra} 
The convergence of spectra of finite size approximations of Laurent and Toeplitz matrices to their infinite counterpart, is a topic of central interest in numerical analysis \cite{LTTM}.
For Lindbladians this subject is, to our knowledge, still untouched. In Corollary \ref{cor:general_rank} for the class of models in question, convergence  $\sigma(\Li_N^{\per}) \to \sigma(\Li) $ can be reduced to the better studied
case of Laurent and Toeplitz matrices.

To study convergence of subsets of $\C$ we use the Hausdorff metric which is a measure of distance between subsets $X, Y \subset \C$ defined by
$
d_{H}(X,Y) = \max \left\{\sup_{x \in X}d(x,Y),\sup_{y \in Y}d(X,y)\right\}, 
$
where $d$ is the distance in $\C$. However, we will not use the definition, but only its characterization in  \eqref{eq:thm:Hausdorff} below. 
Following \cite{bottcher2002spectral} for a sequence of non-empty subsets of the complex plane $\{S_n \}_{n\in \N}$, define $\liminf_{n \to \infty} S_n$ as the set of all $\lambda \in \C$ that are limits of a sequence $\{\lambda_n  \}_{n \in \N}$ which satisfies $\lambda_n \in S_n$.
Conversely,  $\limsup_{n \to \infty} S_n$ is defined as all subsequential limits of such sequences  $\{\lambda_n  \}_{n \in \N}$ with $\lambda_n \in S_n$.
If $S$ and each $S_n$ are nonempty compact subsets of $\C$ then\footnote{See \cite[Sections 3.1.1 and 3.1.2]{hagen2000c} or \cite[Section 2.8]{hausdorff1937set} for a proof.}
\begin{align}\label{eq:thm:Hausdorff}
S_n \overset{d_H}{\to} S \text{  } \Leftrightarrow \text{   } \limsup_{n \to \infty} S_n = S = \liminf_{n \to \infty} S_n.
\end{align}

In  \cite{bottcher2002spectral} convergence is proven for periodic boundary conditions for tridiagonal matrices and diagonal perturbations except that it has not been proven that the symbol curve is fully captured by the finite size approximations.
Figure \ref{free_bc} and  \ref{periodic_bc} in the next section show how periodic boundary conditions are essential. 

Recall that $\T_n =  \left \{ \frac{2 \pi k}{n} \mid k = 1, \dots n  \right \}$. Call a pair ($	\{q_n\}_{n \in \N},\{a_n\}_\N$) with $\{q_n\}_{n\in\N} \subset [0,2\pi], \{a_n\}_{n \in \N} \subset \N$ \emph{consistent} if $a_n \to \infty$ and $q_n \in \T_{a_n}$ for every $n \in \N$. In \Cref{sec:finite_systems_def} the finite volume generator is explicitly constructed and its representation as a direct sum of circulant matrices is shown. 

\begin{theorem} \label{thm:convergence_finite}
Suppose that $\Li$ satisfies the conditions of Theorem \ref{main} and let $T(q)$ be the corresponding bi-infinite $r$-diagonal Laurent operator and $F(q)$ the finite-range operator for each $q\in\lbrack0,2\pi\rbrack$ and let $\Li_n^{\per}$ be its corresponding finite volume version with periodic boundary conditions defined in \eqref{eq:Lper}. 
Assume that
\begin{enumerate}[label=\roman*)] 
    \item $q \mapsto \sigma( T(q) + F(q))$ is continuous with respect to the Hausdorff metric.
    \item If $q_n \to q_0$  and $(\{q_n\}_{n\in\N} , \{a_n\}_{n \in \N})$ is consistent, then $ \sigma \left( T_{a_n}^{\per}(q_n) + F_{a_n}(q_n) \right) \to \sigma \left( T(q_0) + F(q_0) \right). $
\end{enumerate}  
Then as $n \to \infty$ 
$$
\sigma(\Li_n^{\per}) \to \sigma(\Li).
$$
\end{theorem}
\begin{proof}
We use \eqref{eq:thm:Hausdorff} and compactness of spectra  repeatedly. 
To prove that  $\limsup_{N \to \infty}\sigma(\Li_N^{\per}) \subset \sigma(\Li)$, let $\lambda_n \in \sigma( \Li_{a_n}^{\per})$ and suppose that $\lambda_n \to \lambda $. By Theorem \ref{thm:finite_main},
$
\sigma(\Li_{a_n}^{\per}) = \bigcup_{q\in \T_{a_n}} \sigma \left( T_{a_n}^{\per}(q) + F_{a_n}(q) \right), 
$
so there is a sequence $q_n \in  \T_{a_n}$ such that $\lambda_n \in  \sigma \left( T_{a_n}^{\per}(q_n) + F_{a_n}(q_n) \right)$. 
By compactness of $\T$, for some $q_0$ on subsequence $q_n \to q_0$ and by ii),
$$
 \sigma \left( T_{a_n}^{\per}(q_n) + F_{a_n}(q_n) \right) \to  \sigma \left( T(q_0) + F(q_0) \right)\subset \sigma(\Li). 
$$

To prove $\sigma(\Li) \subset \liminf_{N \to \infty} \sigma(\Li_N^{\per})$, let $\lambda\in \sigma(\Li) $. By Corollary \ref{cor:general_rank}, $\lambda \in  \sigma( T(q_0) + F(q_0))$ for some $q_0 \in \lbrack 0, 2\pi \rbrack$. Find a sequence $q_n \in \bigcup_{N \in \N}\T_N $ such that $q_n \to q_0$.
By i), there is a sequence $\lambda_n \in \sigma( T(q_n) + F(q_n))$ such that $\lambda_n \to \lambda$.

For each $q_n$ there is a sequence $\{a_m \}_{m \in \N} \subset  \T_{a_m}$ such that $a_m \to \infty $ and $q_n \in \T_{a_m} $ for each $m \in \N$. By ii) 
there is a sequence $\lambda_n^m \in \sigma \left( T_{a_m}^{\per}(q_n) + F_{a_m}(q_n) \right)$ such that $\lambda_n^m \to \lambda_n$ as $m \to \infty$. For each $n$ find $k(n) \in \N$ such that
$\abs{\lambda_n^{k(n)}- \lambda_n} \leq \frac{1}{n}$  and $\lambda_n^{k(n)} \in \sigma \left( T_{a_{k(n)}}^{\per}(q_n) + F_{a_{k(n)}}(q_n) \right)$. As $n \to \infty$, 
\begin{equation*}
    \abs{  \lambda_n^{k(n)}- \lambda}\leq \abs{  \lambda_n^{k(n)} - \lambda_n} +  \abs{\lambda_n - \lambda} \to 0. \hfill \qedhere 
\end{equation*}
\end{proof}

\section{Spectra of examples of translation-covariant Lindbladians} \label{sec:examples}\label{applications}
In this section, the decomposition in \Cref{sec:translation_invariance} is used to determine the spectra of Lindbladians  of open quantum systems studied in the literature such as local dephasing and incoherent hopping.  $\mathcal{L}$ satisfies  \eqref{eq:finite-range},\eqref{eq:translation-invariance}, and \eqref{eq:Rank1} with
\begin{align} \label{eq:laplacian}
H = - \tilde \Delta = - \sum_{k \in \Z}\ket{\delta_k} \bra {\delta_{k+1}}+\ket{\delta_{k+1}} \bra{\delta_k}.
\end{align}
Throughout the overall strategy is the same: Using Theorem \ref{main} rewrite $\Li$ as a direct integral of the operators $T(q) +F(q)$ where $T(q)$ is a banded Laurent operator and $F(q)$ is finite range and rank 1.
By Corollary \ref{cor:general_rank} the spectrum of $\Li$ is the union of the spectra of $T(q) +F(q)$ and $F(q) =  \ket{\mathtt{v}(q)}\bra{\mathtt{u}(q)}$ defined in \eqref{eq:main_theorem_rank_one_case}. 
By Corollary \ref{cor:rank_one} find $\sigma(\Li)$ as the union of $\bigcup_{q \in \lbrack 0,2 \pi \rbrack} \sigma(T(q))$ with all solutions $z \in \C$  to the equation
\begin{align}\label{eq:solving_equation}
 \bra{\mathtt{u}(q)},(T(q)-z)^{-1} \ket{\mathtt{v}(q)} = -1.
\end{align}
Since $T(q)$ is translation-invariant, its spectrum is the image of the symbol curve (see \Cref{sec:symbol} for an introduction), which is easy to calculate. However, solving \eqref{eq:solving_equation} can be harder as it involves inverting $T(q)-z$. But as most $T(q)$ considered are tridiagonal the inverse can be (almost) explicitly calculated, see \Cref{sec:invertibility_tridiagonal}.
If $T(q)$ is tridiagonal with $\alpha, \beta, \gamma: \lbrack 0, 2 \pi \rbrack \to \C $ on the diagonals. Then the symbol curve is, by Lemma \ref{lemma:transform_Laurent}, the (possibly degenerate) ellipse,  
\begin{align*}
a(z)=  \alpha z^{-1} + \beta + \gamma z,  z \in \mathbb{T}.
\end{align*}


\subsection{Local dephasing}\label{sec:dephasing}
A simple open system, which has been discussed in the physics literature is local hopping with local dephasing. The spectrum was investigated numerically with free boundary conditions in \cite{Znidaric2015RelaxationTO} and analytically many of the same considerations were made in finite volume with periodic boundary conditions in \cite{Esposito2005EmergenceOD,Esposito2005ExactlySM}. We plot some numerical results in finite volume in Figure \ref{dephasing} and determine the spectrum explicitly: 
  \begin{SCfigure}[0.6][ht] 
  \centering
  \includegraphics[scale =0.5]{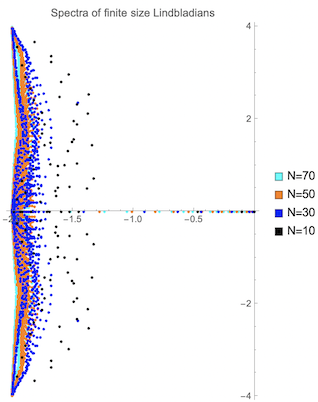}
    \caption{ Spectrum of the Lindbladian corresponding to local dephasing (see \eqref{eq:local_dephasing})  restricted to $N=10, 30, 50, 70$ lattice sites for $G=2$. The spectrum in the infinite volume limit is calculated in \Cref{theorem:dephasing_spectrum}.
    Larger values of $N$ plotted first and they therefore appear behind smaller values of $N$. Notice how it seems that most of the spectrum converges to  $-2 + i \lbrack -4,4 \rbrack$ as $N$ gets larger, but that we also see some spectrum in the interval $\lbrack -2,0\rbrack$.  \label{dephasing}}
\end{SCfigure}

\begin{theorem}[Full spectrum for local dephasing] \label{theorem:dephasing_spectrum}
Let $\Li \in \mathcal{B}(\HS(\ell^2(\Z)))$ of the form \eqref{Lindblad_form} be defined by $H = - \tilde \Delta$
and local dephasing Lindblad operators,  $\label{eq:local_dephasing}
L_k=\ketbra{\delta_k}{\delta_k},  
$ for each $k \in \Z$. 
Then 
\begin{align*}
    \sigma( \mathcal{L} ) =  \left( - G +  i  \lbrack -4,  4 \rbrack \right)  \cup \begin{cases}
                               \lbrack -G , 0\rbrack, & \mbox{if } G \leq 4. \\
                              \lbrack -G + \sqrt{G^2-16} , 0\rbrack, & \mbox{otherwise}.
                            \end{cases} \;
\end{align*}
\end{theorem}
\begin{proof}
The Lindbladian $\Li$ satisfies \eqref{eq:translation_covariant}, \eqref{eq:finite-range} and \eqref{eq:Rank1}. 
Since  $\sum_{k}L_k^* L_k = \id$ by \eqref{eq:T(q)_in_main} in Theorem \ref{main}:
\begin{align} \label{ex_form}
  T(q) = \underset{\alpha}{\underbrace{i(1-e^{-iq})}} S + \underset{\gamma}{\underbrace{i(1-e^{iq})}}S^* \underset{\beta}{\underbrace{-G}} \;.
\end{align}
For fixed $q$ this tridiagonal Laurent operator has symbol curve
$T(q,\theta) = i(1-e^{-iq}) e^{i \theta} + i(1-e^{iq}) e^{-i\theta} - G$. By Theorem  \ref{direct_continuity} and  Theorem \ref{range_of_symbol}, combined with the explicit form \eqref{ex_form},  
\begin{align}\label{eq:NHE_dephasing}
\sigma(\mathcal{T}) = \bigcup_{q\in \lbrack 0, 2\pi \rbrack}\hspace{-0.7em}\sigma(T(q)) = \bigcup_{q\in \lbrack 0, 2\pi \rbrack}\hspace{-0.7em} \left \{T(q,\theta)\mid \theta \in \lbrack 0, 2 \pi \rbrack \right \} = \bigcup_{q\in \lbrack 0, 2\pi \rbrack}\hspace{-0.7em} \left \{  - G + 2 i( \cos(\theta) - \cos( \theta -q)) \right \} =  - G +  \lbrack - 4i, 4i \rbrack. 
\end{align}
By Corollary \ref{cor:general_rank}, 
$
 - G +  \lbrack - 4i, 4i \rbrack = \sigma(\mathcal{T})  \subset \sigma(\Li).
 $
From \eqref{eq:main_theorem_rank_one_case} in Theorem  \ref{main}, $F(q) = G \ketbra{0}$, independent of $q$, so we may choose $\ket{\mathtt{v}}= \ket{\mathtt{u}} =  \sqrt{G} \ket{\delta_0}$.
By Corollary \ref{cor:rank_one} additional spectrum is given by solutions to
\begin{align}\label{eq:GT_eq} 
G\bra{\delta_0},(T(q) -z)^{-1}\ket{\delta_0} = -1. 
 \end{align}

Define $\lambda_+$ and $\lambda_-$ by (with a convention for square roots of complex numbers elaborated on in Appendix \ref{sec:invertibility_tridiagonal})
\begin{align} \label{eq:midnat}
\lambda_{\pm} = - \frac{ \beta}{2 \gamma} \pm \sqrt{ \left( \frac{\beta}{2\gamma} \right)^2 -  \frac{\alpha}{\gamma} }.
\end{align}
Let further, $\abs{\lambda_2} \leq \abs{\lambda_1} $ such that $\{\lambda_1, \lambda_2 \} = \{\lambda_+, \lambda_- \}$.
The conditions of Lemma \ref{eig} are satisfied so
$ \abs{ \lambda_2}< 1 < \abs{ \lambda_1} $. Thus, by the formula the matrix elements of the inverse of the Laurent operator in Lemma \ref{inverse}, 
\begin{align} \label{frq}
- 1= (-1)^{ \id \lbrack \abs{\lambda_-}<1<\abs{\lambda_+}\rbrack} \frac{G}{ \sqrt{ (\beta - z) ^2 - 4 \alpha \gamma}}.
\end{align}
Our strategy is to square the equation, solve to find a set of possible $z$ and then reinsert into \eqref{frq} to see which sign is correct. The potential solutions satisfy for $q\neq 0$ (if $q = 0$ then $z=0$ is the only solution to \eqref{eq:GT_eq})
\begin{align}\label{eq:dephasing_squared}
z = \beta \pm \sqrt{ G^2 + 4 \alpha \gamma} = -G  \pm \sqrt{ G^2 + 8( \cos(q) -1) }.
\end{align}

\noindent \textit{In the case $G <  4 $:}  we have  $ - 16+ G^2 < 0$. 
Thus, $$ \bigcup_{q\in\lbrack 0,2 \pi \rbrack}\hspace{-0.7em}\{\sqrt{G^2+  8( \cos(q) -1)}\} = i  \lbrack 0,  \sqrt{16 - G^2} \rbrack  \cup \lbrack 0, G\rbrack.$$
Therefore, the potential values of $z$ are
\begin{align*}
z \in \lbrack - G, 0 \rbrack \cup \lbrack - 2G, -G \rbrack \cup \left(  - G +  i \left  \lbrack - \sqrt{16 - G^2} , \sqrt{16 - G^2}  \right \rbrack  \right).
\end{align*}
Notice that since $  \left  \lbrack - \sqrt{16 - G^2} , \sqrt{16 - G^2}  \right \rbrack  \subset \lbrack - 4i , 4i \rbrack$ this does not give additional spectrum.

It remains to check the sign of the remaining possible solutions. The square root in (\ref{frq}) must yield a (positive) real number and simultaneously $\abs{\lambda_+}< 1 < \abs{\lambda_-}$, so which leaves only the case $0 \leq G^2+  8( \cos(q) -1) \leq G^2 $. Then
\begin{align*}
\beta - z =  \mp_z   \sqrt{ G^2 + 8( \cos(q) -1) }.
\end{align*}
Using $ \mp_z $ and $ \mp_\lambda$ to denote two, on the outset, independent signs we obtain from \eqref{eq:midnat} that
 \begin{align*}
2 \gamma  \lambda_{\pm} =  - ( \beta - z)  \pm_\lambda \sqrt{ \left( \beta - z \right)^2 - 4 \alpha \gamma} = \pm_z   \sqrt{ G^2 + 8( \cos(q) -1) }   \pm_\lambda \sqrt{G^2}.
 \end{align*}
If $ \pm_z = +$ then $\abs{\lambda_-}< \abs{\lambda_+}$ and $ \pm_z = - $ then $\abs{\lambda_+}<\abs{\lambda_-}$. Thus, from (\ref{frq}) we see that $ \pm_z = + $ is the only valid solution. Thus, only $z \in \lbrack - G, 0 \rbrack $ are valid solutions.

\textit{In the case $G \geq 4$:} Since $ - 16+ G^2 \geq 0$ there is only one segment
$ \lbrack \sqrt{G^2-16}, G \rbrack$, meaning
\begin{align*}
z \in \lbrack -G + \sqrt{G^2-16}, 0 \rbrack \cup \lbrack  -2 G,  -G -  \sqrt{G^2-16} \rbrack.
\end{align*}
Using a similar argument as above one finds that only the part $ \lbrack -G + \sqrt{G^2-16}, 0 \rbrack$ has the correct sign.\end{proof}

\begin{remark}[Emergence of two timescales] From \Cref{theorem:dephasing_spectrum} note that if  $G$ changes from a value below 4 to a value above 4 the spectrum transitions from being connected to consist of two connected components. The connected component of the spectrum containing $\{0 \} $ shrinks when $G$ increases. This indicates the emergence of two timescales in the dynamics. The first corresponds to fast decay at rate $e^{- t G}$ and second has much slower decay.
On the infinite lattice it is difficult to discuss the density of states, but it is noticed numerically in \cite{Znidaric2015RelaxationTO}, that there are of the order of $L$ eigenvalues on the real axis close to $0$ and $L^2 -L$ eigenvalues with real part close to $-G$. In general, such a phenomenon can be explained from symmetry as in \cite[Appendix B.9]{PhysRevLett.123.234103}. 
\end{remark}

\begin{remark}[Heuristic calculation of the finite-volume spectral gap]
Naively letting $q = \frac{2 \pi}{N} $ in \eqref{eq:dephasing_squared} yields upon approximating $\cos(\frac{2 \pi}{N}) -1 \approx  \frac{1}{2} (\frac{2 \pi}{N})^2 = \frac{2 \pi^2}{N^2} $ and $\sqrt{1+x} \approx 1+ \frac{x}{2}$ for small $x$, 
$$
z \approx -G + \sqrt{ G^2 + 8 \frac{2 \pi^2}{N^2} } \approx - G + G \left(1+ \frac{16 \pi^2}{G^2 N^2} \right) =  \frac{16 \pi^2}{G N^2}.
$$
This formula was also obtained in finite volume with periodic boundary conditions by Znidanic in \cite[(8)]{Znidaric2015RelaxationTO}. 
\end{remark}

\subsection{Non-normal dissipators\label{non_normal_dissipators}}
\begin{figure}
\centering
\begin{minipage}{.5\textwidth}
  \centering
  \includegraphics[scale =0.55]{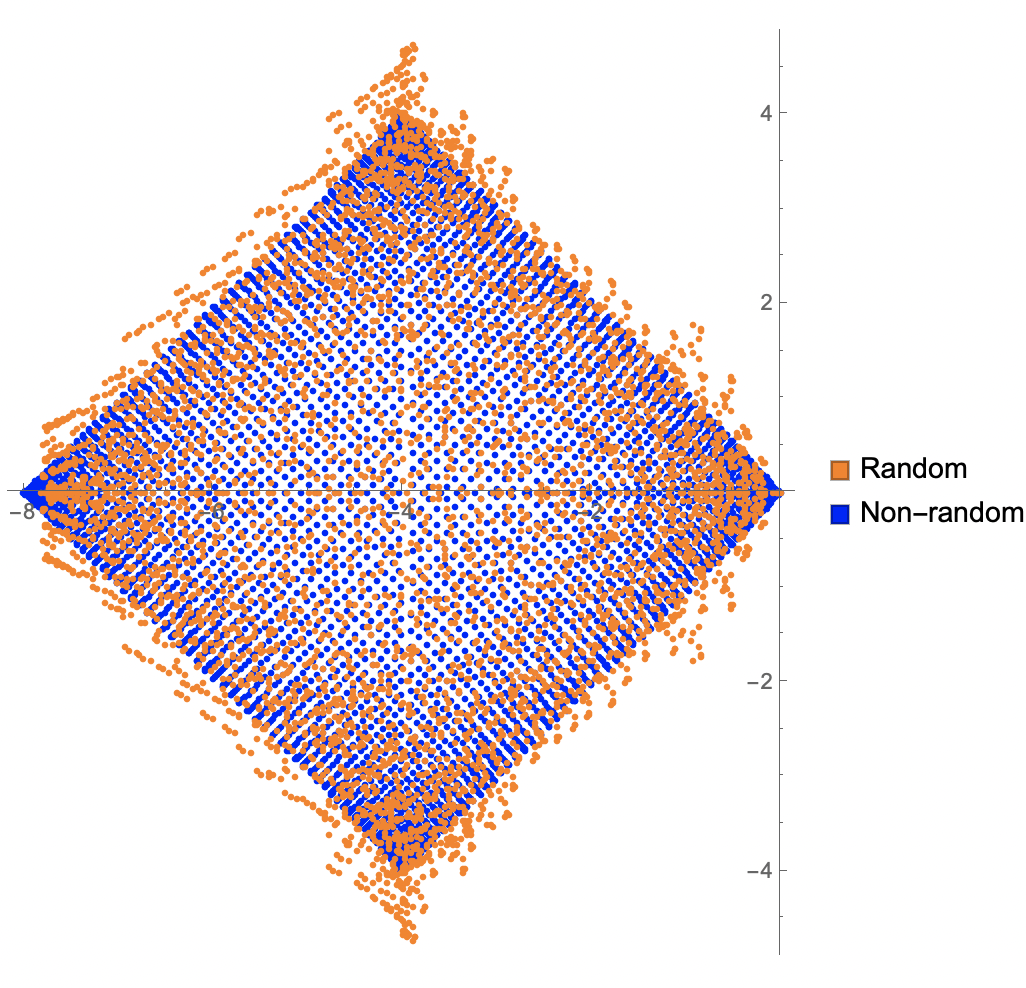}
  \captionof{figure}{Spectrum $\Li$ with non-normal Lindblad operators given by \eqref{eq:non_normal_dissipators} with a random potential in orange and without in blue in the complex plane. In the right picture blue points are plotted on the top and on the left it is the orange points. In this case $G=1$, the lattice size $n = 70$ and the support of the distribution of the strength of the external potential $V = 2$. Notice how the blue points are ordered very regularly except that there is a vertical hole in the middle and those eigenvalues tend to collapse to the real axis. It seems that the main effect of the external field is to push the eigenvalues vertically. }
  \label{fig:test1}
\end{minipage}%
\begin{minipage}{.5\textwidth}
  \centering
  \includegraphics[scale =0.55]{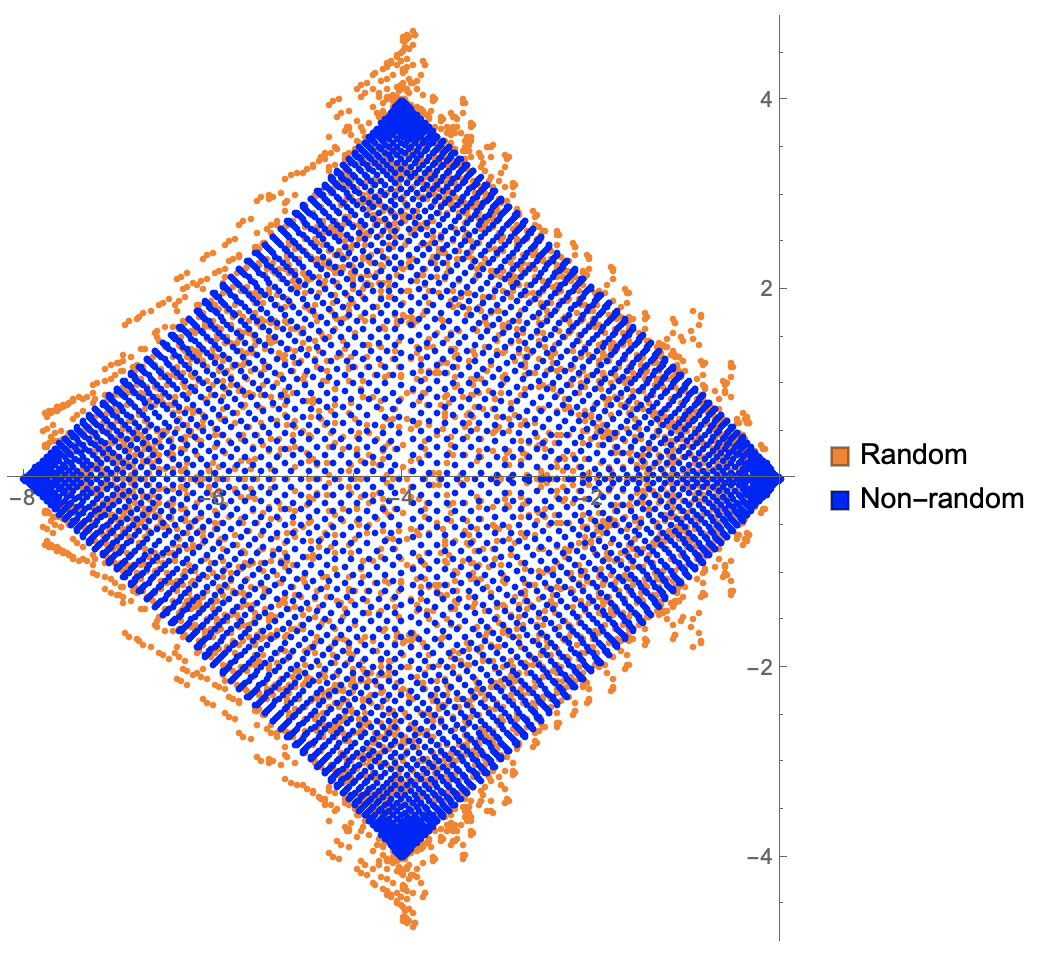}
  \label{fig:test2}
\end{minipage}
\end{figure}

\begin{SCfigure}[0.81][ht] 
  \centering
  \includegraphics[scale =0.35]{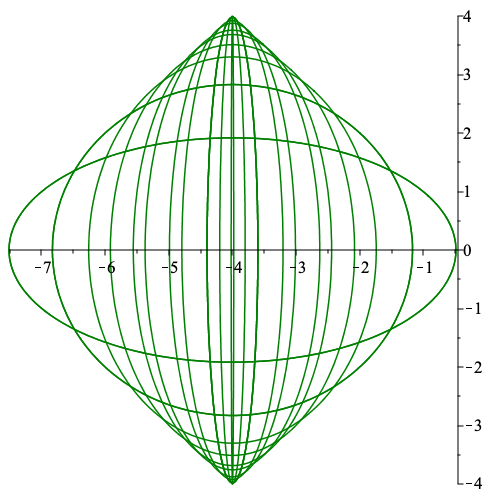}
    \caption{The ellipses corresponding to the spectrum of the Laurent matrices $T(q)$ in the model with non-normal Lindblad operators see for example (\ref{ellipses_make}). For the plot, we chose $\delta = 0$ corresponding to Proposition \eqref{specNHE}. Notice how the union of the ellipses make up the quadrilateral described in \eqref{ellipses_make}, compare also to the shape in Figure \ref{fig:test1}.}
  \label{ellipses}
\end{SCfigure}

The following family of dissipative models was studied in \cite{Diehl, LocalizationinOpenQuantumSystems, vershinina2017control,xu2018interplay}. Let the Hamiltonian $H = - \tilde \Delta$,  the discrete Laplacian from \eqref{eq:laplacian}.
Define Lindblad operators for some $\delta \in \lbrack 0, 2 \pi \rbrack $ and $l \in \N$, 
\begin{align} \label{eq:non_normal_dissipators}
L_k = \left(\ket{\delta_k}+ e^{i \delta}\ket{\delta_{k+l}} \right)\left(\bra{\delta_k}- e^{-i \delta}\bra{\delta_{k+l}}\right). 
\end{align}
Note that $L_k$ is not normal.
Since
$
\sum_{k\in \Z} L_k^* L_k  = 4 \id - 2 (e^{i \delta}S_l + e^{-i\delta}S_{-l}),
$
reading off \eqref{eq:T(q)_in_main} in Theorem \ref{main} yields, 
\begin{align}\label{eq:5_diagonal}
T(q) = G  e^{-i \delta} \left(1 +   e^{i q l}\right)  S_{-l}  +  e^{ i \delta}  \left(    1 +  e^{-ilq} \right) S_l - 4G \id \otimes \id +  i(1-e^{-iq}) S + i(1-e^{iq})S^*. 
\end{align}
The operator $T(q)$ has non-zero entries in 5 diagonals if $l > 1$.  For $l=1$, the case mainly studied in \cite{LocalizationinOpenQuantumSystems} and \cite{Diehl}, $\sigma(\mathcal{T})$ is a square as illustrated in Figure \ref{fig:test1} and \ref{ellipses}. 
\begin{proposition} \label{specNHE}
For $\Li$ defined by  \eqref{eq:non_normal_dissipators} and $H = - \tilde \Delta$, $l=1, \delta=0$ with non-Hermitian evolution $\mathcal{T}$. Then
\begin{align*}
\sigma(\mathcal{T}) = \bigcup_{q, \theta \in \lbrack 0,2 \pi \rbrack }\hspace{-0.7em}\{- 4 G + 2i \cos( \theta) - 2i \cos( q- \theta) + 2G \cos( \delta + \theta) + 2G \cos(q-\delta - \theta)  \}.
\end{align*}
For $\delta = 0$ and $ \delta = \pi $ this set is the convex envelope of the points $-8G , - 4i , 0 , 4i$.
\end{proposition}
\begin{proof}
Since $l=1$, $T(q)$ in \eqref{eq:5_diagonal} reduces to a tridiagonal operator with diagonals
\begin{align*}
\alpha = i( 1- e^{iq}) + G e^{-i \delta}\left(1 +   e^{i q}\right) & & \beta = - 4 G  & & \gamma =  i(1- e^{-iq}) + G e^{ i \delta}\left(1 +   e^{-iq}\right).
\end{align*}
As before, by Corollary \ref{cor:general_rank} and \Cref{range_of_symbol} (spectrum of a Laurent operator is the symbol curve), 
\begin{align} \label{ellipses_make}
\sigma(\mathcal{T}) = \bigcup_{q\in \lbrack 0, 2 \pi \rbrack}\hspace{-0.7em}\sigma(T(q)) = \bigcup_{q\in \lbrack 0, 2 \pi \rbrack}\hspace{-0.7em} \left \{\alpha(q) e^{-i \theta}+ \beta + \gamma(q) e^{i \theta} \mid \theta \in \lbrack 0, 2 \pi \rbrack \right \}.
\end{align}
For $\delta = 0$ and $ \delta =\pi $ it reduces to
\begin{align*}
\sigma(\mathcal{T}) & = \bigcup_{q, \theta \in \lbrack 0,2 \pi \rbrack }\hspace{-0.7em}\{- 4 G +  \cos( \theta) \left(  2i  \pm  2G \right) +   \cos( q- \theta) \left( - 2i \pm 2G  \right)  \} 
 = \text{conv}( -8G , - 4i , 0 , 4i). \hfill \qedhere
\end{align*}
\end{proof}

%
%
Thus, $\mathcal{T}$ is gapless and so is $\Li$ by \Cref{cor:rank_one}. As before, solutions to \eqref{eq:solving_equation}  could increase the spectrum.
The vectors $\ket{\mathtt{v}}$,  $\bra{\mathtt{u}}$ from Theorem \ref{main} have all entries zero except for the $-1, 0, 1$st which in the basis $\{ \ket{-1}, \ket{\delta_0}, \ket{1} \}$  are
\begin{align*}
\ket{\mathtt{v}} =  \sqrt{G} \begin{pmatrix} e^{iq} \\ 1+ e^{iq} \\ 1  \end{pmatrix} \text{ as well as } 
\bra{\mathtt{u}} = \sqrt{G} \begin{pmatrix} - 1 & 1+ e^{-iq} & -e^{-iq} \end{pmatrix}. 
\end{align*}
Letting $\beta = - 4G - z$ and $ \omega = \sqrt{\beta^2 - 16 \alpha \gamma}$ and then from Lemma \ref{inverse} the equation \eqref{eq:solving_equation}  reduces to
\begin{align*}
- \frac{\omega}{G} =  \frac{\bra{\mathtt{u}}T^{-1}\ket{\mathtt{v}}}{G}  
 = \begin{pmatrix} - 1 & 1+ e^{-iq}& -e^{-iq} \end{pmatrix}\begin{pmatrix} 1 & \frac{1}{ \lambda_1} & \frac{1}{ \lambda_1^2} \\ \lambda_2 & 1 & \frac{1}{ \lambda_1} \\ \lambda_2^2 & \lambda_2 & 1  \end{pmatrix} \begin{pmatrix} e^{iq} \\1+ e^{iq}\\ 1\end{pmatrix}
= 2 + 2 \left( \frac{1}{ \lambda_1} - \lambda_2 \right) \sin(q) - \left( \frac{1}{ \lambda_1^2} + \lambda_2^2 \right), 
\end{align*}
where $\lambda_1, \lambda_2$ are defined right after \eqref{eq:midnat}.
This equation in $z$ can be numerically solved for fixed $q$ (and $G$).
Numerical evidence (see finite volume in Figure \ref{fig:test1}) suggests that the curve stays inside the quadrilateral $\conv( -8G, - 4i, 0 , 4i)$:
\begin{conjecture}
For $\Li$ defined by  \eqref{eq:non_normal_dissipators} and $H = - \tilde \Delta$, $l=1, \delta=0$, 
$
\sigma( \Li ) = \sigma(\mathcal{T}) = \conv( -8G , - 4i , 0 , 4i).
$
\end{conjecture}
Independently of the conjecture, both $\Li$ and $\mathcal{T}$ are gapless which may be related to the dynamical behavior in a random  potential observed in  \cite{LocalizationinOpenQuantumSystems} which motivated  \cite{klausen2025decoherence}. Random systems are studied in more detail in \Cref{sec:Lindblad_random}. 

\subsection{Incoherent hopping} \label{sec:incoherent_hopping}
\begin{figure}
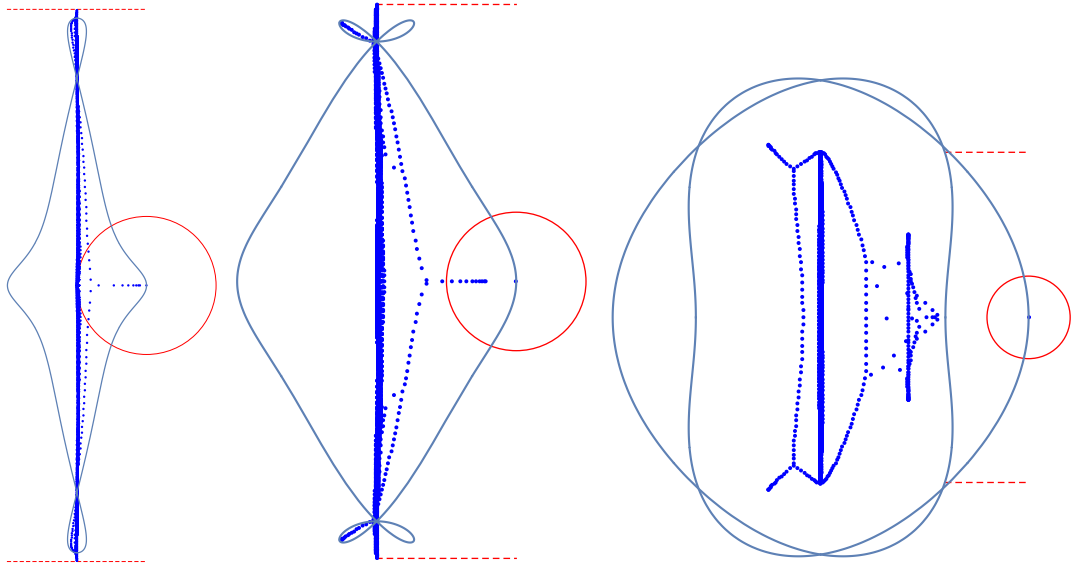

  \centering
    \includegraphics[scale =0.27]{G1N70.png}
      \includegraphics[scale =0.45]{G2N70.png}
            \includegraphics[scale =0.45]{G5N70.png}
    \caption{Exact diagonalization of the Lindbladian $\Li$ in Section \ref{sec:incoherent_hopping} for $l=1$,  $G = 1,2,5 $ and $ N=70$ comparison of the predicted curve with numerics with free boundary conditions. The red circle is the unit circle.   Notice how the two curves do not match due to the non-Hermitian Skin effect. \label{free_bc}}
\end{figure}

Now turn to an incoherent hopping studied numerically in \cite{Znidaric2015RelaxationTO} where $L_k = \ket{\delta_k} \bra{k+l}$. 
 The numerical finding for finite sections with free boundary conditions of the lattice \cite{Znidaric2015RelaxationTO} is that the gap is uniformly positive as the length of the lattice increases, see  Figure \ref{free_bc}. 
 We find the spectrum for periodic boundary conditions. 

\begin{theorem} \label{thm:incoherent_hopping}
Let $l \in \mathbb{Z}$. Define $\Li$ through $H= - \tilde \Delta$ and $L_k =  \ketbra{\delta_k}{\delta_{k+l}}$ then
\begin{align*}
\sigma(\Li ) =   (- G + i\lbrack - 4, 4 \rbrack)\cup \bigcup_{q \in \lbrack 0, 2 \pi \rbrack}\hspace{-0.7em}\left\{  - G \pm_q \sqrt{e^{- 2i ql }  G^2 +  8(\cos(q) - 1)}  \right \}
\end{align*}
where $\pm_q$ is either $+$ or $-$ for each $q\in  \lbrack 0, 2 \pi \rbrack$.
\end{theorem}
\begin{proof}
 Since $L_k^* L_k =  \ketbra{\delta_{k+l}}{\delta_{k+l}}$, $\sum_k L_k^* L_k  = \id$, 
hence the Thm. \ref{main} $T(q)$ is \eqref{ex_form},
and so $- G + i\lbrack - 4, 4 \rbrack   \subset \sigma(\Li)$ by \eqref{eq:NHE_dephasing}.
It remains to solve \eqref{eq:solving_equation}. Inserting 
 $\alpha_r = \delta_{r,0} $ and $\beta_r = \delta_{r,1}$
 in \eqref{eq:main_theorem_rank_one_case} yields $
\ket{\mathtt{v}}  =\sqrt{G} \ket{\delta_0} $ and $
\bra{\mathtt{u}}= \sqrt{G} e^{-i ql} \bra{\delta_0}, 
$
so
\begin{align}\label{with_hop} 
-1 =\bra{\mathtt{u}},\frac{1}{T - z} \ket{\mathtt{v}} = e^{-i ql}  G \bra{\delta_0},\frac{1}{T-z} \ket{\delta_0} =  \frac{ (-1)^{ \id \lbrack \abs{\lambda_-} < 1 < \abs{\lambda_+} \rbrack} e^{-i ql}  G}{ \sqrt{ (\beta -z)^2 - 4 \alpha \gamma}}.
\end{align}
Squaring yields
$
1 =    \frac{ e^{- 2i ql}  G^2}{ (\beta -z)^2 - 4 \alpha \gamma},
$
so solving for $z$ gives
$
 z =
 - G \pm \sqrt{ e^{- 2i ql}  G^2 +  8(\cos(q) - 1)},
$
where we again, as in the proof of \Cref{theorem:dephasing_spectrum},  have to throw some solutions away to get the correct sign in (\ref{with_hop}). \end{proof}

\begin{figure}
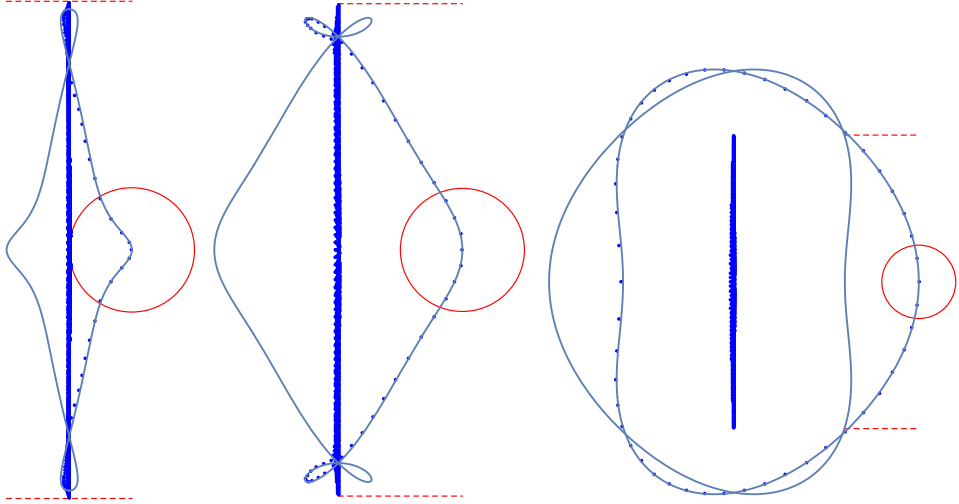

  \centering
    \includegraphics[scale =0.40]{G1N50per.png}
      \includegraphics[scale =0.40]{G2N50per.png}
            \includegraphics[scale =0.40]{G5N50per.png}
    \caption{Exact diagonalization of the Lindbladian corresponding to \Cref{thm:incoherent_hopping} in Section \ref{sec:incoherent_hopping}  for $l=1$,  $G = 1,2,5 $ and $ N=50$ comparison of the predicted curve with numerics for periodic boundary conditions. The red circle is the unit circle.  Notice how the numerics and the analytical spectrum in the infinite volume fit. \label{periodic_bc} }
\end{figure}

 Figure \ref{free_bc} and \ref{periodic_bc} explicitly shows the solutions as a function of $q$ as well as the spectra obtained by exact diagonalization of $\Li$ in finite volume.
For periodic boundary conditions the predicted spectrum fits well with the numerical spectra for finite size systems as is consistent with Theorem \ref{thm:convergence_finite}. However, for open boundary conditions, the picture is dramatically different. This dependence on boundary conditions is sometimes called the non-Hermitian Skin effect \cite{lee2016anomalous, bergholtz2021exceptional, song2019non}, similarly to difference between Toeplitz and Laurent operators. It is a feature of the non-Hermitian skin effect that the spectrum is pushed inwards and real eigenvalues start to appear \cite{okuma2020topological} as seen in Figure \ref{free_bc}.

\subsection{Single particle sector of a quantum exclusion process}
In a many-body setting a model with $ L_k = \ketbra{\delta_k}{\delta_{k+1}}$ and $L_k' = \ketbra{k+1}{k}$ was studied analytically in \cite{Eisler_2011}. Let us derive $\sigma(\Li)$ in the single particle sector: 
 \begin{theorem}[Hopping both ways]
Let $\Li$ be defined by $H = -\tilde \Delta$, and $ L_k = \ket{\delta_k}\bra{\delta_{k+1}}$ and $L_k' = \ket{\delta_{k+1}}\bra{\delta_k}$ then
\begin{align*}
\sigma( \mathcal{L} ) =     \lbrack -2G , 0\rbrack  \cup \{- 2G +  i  \lbrack -4,  4 \rbrack \}.
\end{align*}

\end{theorem}
\begin{proof}
Going through the proof of Theorem \ref{main},the $L_k$ and $L_k'$ give two independent contributions that diagonalize in the same way.
Since $H = -\tilde \Delta$,  $\alpha =  i(1- e^{iq})$ and $\gamma = i( 1- e^{-iq})$.  On the other hand, the sum $\sum_k L_k^* L_k = \id$ appears twice, which means that  $\beta = - 2G$. Therefore  from \eqref{eq:NHE_dephasing},  $\sigma(\mathcal{T}) =  - 2 G + \lbrack - 4i, 4i \rbrack$.
The quantum jump terms are
$
F_1(q) = G\ketbra{\delta_0}{\delta_0} e^{-iq} $
and $
F_2(q) = G \ketbra{\delta_0}{\delta_0} e^{iq}.
$
The total the quantum jump contribution is
\begin{align*}
F(q) = F_1(q) + F_2(q) = 2 G \cos(q) \ketbra{\delta_0}{\delta_0},
\end{align*}
which is still rank-one (and real). 
The deciding equation \eqref{eq:solving_equation} has the same form as \eqref{eq:GT_eq},  
$
-1 
= \pm \frac{ 2 \cos(q) G}{\sqrt{ (\beta-z)^2 - 4 \alpha \gamma}},
$
squaring, solving for $z$ and inserting $\alpha, \beta, \gamma$ yields that
\begin{align*}
z = - 2G \pm \sqrt{4 \cos^2(q) G^2 +   8(\cos(q) - 1)}.
\end{align*}
The function $f(q) = 4 \cos^2(q) G^2 +   8(\cos(q) - 1)$ has extremal points $q$ satisfying
$
 0 = 8 \cos(q) \sin(q) G^2 +   8 \sin(q).
$
If $q \not \in \{ 0, \pi\} $ the equation reduces to 
$
   \cos(q) =  - \frac{1}{G^2}
$
which has a solution if and only if $G \geq 1$.

If $G < 1$ then $\{0, \pi \}$ are the only extremal points. As
$f(0) = 4 G^2 $ and $ f(\pi) = 4 G^2 - 16$ by the intermediate value theorem,  the range of $f$ is
$\lbrack 4 G^2 - 16, 4 G^2 \rbrack  = \lbrack 4 G^2 - 16, 0 \rbrack \cup \lbrack 0, 4 G^2 \rbrack $. 
The first interval corresponds to the segments $\pm i \lbrack 0, 2 \sqrt{4-G^2} \rbrack \subset i\lbrack - 4, 4\rbrack $. The second one to
$\lbrack 0, 2G \rbrack$ analogously to the dephasing case from Section \ref{sec:dephasing}.

For $G \geq 1$ there is in addition the solution   $\cos(q) =  - \frac{1}{G^2}$ which has values $ - \frac{4}{G^2}  - 8$. Since $G \geq 1$,  the potential values of $f$ are extended to the interval $ \lbrack - \frac{4}{G^2}  - 8, 0 \rbrack \cup \lbrack 0, 4 G^2 \rbrack $ which corresponds to solutions $i \left \lbrack - \sqrt{8 + \frac{4}{G^2}}, +\sqrt{8 + \frac{4}{G^2}}  \right \rbrack $ and $ \lbrack 0, 2G \rbrack$ respectively. Yet
$i \left \lbrack - \sqrt{8 + \frac{4}{G^2}}, +\sqrt{8 + \frac{4}{G^2}}  \right \rbrack \subset i\lbrack - 4, 4\rbrack$, so the  spectrum is not extended in that case.
\end{proof}

\section{Outlook and further questions}
\subsection{Open systems with disorder}\label{sec:Lindblad_random}
Although, some results exist   \cite{LocalizationinOpenQuantumSystems, frohlich2016quantum, klausen2025decoherence}, determining the effects of disorder in an open quantum system are is still an open problem. 
In addition, there has recently been interest in random Lindblad systems from the point of view of random matrix theory \cite{PhysRevLett.123.140403, CanRMT}.

Disorder is modeled by a random (Anderson) potential $V = \sum_{n \in \Z} V(n) \ketbra{n}{n}$ added to the Hamiltonian $H$. Suppose here that $(V(n))_{n \in \Z}$ is i.i.d. uniformly distributed potential in some range $\lbrack - \lambda, \lambda \rbrack $ for some $\lambda >0$.
Let $\mathcal{E}_V$, be the action of $V$ in the commutator defined by
$\mathcal{E}_V( \rho ) = - i \lbrack V, \rho \rbrack$.
The numerical range of an operator $A$ is defined by,
$
W(A) = \{ \langle v, A v \rangle \mid \norm{v} = 1 \}.
$
For example $W(\Ei_V ) = - i \lbrack \lambda, \lambda \rbrack$.
The following is a Lindbladian version of the Kunz--Soulliard Theorem \cite{kunz1980spectre} generalized in \cite{kirsch1982spectrum}. Recall that \Cref{thm:approx_point} proved that \eqref{eq:translation-invariance}, \eqref{eq:finite-range} and \eqref{eq:Rank1} implies that the spectrum is approximate point.  

\begin{proposition}
\label{kunz}
Assume $\Li_0$ is a translation-covariant (cf. \eqref{eq:translation_covariant}) operator on $\HS( \ell^2(\Z))$ satisfying $\sigma_{\appt}(\Li_0) = \sigma(\Li_0)$.  Define
$
\Li = \Li_0 + \Ei_V \in \mathcal{B}(\HS( \ell^2(\Z)))$ for a random potential $V$ as above.
Then almost surely, 
	\begin{align*}
	\sigma(\Li_0)  \subset  \sigma(\Li) \subset \overline{ W(\Li_0 ) + W(\Ei_V ) }. 
\end{align*}
\end{proposition} 
\begin{proof}
First by the Toeplitz--Hausdorff Theorem \cite{toeplitz1918algebraische, hausdorff1919wertvorrat} it holds (surely) that
$\sigma(\Li) \subset \overline{W(\Li )} \subset  \overline{W(\Li_0 ) + W(\Ei_V ) }.
$	

For the first inclusion, consider $\lambda \in \sigma_{\appt}(\Li_0)$ and let $\{\rho_n \}_{n \in \N}$ be a  Weyl sequence corresponding to $\lambda$ for $\Li_0$.
Each $\rho_n$ can be assumed compactly supported: $ \rho_n(x,y)   = 0 $ for $(x,y) \not \in \Lambda_{R_n} \subset \Z^2$ for some large but finite box $\Lambda_{R_n} \subset \Z^2$.
Then define the probabilistic event
\begin{align*}
\Omega_n  = \left \{  V \mid \text{ for some  } j_n \in  \Z : \sup_{(x,y) \in \text{supp}(\rho_n) } \abs{ (-i) (V(x+j_n) - V(y+j_n))  } \leq \frac{1}{n}    \right \},
\end{align*}
and notice that each $\Omega_n$  is a set of full measure\footnote{One way to see this is that for any $k\in \N$ there is almost surely an $a \in \Z$ such that $\abs{V(a+j)} \leq \frac{1}{2n}$ for any $j \in \{1, \dots, k\}$.}. Hence $\cap_{n \in \N} \Omega_n $ must have probability 1 as well, so almost surely there is $\{j_n\}_{n \in \N} \subset \Z$ such that  $ \abs{ (-i) (V(x+j_n) - V(y+j_n))  } \leq \frac{1}{n}$ for each $n \in \N$. Define $\{\gamma_n\}_{n\in\N} \subset  \HS(\ell^2(\Z) )$ by $\gamma_n = \rho_n( \cdot - j_n) $, where $ \rho_n( \cdot - j_n) \in \HS(\ell^2(\Z) )$ is defined by $\bra{\delta_x},\rho_n( \cdot - j_n)\ket{\delta_y} = \rho_n(x - j_n, y - j_n)$, i.e. $\rho_n$ shifted by $j_n$ in both coordinates. By translation covariance of $\Li_0$, 
 \begin{align*}
\n{  \left(\Li -  \lambda   \right)(\gamma_n)} \leq \n{ \left(\Li_0 - \lambda \right)(\rho_n)} + \n{  \Ei_V (\gamma_n) }.
 \end{align*}
The first term is small since $\rho_n$ is a Weyl sequence for $\Li_0$.  For the second term, 
\begin{align*}
\n{ \Ei_V  \left(  \rho_n (\cdot - j_n) \right)}^2  
& =  \sum_{x,y \in \supp \rho_n + j_n } 
\abs{\rho_n(x - j_n, y - j_n) \Ei_V (\ket{\delta_x} \bra{\delta_y})}^2
= \sum_{x,y \in \supp \rho_n + j_n } \abs{V(x) - V(y)}^2\abs{\rho_n (x - j_n, y - j_n)}^2\\
& = \sum_{x,y \in \supp \rho_n }\abs{V(x +j_n) - V(y + j_n)}^2\abs{\rho_n (x, y)}^2 \leq \frac{1}{n^2}\sum_{x,y \in \supp \rho_n }\abs{\rho_n (x, y)}^2 \leq \frac{1}{n^2}. \qedhere
\end{align*}
\end{proof}

	
In Figure \ref{fig:test1} suggests that the random field tends to push eigenvalues vertically. 
However, the effect is much stronger in the bulk of the spectrum, whereas close to $\{0 \}$ it does not seem as if the spectrum changes.  A model that describes this phenomenon which is exactly solvable (even without the developed theory) is the following.
\begin{example}\label{sec:exact_solv}
Define $\Li_0$ through $H=0$ and $L_k = \ket{\delta_k}\bra{\delta_k}$. If $i \neq j$, $\Li_0( \ket{\delta_{i}} \bra{\delta_{j}}) = - G  \ket{\delta_{i}} \bra{\delta_{j}}$ and $\mathcal{E}_V(\ketbra{\delta_i}{\delta_j}) = i(V(i) - V(j)) \ketbra{\delta_i}{\delta_j}.$
$\sigma \left( \Li_0 \right) = \{0, - G\}$ as $\Li_0( \ket{\delta_{i}} \bra{\delta_{i}}) =0$. 
And as $\mathcal{E}_V(\ketbra{\delta_i}{\delta_i}) =0$, $\sigma \left( \Li \right)  =  \{ -G + \lambda i \lbrack - 1, 1 \rbrack \} \cup \{ 0 \}$.

$\Li$ leaves both the diagonal and off-diagonal subspaces invariant and both inclusions in Theorem \ref{kunz} are strict\footnote{Since $\Li_0$ is normal (indeed self-adjoint) the numerical range is the convex hull of the spectrum i.e. $\lbrack - G, 0 \rbrack$.
}.
\end{example}

\begin{SCfigure}[1][ht] 
  \centering
  \includegraphics[scale=0.4]{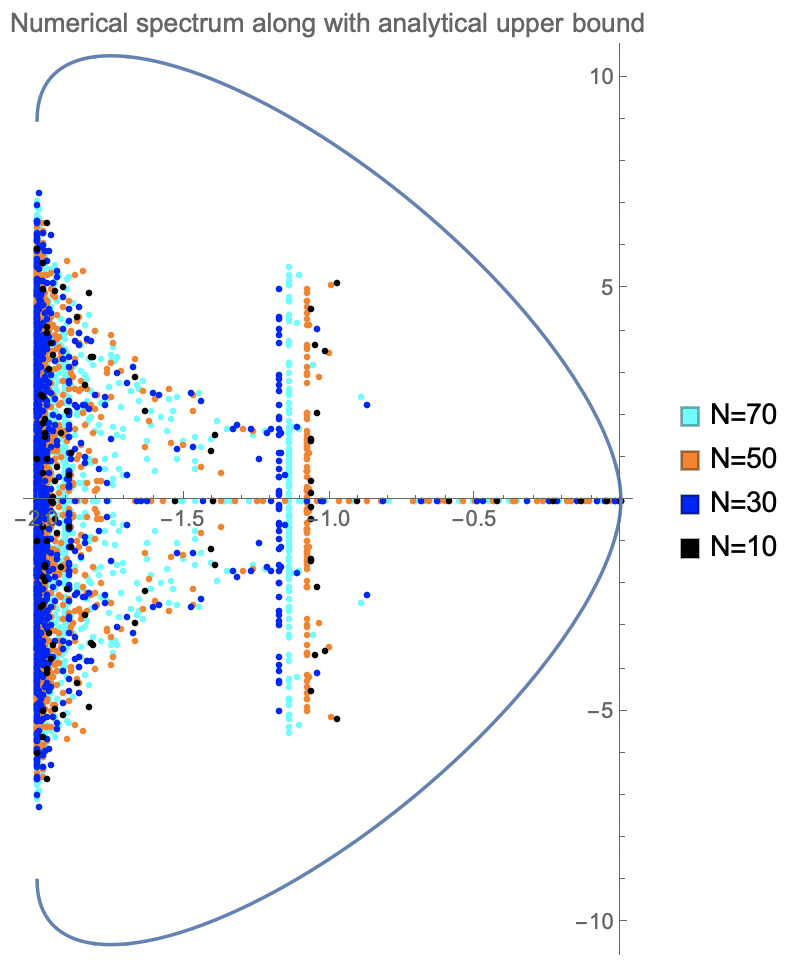}
  \caption{The numerical spectrum of the dephasing Lindbladian $\Li$ in a random potential studied in Section \ref{improving_upper_bound} for $N \in \{10,30,50,70\}, G=2, V=5$ along with the analytically calculated upper bound (from Proposition \ref{theorem:analytical_upper_bound_numerical_range}) for the numerical range (and therefore the spectrum) sketched. Notice the line with real part close to $-1$ which is consistent with, but not predicted by our analysis. Notice further how it seems that the spectrum does not extend much when the real part is less than 1.}
  \label{dephasing_rand}
\end{SCfigure}

For the dephasing Anderson model $H = - \tilde \Delta, L_k = \ket{\delta_k}\bra{\delta_k}$, the two subspaces get mixed, the effects are more complicated and rigorous guarantees are harder to obtain, see also Figure \ref{dephasing_rand} and \Cref{theorem:analytical_upper_bound_numerical_range}.

\begin{remark}[Further discussion of spectral effects of random potentials]
Spectra of random Lindbladians have been studied in random matrix theory approaches in for example \cite{PhysRevLett.123.140403}. There a lemon-like shape of the spectrum was found. This shape is reminiscent of the spectrum in both Figure \ref{fig:test1} and Figure \ref{dephasing_rand} where there seems to be a tendency that the spectrum close to 0 extends less in the direction of the imaginary axis. This is also mimicked in the exactly solvable example in Section \ref{sec:exact_solv} and the discussion in the previous section.
Furthermore, inspecting the non-random spectrum with and without the perturbation one can see how there is a similar phenomenon with eigenvalues jumping from the bulk of the non-Hermitian spectrum and down to the real line  \cite{Znidaric2015RelaxationTO}. An analytical explanation stemming from the symmetries of the Lindbladian is given in \cite{PhysRevLett.123.234103}.
The one also sees in a corresponding RMT model \cite{tarnowski2021random} and the previous example indicates a mechanism for this behavior.
\end{remark}

\subsection{Further questions}\label{sec:discussion}
The methods developed here can be applied to many one-particle open quantum systems. In particular, many of our results have generalizations to $\Z^d$ for $d \geq 1$. Other directions for further study are: 
\begin{itemize}
    \item Justify the conditions of \Cref{thm:convergence_finite} for the models studied in this paper. In particular solve, (\cite[Conjecture 7.3]{bottcher2002spectral}).  As shown by the examples in Figures  \ref{free_bc} and \ref{periodic_bc} using periodic boundary conditions will be important, as it is the case in the corresponding question for Laurent operators compared to Toeplitz operators in \cite{colbrook2020pseudoergodic}. The construction in \cite[Lemma 43]{koekenbier2024transfer} could be relevant.
    \item Strengthen the connection between spectra and dynamics, see also Section \ref{sec:spec_dyn}. In particular, relate the gap/gaplessness in the non-Hermitian Hamiltonian to localization properties \cite{LocalizationinOpenQuantumSystems,vakulchyk2018signatures}. 
    \item Further elucidate how the quantum-jump terms increase the number of real eigenvalues, which was observed in \cite{Znidaric2015RelaxationTO} and one explaination was given using symmetries of the Lindbladian in  \cite[Appendix B.9]{PhysRevLett.123.234103}. The proof there relies on some rather old results, so it would be of value to obtain a self-contained proof. 
    \item  Prove that if $\Li$ is \eqref{eq:finite-range}, \eqref{eq:translation-invariance}, and  \eqref{eq:Rank1} then $0 \in \sigma(\Li)$, thereby strengthening \Cref{thm:no_gap}. 
    \item Find out whether the \eqref{eq:Rank1} assumption can be removed in the theorems in \Cref{sec:general_app}.
    \item Extend the gaplessness result in \Cref{thm:no_gap} to finite system, i.e. prove that with increasing system   size either the gap closes or the kernel dimension increases. 
    \item Expand the spectral theory of disordered open systems and prove a stronger lower bound to Theorem \ref{kunz}.
\end{itemize}

 \section*{Acknowledgments}
The author acknowledges support from the Villum Foundation through the QMATH center of Excellence (Grant No. 10059), the Villum Young Investigator (Grant No. 25452) program and the Carlsberg Foundation CF24-0466.
Warm thanks to Albert Werner, Simone Warzel, Daniel Stilck França, Alex Bols, Jacob Fronk, 
Paul Menczel and Cambyse Rouzé for discussions, Nick Weaver for the useful input \cite{406705, 355246} and referees for useful feedback. A previous substantially different version of this paper is part of the author's PhD thesis \cite{klausen2023random}. 


\begin{appendix}

\section{Postponed theory}

\subsection{Review of Fourier theory}\label{sec:symbol}
 Define\footnote{Following the normalization convention in \cite[(A.9)]{Aizenman2015RandomOD}.} $\mathcal{F}: \ell^2(\Z) \to L^2( \lbrack 0,2 \pi \rbrack) $  by 
$
(\mathcal{F}\psi)(q) = \frac{1}{\sqrt{2\pi}} \sum_{x\in \Z}e^{-iqx}\psi(x),
$
for any $\psi \in \ell^2(\Z)$.  $\mathcal{F}$ is unitary and it maps the standard basis $\{\ket{\delta_{j}}\}_{j \in \Z}$ to the Fourier basis consisting of functions $\{\frac{1}{\sqrt{2\pi}} e^{-ikj}\}_{j\in\Z}$. An operator $T \in \mathcal{B}(\ell^2(\Z))$ is \emph{Laurent} if it is constant on the diagonals, i.e.  $\bra{\delta_i}, T \ket{\delta_j}  = \bra{\delta_{i+n}}, T \ket{\delta_{j+n}}$  for all $i,j,n \in \Z$.
  If $a_{n}= 0$ whenever $\abs{n} > r$ say that $T$ is \emph{r-diagonal}. In that case, with $S_i$ defined in \eqref{eq:shift_operator},
$
  T = \sum_{i = -r}^r a_i S_i 
  $
  and the Fourier transform maps $T$ into a multiplication operator:
\begin{lemma} \label{lemma:transform_Laurent} 
If $T \in \mathcal{B}(\ell^2(\Z) )$ is $r$-diagonal with entries $a_i$ on the $i$-th diagonal then
$
\mathcal{F} T \mathcal{F}^*  \in \mathcal{B}(L^2( \lbrack 0,2 \pi \rbrack) )
$
is the multiplication operator $a_T(q) =  \sum_{j=-r}^r a_j e^{iqj} $. 
In particular,
$ \mathcal{F} S \mathcal{F}^* = e^{-iq}
$
and $ \mathcal{F} S_l \mathcal{F}^* = e^{-iql}.
$
\end{lemma}
The curve $a_T(q) =  \sum_{j=-r}^r a_j e^{iqj} $ can also be viewed as a function from the unit circle upon defining 
$z = e^{iq}$, we call this curve the \emph{symbol curve}. Since the Fourier transformation is unitary and the essential spectrum is the part of the spectrum that is not isolated eigenvalues with finite multiplicity  \cite[IV.5.33]{kato2013perturbation} we obtain: 
\begin{corollary}\label{range_of_symbol}
For any Laurent operator $T \in \mathcal{B}( \ell^2(\Z))$,
 $\sigma(T) = \{a_{T}(z) \mid z \in \T \} = \sigma_{\ess}(T). $
\end{corollary}

\subsection{Boundedness on Schatten spaces}\label{sec:bound_op}
Denote the space of trace-class operators by $\TC( \mathcal{H})$, the Hilbert-Schmidt operators by $\HS(\mathcal{H})$ and the space of bounded operators by $ \mathcal{B}( \mathcal{H})$. All three spaces are Banach spaces with regards to their respective norms and $\HS(\mathcal{H})$ is also a Hilbert space with the inner product
$
\langle X, Y \rangle = \Tr(X^* Y). 
$
Define for $p \in (1, \infty)$ the Schatten-$p$-norm of $A \in \mathcal{B}( \mathcal{H})$ via
$
\norm{A}_p = \Tr( \abs{A}^p)^{ \frac{1}{p}},
$
where $\abs{A} = (AA^*)^{\frac{1}{2}}$. The Schatten-$p$-class $ \mathcal{S}_p$ then consists of all bounded operators with finite $p$-norm.
For $p= \infty$ we set  $ \mathcal{S}_\infty = \mathcal{K}(\Hi)$, the compact operators on $\Hi$ and $\mathcal{S}_1 = \TC( \Hi),  \mathcal{S}_2 = \HS( \Hi)$, for more information see e.g. \cite{simon2005trace,Davies2007LinearOA,einsiedler2017functional}. 

The following lemma ensures that the class of Lindblad operators we consider are bounded operators on $\HS(\Hi)$. A similar level of generality was used in  \cite{Falconi2016ScatteringTF, Attal}, for further discussions and similar results see also \cite{olkiewicz1999environment,Perez}.
\begin{lemma} \label{bounded_op}
Suppose that $\Li$ is of the Lindblad form \eqref{Lindblad_form}. Let $N_n = \sum_{k\in \Z:  \abs{k}\leq n} L_k L_k^* $ and $M_n = \sum_{k\in \Z:  \abs{k}\leq n} L_k^* L_k$. Suppose that both $\{N_n\}$ and $\{M_n\}$ converge weakly in $\mathcal{B}(\Hi)$.  Then
\begin{align*}
\Li \in \mathcal{B}( \mathcal{B}{(\mathcal{H})} ), \mathcal{B}( \HS(\mathcal{H}) ), \mathcal{B}( \TC(\mathcal{H})).
\end{align*}
\end{lemma}
\begin{proof}
$\Li \in  \mathcal{B}( \TC(\mathcal{H})) $ by \cite[prop 6.4]{Attal}.
Suppose that $\Li \in \mathcal{B}( \mathcal{B}{(\mathcal{H})} )$, which implies
that  $\Li \in \mathcal{B}( \mathcal{K}{(\mathcal{H})} )$.
 By the non-commutative Riesz-Thorin theorem \cite[Section IX.4]{reed1975ii},  
 $ \displaystyle
\norm{ \Li }_{2 \to 2}\leq \norm{ \Li }_{1 \to 1}^{\frac{1}{2}}  \norm{ \Li }_{\infty \to \infty}^{\frac{1}{2}}, 
$
which means that $\Li \in  \mathcal{B}( \HS(\mathcal{H})).$ 

To see that $\Li\in \mathcal{B}( \mathcal{B}{(\mathcal{H})} )$.
By assumption, the commutator and anti-commutator terms in the Lindbladian are bounded. Thus, we are left with the operator $\mathcal{J}$ defined by $\mathcal{J}(X) = \sum_k L_k X L_k^*$.

Let $N$ be the weak limit of  $\sum_k L_k L_k^* $.   By assumption  $ \norm{N}_\infty  <  \infty$. Since $ \sum_k L_k L_k^*$ is also self-adjoint it follows by the spectral radius theorem for every $\ket{x}\in H$ that
\begin{align*}
\sum_k \norm{ L_k^* \ket{x} }^2 = \sum_k \bra{x} , L_k L_k^* \ket{x} =  \bra{x}  , \sum_k L_k L_k^* \ket{x} \leq  \norm{N}_\infty \norm{x}^2. 
\end{align*}
Then consider $X\in \mathcal{B}{(\mathcal{H})} $ with $X \geq 0$.
Then $X = A A^* $ for some $A\in \mathcal{B}{(\mathcal{H})}$ and 
$
  \sum_k L_k  X L_k^*  =   \sum_k L_k  A A^* L_k^*
$
is a sum of positive operators and therefore positive. We bound the norm
\begin{align*}
\bra{x} , \sum_k L_k  A A^* L_k^* \ket{x}  = \sum_k \bra{x} ,  L_k  A A^* L_k^* \ket{x}  = \sum_k \norm{ A^* L_k^* \ket{x}}^2 \leq  \sum_k \norm{ A^*}^2\norm{ L_k^* \ket{x} }^2  \leq  \norm{ A^*}^2  \norm{N}_\infty \norm{x}^2. 
\end{align*}
Again, by self-adjointness of $\sum_k L_k  A A^* L_k^*$ the spectral radius theorem holds and therefore if the numerical range is bounded then so is the norm. We conclude that
\begin{align*}
\norm{  \sum_k L_k  A A^* L_k^*  } \leq \norm{A}^2  \norm{N}_\infty.
 \end{align*} 

Now, for any element $X\in \mathcal{B}{(\mathcal{H})}$ we write $X = P_1 - P_2 + i P_3 - i P_4$ where $P_1, P_2, P_3, P_4 $ are all positive and satisfy
$ \norm{P_i } \leq \norm{X}$ for each $i = 1, \dots, 4 $ (see \cite[Theorem 11.2 and 9.4]{zhu2018introduction}).
Then using the $C^*$-identity$
  \norm{A_i}^2 = \norm{A_i {A_i}^*} = \norm{P_i}\leq \norm{X}. 
$  
Conclude that  $
\norm{\mathcal{J}(X)}   
 \leq 4 \norm{X}  \norm{N}_\infty.
$ 
\end{proof}

 \subsection{Relation between spectra and dynamics for Lindblad systems}\label{sec:spec_dyn}
 In the Hamiltonian case, the RAGE theorem gives a dynamical interpretation of the spectra and the different types of spectra. However, due to non-normality
of Lindblad operators, the dynamical implications of the spectra are more subtle and the details of the topic are still under discussion in the physics literature \cite{mori2020resolving}. 

\emph{Finite dimensions:} In the finite-dimensional, case the relationship between eigenvalues of $\Li$ and the time evolution is given through the Jordan normal form. I.e.  $\Li = S \Lambda S^{-1}$ where $S$ is invertible and $\Lambda$ is of a certain almost-diagonal form. However, even in the cases where $\Lambda$ is diagonal, $\Li$ is not necessarily normal and so the mathematical guarantees for convergence of the semigroup are much weaker than in the normal case and they scale much worse with the dimension of the system (see e.g. \cite{szehr2015spectral}).
Another peculiarity is the fact that all eigenvectors of $\Li$ are traceless:
\begin{remark} \label{rem:trace}
All eigenvectors of $\Li$ defined by \eqref{Lindblad_form} corresponding to nonzero eigenvalues are traceless: Since $\Tr(\Li(\rho)) =0$, $e^{t \Li} $ is trace-preserving for all $t \in \lbrack 0, \infty)$,
$
\Tr( \rho) = \Tr \left( e^{t \Li} (\rho) \right)    = e^{\lambda t} \Tr( \rho).
$
Thus, if $\Tr( \rho) \neq 0$ then $1=  e^{\lambda t}$ which implies that $\lambda = 0$.
\end{remark}
To give guarantees about the dynamics in terms of the  spectral gap $g(\Li)$  which we define as follows
\begin{align*}
g(\Li) = \sup \left\{ \text{Re}( \lambda) \mid  \lambda \in \sigma(\Li) \backslash \{0\} \right\}.
\end{align*}
In the case of a unique steady state $\rho_\infty$, we can get a dynamical guarantee for the speed of decay towards the steady state in terms of the gap $g(\Li)$. Namely that
$\norm{ e^{t \Li}( \rho) -  \rho_\infty} \leq C e^{tg}$ where $C>0$ is a constant that depends heavily on the size of the Jordan blocks of the systems.

In the literature, the cases where $\Li$ is not diagonalizable are called exceptional points, there is evidence that these points can also lead to faster decay towards the steady state \cite{khandelwal2021signatures}, although the mathematical guarantee gets worse.

\emph{Infinite dimensions:} In infinite dimensions the relationship between spectra and dynamics can break down due to Jordan blocks of unbounded size (and more generally the breakdown of the Jordan normal form), due to the lack of a trace class steady state (a phenomenon seen in most examples in Section \ref{applications}) and due to the lack of a spectral gap (proven for our models in Theorem \ref{thm:no_gap}).

However, if $\sigma(\Li)$ has disconnected parts (as in \Cref{sec:dephasing}). E.g. $ \sigma(\Li) = \Sigma_A \dot \cup \Sigma_B$,  with $\sup \{\Re(z) \mid z \in \Sigma_A \}\leq g$ for some gap $g \in (- \infty, 0)$ and such that there exists a closed continuous curve encircling only $\Sigma_A $.  Define the Riesz projections by contour integration to get a decomposition of $\Hi = \Hi_A \oplus  \Hi_B$ for two orthogonal subspaces $\Hi_A,  \Hi_B$ such that $\Li$ leave each of the two subspaces invariant and decomposes
$\Li = \Li_A \oplus \Li_B$. If $\rho \in \Hi_A$ then, since $ \Li_A$ is the generator of a semigroup by \cite{engel2000one}, 
\begin{align*}
\norm{e^{t \Li}( \rho) } = \norm{e^{t \Li_A}( \rho) }  \leq Ce^{tg}\norm{\rho}
\end{align*}
for some constant $C>0$. Thus, if $\rho = \rho_A + \rho_B$ with $\rho_A \in  \Hi_A$ and $\rho_B \in  \Hi_B$ the part $\rho_A$ decays quickly. 

It is left to future work to establish a stronger relationship between spectra and dynamics in the infinite-dimensional case. In particular, one cannot use our work to gain many rigorous guarantees about the evolution of infinite open quantum systems, but we consider the results presented as steps towards such rigorous guarantees. Furthermore, due to the apparent convergence of the spectra of some finite-dimensional Lindbladians (see Theorem \ref{thm:convergence_finite} and the discussion in Section \ref{sec:discussion}) one can also view the method presented here as a way to compute large volume approximations to finite systems (which have discrete spectra and where the relation between spectra and dynamics is clearer).


\section{Finite systems with periodic boundary conditions}

\subsection{Direct sum decomposition for finite systems with periodic boundary conditions} \label{sec:finite_systems_def}
In this section, the decomposition from \Cref{main} is reviewed in finite volume with periodic boundary conditions.
Let $\{1, \dots , n \} = \lbrack n\rbrack $, $\T_n =  \left \{ \frac{2 \pi k}{n} \mid k = 1, \dots n  \right \}$, $[a]_n = a \pmod{n}$, and $\Hi_n = \ell^2( \lbrack n \rbrack)  = \text{span} \{\ket{\delta_{j}}, j = 0 \dots n-1 \}$, with shift 
\begin{align}
S^{(n)}  \ket{\delta_{j}} = \ket{\delta_{[j+1]_n}}, 
\end{align} which is a unitary on $\Hi_n $ and satisfies 
$ \label{eq:shift_period} 
(S^{(n)})^n = \id.
$
An $n \times n$ matrix satisfying $C_n^{\per} = S^{(n)}C_n^{\per} (S^{(n)})^*$ is \emph{circulant}. 
Denote circulant matrices with $a_i$ on the $i$'th diagonal by $C_n^{\per}\lbrack a_0, a_1, \dots, a_n \rbrack$. They are diagonalized using the discrete Fourier transform $\Ft^{(n)}$, which is the matrix where the $i$'th column is  $(1,\omega_n^i,\omega_n^{2i}, \dots \omega_n^{(n-1)i})$, where $\omega_n = e^{\frac{2\pi i}{n}}$. 
Then, $
D = \Ft^{(n)}C_n^{\per}\Ft^{(n)*}
$ is diagonal with $\{a_{C_n^{\per}}(z), z \in \T_n \} $ on the diagonals, where $a_{C_n^{\per}}(z)$ is the  \emph{symbol curve} defined by 
$
a_{C_n^{\per}}(z) = \sum_{i=0}^n a_i z^i 
$
for $z \in \T$ (the unit circle). 
The inverse of an invertible circulant $C_n$ diagonalized by $D$ is
$
C_n^{-1}  =  \Ft_n D^{-1}  \Ft_n^*, 
$
 The matrix elements of $C_n^{-1}$ are given by  (cf. \cite{bottcher2002spectral}) 
\begin{align}\label{eq:inverse_finite_fourier}
\bra{\delta_{j}}, C_n^{-1}\ket{\delta_{k}}  = \frac{1}{n}\sum_{l=0}^{n-1}  \frac{ \bar{\omega}_n^{l(j-1)}\omega_n^{l(k-1)}}{a_{C_n}^{\per}(\omega_n^l)}. 
\end{align}

The finite volume vectorization analogue $\vectorize^{(n)}: \HS( \ell^2( \lbrack n \rbrack)) \to  \ell^2( \lbrack n \rbrack) \otimes  \ell^2( \lbrack n \rbrack)$ satisfy a version of \eqref{eq:vectorization}.   
Analogous to \eqref{eq:Cdef} define $C^{(n)}: \ell^2 \left( \lbrack n\rbrack \right)\otimes \ell^2 \left( \lbrack n \rbrack \right)\to  \ell^2 \left( \lbrack n\rbrack \right)\otimes \ell^2 \left( \lbrack n \rbrack \right)$ by 
$$
C^{(n)}\ket{\delta_{j}}\ket{\delta_{k}} = \ket{\delta_{j}} \ket{\delta_{[k-j]_n}}. 
$$ 
 $f \hspace{-2pt}:\lbrack n \rbrack  \to  \lbrack n\rbrack $ given by $f(j,k) = (j,[k-j]_n)$ is a bijection, so $C^{(n)}$ is unitary with inverse,
$$ (C^{(n)})^*\ket{\delta_{j,k}} = (C^{(n)})^{-1}\ket{\delta_{j,k}} = \ket{\delta_{j,[k+j]_n}}.$$
The finite-dimensional analogue of Lemma \ref{lemma:C_relations} follows with the same proof. 
\begin{lemma}\label{lemma:C_relations_finite}
The operator $\Ad_{C^{(n)}}$ satisfies the following relations
\begin{align*}
\text{(i) } \Ad_{C^{(n)}}(\id \otimes S^{(n)})  = \id\otimes S^{(n)}, 
\text{  (ii) }\Ad_{C^{(n)}}(S^{(n)} \otimes \id) = S^{(n)}\otimes (S^{(n)})^*,
\text{  (iii) } \Ad_{C^{(n)}}(S^{(n)} \otimes S^{(n)}) = S^{(n)}\otimes \id.
\end{align*} 

\end{lemma}

Suppose that $H$ is a range-$r$, (self-adjoint) Laurent operator with diagonals $h_{-r}, h_{-r+1}, \dots, h_{r-1}, h_{r}$. 
Let $n \geq 2r$ and define the $n$ dimensional version of $H$ with periodic boundary conditions as 
$$
H_n^{\per} = C_n^{\per} \lbrack h_{-r}, h_{-r+1}, \dots, h_{r-1}, h_{r}\rbrack. 
$$
Given $L_0$ with range\footnote{This operator, even though strictly speaking infinite-dimensional, can be viewed as a $n \times n$ matrix (padded with zeros appropriately).} at most $r$, define for every $k \in \Z$, 
$
L_k = (S^{(n)})^k L_0  (S^{(n)*})^k 
$
By \eqref{eq:shift_period}
$L_k = L_{[k]_n} $. 
Given $ \lbrack h_{-r}, h_{-r+1}, \dots, h_{r-1}, h_{r}\rbrack $ and $L_0$ define 
$
\Li_n^{\per}\hspace{-2pt}: \HS\left(\ell^2 \left( \lbrack n \rbrack \right)\right)  \to \HS\left(\ell^2 \left( \lbrack n \rbrack \right)\right) 
$
by 
\begin{align} \label{eq:Lper} 
\Li_n^{\per}(\rho) = - i \lbrack H_n^{\per}, \rho \rbrack + G \sum_{k \in \lbrack n \rbrack}L_k  \rho L_k^* - \frac{1}{2} \{ L_k^* L_k, \rho \}. 
\end{align}

Define $\Ft^{(n)}_1 = \Ft^{(n)} \otimes \id$ to be the discrete Fourier transform in the first coordinate. Consider a map $I^{(n)}: \C^n \otimes V \to \oplus_{i=0}^{n-1}V^{i} $ (where each $V^{i}$ is a copy of $V$) defined by 
$
I^{(n)}( \ket{i}\otimes \ket{v} ) =\underbrace{0 \oplus \dots \oplus 0}_{i-1} \oplus v \oplus 0 \oplus \dots \oplus 0, 
$
 e.g. $I^{(n)}$ maps $\ket{i}\otimes \ket{v}$ to a $v$ in the $i$'th direct summand. If $V$ is a Hilbert space then  $I^{(n)}$ is an isometric isomorphism. 
Then, as \eqref{eq:lemma:isomorphisms} above,
 $
\HS\left(\ell^2 \left( \lbrack n \rbrack \right)\right)  \overset{\vectorize^{(n)}}{\cong} \hspace{-4pt}\ell^2 \left( \lbrack n\rbrack \right)\otimes \ell^2 \left( \lbrack n\rbrack \right) \overset{ \Ft_1^{(n)}}{\cong}  \hspace{-4pt} L^2\left( \T_n \right) \otimes  \ell^2( \lbrack n \rbrack) \overset{I^{(n)}}{\cong}\hspace{-4pt} \bigoplus_{ q\in \T_n }\ell^2_q( \lbrack n\rbrack).
$
Define  $\mathcal{J}^{(n)}\hspace{-2pt} : \HS\left(\ell^2 \left( \lbrack n\rbrack \right) \right) \to \bigoplus_{ q\in \T_n }\ell^2_q( \lbrack n\rbrack)$ by
$
\mathcal{J}^{(n)} =   I^{(n)}\circ \Ft^{(n)}_1\circ C^{(n)}\circ \text{vec}^{(n)}.
$
Mimicking the computation in Theorem \ref{main}:
\begin{theorem} \label{thm:finite_main}
Suppose that  $n \geq 2r$ and $\Li_n^{\per}$ is of the form \eqref{eq:Lper}. Then
$
\Ad_{\mathcal{J}^{(n)}}(\Li ) =
\bigoplus_{q \in \T_n} \left( T^{\per}_n(q) + F_n(q)  \right),
$
with $T_n^{\per}(q)$ an $r$-diagonal circulant $n\times n$ matrix and $F_n(q)$ a finite-rank operator with finite range (uniformly in $n$). 

If $H_{\eff} =   H_n^{\per}  -  \frac{iG}{2} \sum_{k\in\lbrack n\rbrack}  L_k^* L_k =  \sum_{l=-r}^r h_l   (S^{(n)})^l  $ then
$
 T_n^{\per}(q)  =  -i \sum_{l=-r}^r h_l   e^{iql}   (S^{(n)*})^l  + i\sum_{l=-r}^r \overline{h_l}   (S^{(n)})^l . 
$
Moreover, in the case $L_0 =  \ket{\phi}\bra{\psi} $ is rank-one with coefficients $\ket{\phi}= \sum_{l}\alpha_l \ket{\delta_l}$ and $ \ket{\psi}= \sum_{l} \beta_l \ket{\delta_{l}}$ for complex numbers $\alpha_l, \beta_l \in \C$ then  $F_n(q)$ is rank-one and
\begin{align*}
F_n(q) = G  \left( \sum_{l_1,l_2}\alpha_{l_1} e^{iql_1}\overline{\alpha_{l_2}}\ket{\delta_{l_2- l_1}}  \right)  \left( \sum_{l_1',l_2'}\beta_{l_1'} e^{-iql_1'}\overline{\beta_{l_2'}}\bra{\delta_{l_2'- l_1'}} \right).
\end{align*}
\end{theorem}

\subsection{Spectral consequences in finite dimensions}
By Theorem \ref{thm:finite_main}, 
$
\sigma\left( \Li_n^{\per} \right) = \bigcup_{q \in \T_n}   \hspace{-0.3 em} \sigma \left( T^{\per}_n(q) + F_n(q)  \right). 
$ 
Whenever $F_n(q) = \ket{ \mathtt{v}} \bra{\mathtt{u}}$, by rank-one update (cf. Appendix \ref{sec:rank_one_pert}), 
\begin{align}\label{eq:finite_rank_one}
\left\{\lambda \in \C \mid   \bra{\mathtt{u}},(T^{\per}_n-\lambda)^{-1} \ket{\mathtt{v}} = -1 \right \} \backslash  \sigma( T^{\per}_n)    \subset \sigma \left( T^{\per}_n+ F_n  \right)  \subset  \sigma( T^{\per}_n)   \cup  \left\{\lambda \in \C \mid   \bra{\mathtt{u}},(T^{\per}_n-\lambda)^{-1} \ket{\mathtt{v}} = -1 \right \}.
\end{align}
The formula motivates efforts finding the matrix elements of inverses of tridiagonal circulants. Let the entries of the three main diagonals be $\alpha,\beta, \gamma$. The solutions $\lambda_1, \lambda_2$ to the quadratic equation 
\begin{align}  \label{eq:eigenvalue_equation2}
\alpha + \beta z + \gamma z^2 = 0
\end{align}
 and knowledge of whether the two solutions satisfy
$
\abs{ \lambda_1}< 1 < \abs{\lambda_2} 
$
is important, see Appendix \ref{sec:invertibility_tridiagonal} for details. 

The formula for the matrix elements of the inverse is tedious calculation, which may be of some independent interest.  To state it more concisely, let  $\lbrack a \rbrack_n$ be the representative between $0$ and $n-1$ of $a$.
\begin{lemma} \label{lemma:technical_sums} 
Suppose that $C_n$ is an $n \times n$ circulant with $\alpha, \beta, \gamma$ on the three main diagonals such that $\gamma \neq 0$.  Suppose that $\lambda_1$ and $\lambda_2$ are solutions to \eqref{eq:eigenvalue_equation2} such that $\abs{\lambda_2} < 1 < \abs{\lambda_1}$. 
Then
\begin{align}\label{eq:inverse_formula}
\bra{\delta_{j}}, C_n^{-1}\ket{\delta_{k}}  = \frac{1}{\gamma(\lambda_1 - \lambda_2)} \left(  \frac{1}{1- \lambda_1^{-n}}  \left(  \id\lbrack j \neq k \rbrack \left( \frac{1}{\lambda_1}\right)^{\lbrack j-k \rbrack_n} +  \id\lbrack j = k \rbrack \frac{1}{\lambda_1^n} \right)+  \frac{1}{1- \lambda_2^{n}}  \lambda_2^{ \lbrack k-j \rbrack_n } \right).
\end{align}
\end{lemma}
\begin{proof}
The polynomial from the symbol curve has roots $\lambda_1$ and $\lambda_2$:
$a(z) = z^{-1} \alpha + \beta + \gamma  z = \gamma z^{-1}( \lambda_1 - z ) (\lambda_2 - z). 
$
So,
$$
\frac{1}{a(z)} = \frac{z}{\gamma}\frac{1}{\lambda_1 - \lambda_2} \left( \frac{1}{z-\lambda_1}- \frac{1}{z- \lambda_2}\right) 
= \frac{1}{\gamma(\lambda_1 - \lambda_2)} \left( \sum_{m=1}^{\infty} \left(\frac{z}{\lambda_1} \right)^m  + \sum_{m=0}^{\infty} \left(\frac{\lambda_2}{z} \right)^m \right). 
$$
So consider 
$
\sum_{m=1}^{\infty} \left(\frac{\omega_n^l}{\lambda_1} \right)^m. 
$
If $m=jn+r$ for $0 \leq r \leq n-1$, then $\omega_n^{lm} = \omega_n^{rl}$ and thus
$$
\sum_{m=1}^{\infty} \left(\frac{\omega_n^l}{\lambda_1} \right)^m  = \sum_{r=1}^{n}  \frac{\omega_n^{lr}}{\lambda_1^r}  \sum_{j=0}^{\infty}\lambda_1^{-nj}  
=  \sum_{r=1}^{n}  \frac{\omega_n^{lr}}{\lambda_1^r} \frac{1}{1- \lambda_1^{-n}} = \frac{1}{1- \lambda_1^{-n}}  \frac{\omega_n^{l}}{\lambda_1}  \sum_{r=0}^{n-1} \omega_n^{lr}\lambda_1^{-r}. 
$$
Analogously, $
\sum_{m=0}^{\infty} \left(\omega_n^{-l} \lambda_2 \right)^m 
= \frac{1}{1- \lambda_2^{n}}   \sum_{r=0}^{n-1}  \left(\omega_n^{-l}\lambda_2\right)^m
$
and the combination of the two, 
$$
\frac{1}{a(\omega_n^{l})} =   \frac{1}{\gamma(\lambda_1 - \lambda_2)}   \sum_{r=0}^{n-1} \frac{1}{1- \lambda_1^{-n}}  \frac{\omega_n^{l}}{\lambda_1}   \left( \frac{\omega_n^{l}}{\lambda_1}\right)^r + \frac{1}{1- \lambda_2^{n}} \left(\omega_n^{-l}\lambda_2\right)^r. 
$$
Inserting that in \eqref{eq:inverse_finite_fourier} yields, 
$$
\bra{\delta_{j}}, C_n^{-1} \ket{\delta_{k}}  = \frac{1}{n} \sum_{l=0}^{n-1}  \frac{\omega_n^{l(k-j)} }{a(\omega_n^l)} =   \frac{1}{n}    \frac{1}{\gamma(\lambda_1 - \lambda_2)}  \sum_{r=0}^{n-1}  \sum_{l=0}^{n-1} \left(\frac{1}{1- \lambda_1^{-n}}  \frac{\omega_n^{l}}{\lambda_1}   \left( \frac{\omega_n^{l}}{\lambda_1}\right)^r + \frac{1}{1- \lambda_2^{n}} \left(\omega_n^{-l}\lambda_2\right)^r \right)
\omega_n^{l(k-j)}. 
$$
Since $
\sum_{l=0}^{n-1} \omega_n^{l(1+r+k-j)}  = n \id \lbrack r = [j-k-1]_n \rbrack, 
$
the first term can be written as
\begin{align*}
  \sum_{r=0}^{n-1} \frac{\lambda_1^{-r-1}}{1- \lambda_1^{-n}} \sum_{l=0}^{n-1} \omega_n^{l} \omega_n^{lr}\omega_n^{l(k-j)} 
   &= n  \frac{\lambda_1^{ -\lbrack j-k-1 \rbrack_n}}{1- \lambda_1^{-n}}   \frac{1}{\lambda_1} = n  \frac{1}{1- \lambda_1^{-n}}  \left(  \id \lbrack j \neq k \rbrack \left( \frac{1}{\lambda_1}\right)^{\lbrack j-k \rbrack_n} +  \id\lbrack j = k \rbrack \frac{1}{\lambda_1^n} \right).
   \end{align*}
\noindent The similar expression for the  second term finishes the proof.  
$
\sum_{r=0}^{n-1} \frac{1}{1- \lambda_2^{n}}  \lambda_2^r \sum_{l=0}^{n-1} \omega_n^{-lr}   \omega_n^{l(k-j)}  = n  \frac{1}{1- \lambda_2^{n}}  \lambda_2^{ \lbrack k-j \rbrack_n}. 
$
\end{proof}

\subsubsection*{Catching $n-1$ eigenvalues of $C_n^{\per}$ close to $\sigma(T)$.}
For tridiagonal Laurent operators $T$ consider functions $R_T: \C \backslash \sigma(T) \to \C$ defined by $R_T(z) =  \bra{\delta_{0}}, (T-z)^{-1} \ket{\delta_{0}} +1$, which is holomorphic \cite[Theorem III-6.7]{kato2013perturbation}. 
\begin{proposition}
Let $T$ be a tridiagonal Laurent operator 
 and let $C_n^{\per}$ be the corresponding $n\times n$ circulant. 
Suppose that there exists exactly one $z_0 \in \C \backslash \sigma(T)$ such that $ \bra{\delta_{0}}, (T-z_0)^{-1} \ket{\delta_{0}} = -1$, and where $z_0$ is a simple pole of $R_T$.  Suppose that the solutions $\lambda_1, \lambda_2$ to  \eqref{eq:eigenvalue_equation2} with $\beta$ replaced by $\beta - z_0$ satisfy $\abs{\lambda_2} < 1 < \abs{\lambda_1}$. 
Let $K$ be any compact subset of $\C \backslash \sigma(T)$ containing $z_0$. Then for sufficiently large $n$,  exactly $n-1$ eigenvalues of $C_n^{\per} + \ket{\delta_{0}}\bra{\delta_{0}}$ are outside $K$ and exactly one eigenvalue is inside $K$ and it converges to $z_0$ when $n \to \infty$. 
\end{proposition}
\begin{proof}
For every $z \in K$ consider the Laurent operator with $\alpha, \beta-z, \gamma$ on the diagonals and consider the solutions $\lambda_1(z), \lambda_2(z)$ to  \eqref{eq:eigenvalue_equation2} with $\beta$ replaced by $\beta - z$. If $\lambda_i(z)$ solves $\eqref{eq:eigenvalue_equation2}$ and $\abs{\lambda_i(z)}=1$, then $z \in \sigma(T)$. Therefore the two solutions (which for a suitable choice of labelling are analytic functions of $z$) must satisfy $\abs{\lambda_2(z)} < 1 < \abs{\lambda_1(z)}$ for every $z \in K$. So $z\mapsto \abs{\lambda_1(z)}, \abs{\lambda_2(z)}$ are
well-defined continuous functions on $K$, and 
the extreme value theorem implies that $\abs{\lambda_1}  \geq c_1 > 1$ and $\lambda_2 \leq c_2 < 1$ for some constants $c_1,c_2 > 0$ uniformly on $K$.  

For each $n \in \N$, consider the holomorphic functions $R_{C_n}$, $R_T$. 
By the explicit form of the functions in Lemma \ref{lemma:technical_sums} and Lemma \ref{inverse}, $R_{C_n} \overset{\lcu}{\to} R_{T}$. 
Suppose that $\{ z_n \}_{n\in\N}\subset K$ and $R_{C_n}(z_n) =0$. Since $K$ is compact take a convergent subsequence  $z_n \to \tilde z$. Since $\abs{\lambda_1}  \geq c_1 > 1$ and $\abs{\lambda_2} \leq c_2 < 1$ and the formula \eqref{eq:inverse_formula},  
$$
- 1 = \frac{1}{\gamma (\lambda_1(\tilde z) - \lambda_2(\tilde z))}. 
$$
Since $z_0$ was the unique solution to  $-1= \bra{\delta_{0}}, (T-z)^{-1} \ket{\delta_{0}}$, by Lemma \ref{inverse} that  $\tilde z = z_0$. 

If (on a subsequence) there exists $z_n^{1}\neq z_n^{2}$ satisfying $R_{C_n}(z_n^{1})=0=R_{C_n}(z_n^{2})$ for each $n \in \N$ then it contradicts Hurwitz' theorem, since $z_0$ was a simple pole of $R_T$.  
Thus, for sufficiently large $n$  only one $z_n\in K$ satisfies $R_{C_n}(z_n) = 0 $ and  (since any convergent subsequence converges to $z_0$) $z_n \to z_0$. 
As $C_n^{\per} +F_n$ has $n$ eigenvalues counted with multiplicity at least $n-1$ of them must be inside $K$.
\end{proof}


\begin{remark}
The argument may potentially shed light on whether the numerical observation that many eigenvalues lie on the real axis comes from a symmetry constraint (as argued in  \cite[Appendix B.9]{PhysRevLett.123.234103})  or that the eigenvalues in the periodic system tend to the eigenvalues of the full Lindbladian exponentially fast (cf. Lemma \ref{lemma:technical_sums}). 
\end{remark}

\section{Proof details}
\subsection{Measurability in the proof of Theorem \ref{directint}} \label{sec:measurability}
In this appendix, we prove that for each fixed $n\in \N$ there exists a measurable choice of the vectors $q \mapsto v_{q,n}$ in the proof of Theorem  \ref{directint}.
We do that with inspiration from \cite{measurability_reference} and let $\{ a_m \}_{m\in\N}$ be a countable dense subset of $\Hi$ which does not contain $0$. 
Define $b_m = \frac{a_m}{\norm{a_m}}$. Recall that $I_n$ is a set such that $\abs{I_n} > 0$ with $ \norm{(A(q) - \lambda)^{-1}}\geq n$. Now, define $N\hspace{-1pt}: I_n \to \N$  by
 \begin{align*}
 N(q) = \min \left \{m\in \N \mid \norm{(A(q) - \lambda)^{-1} b_m} \geq  \frac{n}{2}  \right \}.
 \end{align*} 
 Notice that $N$ is well-defined since $(A(q) - \lambda)^{-1}$ is bounded and hence continuous and by density of  $\{ a_m \}_{m\in\N}$.
 We claim that $N$ is also $\mathcal{B}\left(I_n \right) \-- \mathcal{P}(\N) $ measurable, where $\mathcal{B}\left(I_n \right) $ is the Borel $\sigma$-algebra on $I_n$.
 To see that, notice first that since $q \mapsto (A(q) - \lambda)^{-1}$ is measurable then also $q \mapsto \norm{(A(q) - \lambda)^{-1} b_m} $ will be measurable for each $m$. Thus, the set
$
 \left\{q\in I_n  \mid \norm{(A(q) - \lambda)^{-1} b_m} \geq   \frac{n}{2} \right\}
$ is measurable and
 \begin{align*}
 N^{-1}( \{1, \dots ,k \}) = \bigcup_{m=1}^k  \left\{q\in I \mid \norm{(A(q) - \lambda)^{-1} b_m} \geq  \frac{n}{2} \right \}.
 \end{align*}
This is sufficient to prove measurability of $N$. Now, since any function from $( \N,  \mathcal{P}(\N))$ to any measure space is measurable and compositions of measurable functions are measurable the map
$q \mapsto b_{N(q)} $ is measurable and so picking $v_{q,n} = b_{N(q)}$ is a measurable choice in the proof of Theorem \ref{directint}.

\subsection{Rank 1 update} \label{sec:rank_one_pert}
In this appendix, we prove the following relation that was used in the proof of Corollary \ref{cor:rank_one}.
\begin{align*}
 \sigma( T(q) + F(q)) = \sigma( T(q))   \cup  \left\{\lambda\in \C \mid   \bra{\mathtt{u}}(T(q)-\lambda)^{-1} \ket{\mathtt{v}}  = -1  \right\}. 
\end{align*}
$"\supset:"$ The proof of Corollary \ref{cor:general_rank} shows $ \sigma( T(q))\subset \sigma( T(q) + F(q))$. So let $\lambda \not\in \sigma( T(q))$ and  $ \bra{\mathtt{u}}(T(q)-\lambda)^{-1} \ket{\mathtt{v}} = -1$.  Assume for contradiction that $T(q)+F(q) - \lambda$ is invertible.  Let for ease of notation  $T = T(q) - \lambda$. The resolvent equation gives
$
\frac{1}{T + \ket{\mathtt{v}}\bra{\mathtt{u}}} = \frac{1}{T }  - \frac{1}{ T + \ket{\mathtt{v}}\bra{\mathtt{u}}} \ket{\mathtt{v}}\bra{\mathtt{u}}\frac{1}{T}.
$
If $\bra{\mathtt{u}},\frac{1}{T} \ket{\mathtt{v}} =  -1$ then
multiplying the resolvent equation with $\ket{\mathtt{v}}$ from the right yields that $T^{-1} \ket{\mathtt{v}} = 0$, which is a contradiction.

\noindent $"\subset:"$ Assume $T$ invertible and    $\bra{\mathtt{u}} \frac{1}{T} \ket{\mathtt{v}}  \neq -1$.
Two multiplications show that 
$
(T +  \ket{\mathtt{v}}\bra{\mathtt{u}})^{-1} = \frac{1}{T}- \frac{1}{\bra{\mathtt{u}} \frac{1}{T} \ket{\mathtt{v}}+1}\frac{1}{T}\ket{\mathtt{v}}\bra{\mathtt{u}} \frac{1}{T}. 
$

\subsection{Proof of Lemma \ref{lemma:assumptions_satisfied}}\label{sec:resolvent_norm_estimates}

First, by continuity of the (holomorphic) resolvent and compactness of $V$ for any $q\in I$
\begin{align}\label{eq:sup_over_V}
    \sup_{z\in V}\norm{(T(q)-z)^{-1}} \leq C(q) < \infty. 
\end{align} 

\begin{claim}\label{claim:sup_over_q}
If $\{T(q)\}_{q\in I}$ is norm-continuous, $V \subset \cap_{q \in I} \rho(T(q))$ is compact, then $\forall z\in V$,
$
\sup_{q\in I}\norm{(T(q) -z)^{-1}}< \infty. 
$
\end{claim}
\begin{proof}
Suppose for contradiction that for some sequence $\{q_n\}$ in $I$ that
$
\lim_{n\to \infty}\norm{ (T(q_n) -z)^{-1}}= \infty.  
$
Let a subsequence converge to $q_0$.  Then for every $\varepsilon > 0$, there is a $n$ large enough so that by  \Cref{lemma:double_essential_spectrum},
$$
z\in \sigma_{\varepsilon}( T(q_n)) \subset   \bigcup_{q\in \lbrack 0, 2\pi \rbrack}^{\ess} \hspace{-2pt}   \sigma_{2\varepsilon}(T(q)), 
$$
as this holds for any $\varepsilon>0$, $z\in \sigma(\mathcal{T}) = \cup_{q\in I} \sigma(T(q))$, which contradicts the combination of \Cref{directint} and \ref{direct_continuity}. \end{proof}

\begin{claim}If $\{T(q)\}_{q\in I}$ is norm-continuous, $V \subset \cap_{q \in I} \rho(T(q))$ is compact, then 
$\sup_{z\in V, q\in I} \norm{(T(q)-z)^{-1}} < \infty$.
\end{claim}
\begin{proof} 
Assume for contradiction that for some $(q_0,z_0)\in I \times V$ and a (sub)-sequence $(q_n,z_n) \to (q_0,z_0)$, satisfying 
$\sup_{z\in V, n\in \N} \norm{(T(q_n)-z_n)^{-1}} = \infty$. 
As $\norm{ (T(q_0)- z_0)^{-1}}< \infty$, for any $(q,z)\in \lbrack 0, 2\pi \rbrack \times V$ by the resolvent equation \eqref{eq:resolvent_bound}, \eqref{eq:sup_over_V}, Claim \ref{claim:sup_over_q},
\begin{align*}
\abs{ \norm{(T(q_0)- z_0)^{-1}}-  \norm{(T(q)-z)^{-1}}} 
& \leq \norm{(T(q_0)- z_0)^{-1} -(T(q)- z)^{-1}} \\
& \leq  \norm{(T(q_0)- z_0)^{-1} -(T(q)- z_0)^{-1}}  +  \norm{ (T(q)- z_0)^{-1} -(T(q)- z)^{-1}}\\ 
& \leq  \norm{(T(q_0)- z_0)^{-1}} \norm{ T(q_0) - T(q)} \norm{(T(q)- z_0)^{-1}}  \\
& + \norm{(T(q)- z_0)^{-1}} \abs{z- z_0}\norm{(T(q)- z)^{-1}} \\
& \leq C(z_0)^2  \norm{T(q_0) - T(q)}+ C(z_0)\abs{z- z_0}\norm{(T(q)- z)^{-1}}. 
\end{align*}
Thus, for any $z$ with $ \abs{z- z_0} \leq \frac{1}{2C(z_0)}$ and $q\in I$: 
\begin{align*}
 \norm{(T(q_0)- z_0)^{-1}} & \geq  \norm{(T(q)-z)^{-1}} - \abs{ \norm{(T(q_0)- z_0)^{-1}} -  \norm{(T(q)-z)^{-1}}}  \\ 
 & \geq   \norm{(T(q)-z)^{-1}}  - C  \norm{T(q_0) - T(q)}  - \frac{1}{2} \norm{(T(q)-z)^{-1}} \\ 
 & = \frac{1}{2} \norm{(T(q)-z)^{-1}} - C  \norm{ T(q_0) - T(q)}. 
\end{align*}
So if $(q_n,z_n) \to (q_0,z_0)$ and  $\norm{(T(q_n)-z_n)^{-1}} \to \infty$ then  $ \norm{(T(q_0)- z_0)^{-1}} =\infty$ which contradicts  $z_0\in \sigma( T(q_0))$. 
\end{proof}
For completeness, we write out the norm continuity of $q \mapsto T(q)$ in the case that is relevant to us.
\begin{lemma}\label{lemma:trivial_norm_continuity}
Let $r\in \N$ and $T(q)$ be an $r$-diagonal Laurent operator with continuous functions $a_i: \lbrack 0, 2\pi \rbrack \to \C$ on the $i$'th diagonal for $-r \leq i \leq r$. Then $
\norm{T(q)- T(q_n)} \to 0
$ as $q \to q_n$. 
\end{lemma}
\begin{proof}
Writing $T = \sum_{i = -r}^r a_i S_i$, by the triangle inequality and continuity: 
\begin{equation*}   
 \displaystyle \norm{ T(q_n) - T(q)}= \norm{\sum_{i=-r}^r (a_i(q) - a_i(q_n))S^i}\leq \sum_{i=-r}^r \abs{a_i(q) - a_i(q_n)}\to 0. \hfill \qedhere
\end{equation*}
%
\end{proof}
\subsection{Invertibility of bi-infinite tridiagonal Laurent matrices} \label{sec:invertibility_tridiagonal}
In this appendix, an explicit inversion of bi-infinite tridiagonal Laurent operators is performed. It was important in the applications of Corollary \ref{cor:rank_one}. 
%
%
\noindent For a tridiagonal operator $T$ with $\alpha, \beta $ and $\gamma$ on the diagonals, by \Cref{range_of_symbol} the spectrum $\sigma(T)$ is the range of the symbol curve, 
$
a(z)=  \alpha z^{-1} + \beta + \gamma z,  z\in \mathbb{T}, 
$
which is a (possibly degenerate) ellipse. Define square roots using 
  \begin{convention} \label{sign_convention}
  For any $z\in \C$ the two branches $ \pm \sqrt{z}$ are defined such that $\Re( + \sqrt{z} ) \geq 0$ and $\Re( - \sqrt{z} ) \leq 0$. If $ \Re(+ \sqrt{z}) = 0$ then the convention is that $ \Im( +\sqrt{z} )   \geq 0$.
  \end{convention}
The Laurent operator $T$ is invertible if and only if there are no solutions on the unit circle to the equation
\begin{align}  \label{eq:eigenvalue_equation}
\alpha + \beta x + \gamma x^2 = 0.
\end{align}
Whenever $\gamma \neq 0$ denote the two solutions  both with $\lambda_1, \lambda_2$ and $\lambda_\pm$, such that $ \{ \lambda_1, \lambda_2\} =  \{ \lambda_+, \lambda_- \} $ and $\abs{\lambda_2}\leq \abs{\lambda_1} $ and by convention  \begin{align} \label{midnat}
\lambda_{\pm}= - \frac{ \beta}{2 \gamma}\pm \sqrt{ \left( \frac{\beta}{2\gamma} \right)^2 -  \frac{\alpha}{\gamma} }.  
\end{align}
From Vieta's formula, $\lambda_1\lambda_2= \lambda_+ \lambda_- = \frac{\alpha}{\gamma}$ and as $\gamma (\lambda_+ - \lambda_-)= \sqrt{\beta^2 - 4 \alpha \gamma}$ and hence (if $\lambda_2 < \lambda_1$) a sign is introduced, 
\begin{align}\label{eq:sign_equation}
    \gamma(\lambda_2 -\lambda_1) =  (-1)^{ \id \lbrack \abs{\lambda_-}< 1 < \abs{\lambda_+}\rbrack} \gamma(\lambda_+ - \lambda_-) = (-1)^{ \id \lbrack \abs{\lambda_-}< 1 < \abs{\lambda_+}\rbrack}\sqrt{\beta^2 - 4 \alpha \gamma}
\end{align}
 vice versa or if the solutions change sign. A continuity argument shows that in our cases of interest there is always one solution inside the unit circle and one outside.
 \begin{lemma}\label{eig}
 Suppose that $T(q)$ is a family of invertible tridiagonal Laurent operators with continuous functions $\alpha, \beta, \gamma : S^1 \to \C$ on the diagonals. 
 Assume that $\gamma(q_2) = 0 $ for at most one $q_2$ and that there exists $q_0, q_1$ such that
$$
   \abs{\alpha(q_0)}\geq \abs{\gamma(q_0)}  \text{  and  } \abs{\alpha(q_1)}\leq \abs{\gamma(q_1)}.
$$
Let $\lambda_1(q) , \lambda_2(q)$ be the two solutions of \eqref{eq:eigenvalue_equation}
such that
 $
\abs{ \lambda_2} \leq \abs{\lambda_1}
$.  Then for all $q \neq q_2$,
 \begin{align*}
\abs{ \lambda_2(q)}< 1 < \abs{\lambda_1(q)}.
\end{align*}
 \end{lemma}
 \begin{proof}
Since  $T(q)$ is invertible $\abs{\lambda_1}$ and $\abs{\lambda_2}$ are never equal to 1.  Since $\lambda_1\lambda_2= \frac{\alpha}{\gamma}$, then $\abs{\lambda_1(q_0) \lambda_2(q_0) } \geq 1$ and so either $ \abs{\lambda_1(q_0)}\geq \abs{\lambda_2(q_0)} > 1$ or $ \abs{\lambda_1(q_0)}< 1  < \abs{\lambda_2(q_0)}$. 
Similarly, as $\abs{\lambda_1(q_1) \lambda_2(q_1) } \leq 1$  either $ \abs{\lambda_1(q_1)}< 1  < \abs{\lambda_2(q_1)}$ or $ \abs{\lambda_1(q_1)} \leq  \abs{\lambda_2(q_1)} < 1$.
By (\ref{midnat}), $q \mapsto \abs{\lambda_\pm(q)}$ are continuous functions on the connected set $S^1 \backslash \{q_2\}$ which are never $1$. This is only consistent with the cases above if for all $q \neq q_2$, $
\abs{\lambda_1(q)}< 1  < \abs{\lambda_2(q)}$. \hfill \qedhere
\end{proof}


\begin{lemma} \label{inverse}
Let $T$ is an invertible tridiagonal Laurent operator with $\alpha, \beta, \gamma \in \C$ on the diagonals with $\gamma \neq 0$.
Let $\lambda_1$ and $\lambda_2$ be the solutions to \eqref{midnat} and assume that $\abs{\lambda_2} < 1< \abs{\lambda_1} $ (as ensured by Lemma \ref{eig}).
Then for $k \geq 0$, 
\begin{align*}
\bra{\delta_{n}},T^{-1}\ket{\delta_{n+k}}= \frac{1}{\lambda_1^k \gamma(\lambda_2-\lambda_1)} 
\hspace{1cm}\text{ as well as   }\hspace{1cm}
\bra{\delta_{n}},T^{-1}\ket{\delta_{n-k}}= \frac{\lambda_2^k}{ \gamma(\lambda_2-\lambda_1)}. 
\end{align*}
\end{lemma}
\begin{proof}
First, $\sigma(T)$ is the image of the symbol curve
$ a(z) = \alpha z^{-1} + \beta +  \gamma z $
for $z\in \mathbb{T}$. 
Since $0 \not \in  \sigma(T)$,  $T^{-1}$ is unitarily equivalent to a multiplication with the inverse symbol (see e.g. \cite[Theorem 1.2]{LTTM}) and with $\lambda_{\pm}$ defined in \eqref{midnat}: 
 \begin{align*}
 \frac{1}{a(z)} =  \frac{1}{\frac{\alpha}{ z} + \beta +  \gamma z }  =   \frac{z}{\alpha+ \beta z  +  \gamma z^2} = \frac{z}{ \gamma( \frac{\alpha}{\gamma} +  \frac{\beta}{\gamma}  z  +   z^2) }  =  \frac{z}{ \gamma ( z - \lambda_+) (z- \lambda_-) } = \frac{z}{\gamma ( \lambda_1 - \lambda_2)}\left( \frac{1}{z-\lambda_1}-  \frac{1}{z-\lambda_2} \right).
 \end{align*}
Now, the assumption $\abs{\lambda_2} < 1< \abs{\lambda_1} $ has implications on how we write this up as a geometric series:
\begin{align*}
\frac{\gamma ( \lambda_1 - \lambda_2)}{za(z)} = - \frac{1}{\lambda_1} \frac{1}{1- \frac{z}{\lambda_1} }- \frac{1}{z}  \frac{1}{1-\frac{\lambda_2}{z}}  =  - \frac{1}{\lambda_1} \sum_{n=0}^\infty (\frac{z}{\lambda_1})^n - \frac{1}{z}  \sum_{n=0}^\infty (\frac{\lambda_2}{z})^n 
=  \sum_{n=1}^\infty (\frac{z}{\lambda_1})^n  +   \sum_{n=0}^\infty (\frac{\lambda_2}{z})^n.
\end{align*}
Using \eqref{eq:sign_equation} the solutions may also be expressed in terms of $\lambda_+$ and $\lambda_-$. \hfill \qedhere
\end{proof}

\subsection{An upper bound for the spectrum of the dephasing Anderson model}\label{improving_upper_bound}
The Anderson model with local dephasing has recently attracted attention \cite{hunter2020quantum, rath2020prominent}. The spectrum can be upper bounded using the numerical range. The proof is instructive as it shows the interplay between disorder and dissipation.

\begin{proposition}
\label{theorem:analytical_upper_bound_numerical_range}
Consider the Anderson model $H= - \tilde \Delta + V$ with local dephasing defined in \eqref{eq:local_dephasing} and disorder strength $\lambda$. Then for any $\rho \in \HS(\Hi)$ with $\|\rho\|_2 = 1$ such that
$\sum_{k \in \Z} |\rho(k,k)|^2 = a \in [0,1]$, 
\[
\langle \rho, \Li \rho \rangle
= Ga + \langle \rho, \mathcal{E}_V \rho \rangle + \langle \rho, \mathcal{T} \rho \rangle
\in G(a-1) + i\,[ - f(a,\lambda),\, f(a,\lambda) ],
\]
with
$
f(a,\lambda) \;=\; 4\Big(1-a + 2\sqrt{a}\sqrt{1-a}\Big) + (1-a)\lambda.
$
It follows that
$
\sigma(\Li) \subset \bigcup_{a \in [0,1]}\Big( G(a-1) + i\,[ - f(a,\lambda),\, f(a,\lambda) ] \Big).
$
\end{proposition}

\begin{proof}
Writing $\Li$ as in Theorem \ref{main}, then by \Cref{kunz}, $\sigma(\Li) \subset \overline{W(\mathcal{T}+\mathcal{F}+\mathcal{E}_V)}$, so we bound the numerical range of each term. Let $\rho \in \HS(\Hi)$ with $\|\rho\|_2=1$ and define
\[
\ket{\rho(q)} = (\mathcal{F}_1 \circ C \circ \vectorize(\rho))(q) \in \ell^2(\Z).
\]
Explicit calculation yields
$
\ket{\rho(q)} = \frac{1}{\sqrt{2\pi}} \sum_{k,j\in\Z} \rho(k,j) e^{-iqk}\,\ket{\delta_{j-k}},
$so $
\langle \delta_0,\rho(q) \rangle = \sum_{k\in\Z} \rho(k,k) e^{-iqk}.
$
Hence
\[
\int_{0}^{2\pi} \big|\langle \delta_0,\rho(q) \rangle\big|^2\,dq
= \sum_{k\in\Z} |\rho(k,k)|^2  \equiv a \in [0,1],
\]
which quantifies how “classical’’ $\rho$ is.
For the potential part, since $V$ is supported in $[0,\lambda]$, 
\begin{align*}
\langle \rho, \mathcal{E}_V \rho \rangle
&= \sum_{k,k',j,j'\in\Z} \Tr\!\Big( \overline{\rho(k',j')} \ket{\delta_{k'}}\bra{\delta_{j'}}\, i\big(V(k)-V(j)\big)\, \rho(k,j)\, \ket{\delta_k}\bra{\delta_j} \Big) \\
&= i \sum_{\substack{k,j\in\Z\\ k\neq j}} |\rho(k,j)|^2 \big( V(k) - V(j) \big)
\;\in\; i \,[ -(1-a)\lambda,\ (1-a)\lambda ]. 
\end{align*}

Using \eqref{ex_form}, $T(q)$ is tridiagonal with $\alpha(q),\beta,\gamma(q)$ on the diagonals, $\overline{\alpha(q)}=-\gamma(q)$, and $\beta(q)=\beta$ constant. Then
\begin{align*}
\langle \rho, \mathcal{T} \rho \rangle
= \Big\langle \rho, \int_{[0,2\pi]}^{\oplus} T(q)\, dq\, \rho \Big\rangle
&= \int_{0}^{2\pi} \langle \rho(q), T(q) \rho(q) \rangle\, dq \\
&= \sum_{n\in\Z} \int_{0}^{2\pi} \langle \rho(q) ,\delta_n \rangle \Big( \alpha(q) \langle \delta_{n+1},\rho(q)\rangle
+ \beta\, \langle \delta_n ,\rho(q)\rangle
+ \gamma(q) \langle \delta_{n-1},\rho(q)\rangle \Big)\, dq \\
&= \beta \sum_{n\in\Z} \int_{0}^{2\pi} \big|\langle \delta_n ,\rho(q)\rangle\big|^2\, dq
\;+\; 2i \sum_{n\in\Z} \int_{0}^{2\pi} \Im\!\Big( \gamma(q) \langle \rho(q), \delta_n\rangle \langle \delta_{n-1}, \rho(q)\rangle \Big)\, dq \\
&= \beta \;+\; 2i \sum_{n\in\Z} \int_{0}^{2\pi} \Im\!\Big( \gamma(q)\, l_{q,n}\, \overline{l_{q,n-1}} \Big)\, dq,
\end{align*}
where we set $l_{q,n} := \langle \rho(q), \delta_n \rangle$. Let $l_{q,n}^{\ge j} := l_{q,n}\,\mathbf{1}_{\{n\ge j\}}$ and $l_{q,n}^{\le j} := l_{q,n}\,\mathbf{1}_{\{n\le j\}}$. Using Cauchy--Schwarz and $\|\gamma\|_\infty=2$,
\begin{align*}
\Big| \sum_{n\in\Z} \int_{0}^{2\pi} \Im\!\big( \gamma(q)\, l_{q,n}\, \overline{l_{q,n-1}} \big)\, dq \Big|
&\le \|\gamma\|_\infty \int_{0}^{2\pi}
\Big( \big|\!\sum_{n\le -1} l_{q,n} \overline{l_{q,n-1}}\big|
+ \big|\!\sum_{n\ge 2} l_{q,n} \overline{l_{q,n-1}}\big|
+ |l_{q,0}\overline{l_{q,-1}}|
+ |l_{q,1}\overline{l_{q,0}}| \Big)\, dq \\
&\le 2 \int_{0}^{2\pi} \Big( \|l_q^{\le -1}\|_2 \|l_q^{\le -2}\|_2
+ \|l_q^{\ge 1}\|_2 \|l_q^{\ge 2}\|_2
+ |l_{q,0}||l_{q,-1}|
+ |l_{q,1}||l_{q,0}| \Big)\, dq \\
&\le 2\Big( 1-a + \sqrt{\textstyle\int_{0}^{2\pi} |\langle \delta_0, \rho(q)\rangle|^2 dq}\, \big( \|l_{q,-1}\|_2 + \|l_{q,1}\|_2 \big) \Big) \\
&\le 2\Big( 1-a + 2\sqrt{a}\sqrt{1-a} \Big).
\end{align*}
Thus the imaginary part coming from $\mathcal{T}$ is bounded in magnitude by
$4\big(1-a + 2\sqrt{a}\sqrt{1-a}\big)$.
Finally, for $F(q)=\ket{\mathtt{v}}\bra{\mathtt{u}}$,
\[
\langle \rho, \textstyle\int_{[0,2\pi]}^{\oplus} \ket{\mathtt{v}}\bra{\mathtt{u}}\, dq\, \rho \rangle
= \int_{0}^{2\pi} \langle \rho(q), \mathtt{v}\rangle \langle \mathtt{u}, \rho(q) \rangle\, dq
= G \int_{0}^{2\pi} |\langle \delta_0 , \rho(q)\rangle|^2\, dq
= G \sum_{k\in\Z} |\rho(k,k)|^2
= Ga.
\]
Combining the three bounds yields the stated inclusion with
$f(a,\lambda)=4\big(1-a+2\sqrt{a}\sqrt{1-a}\big) + (1-a)\lambda$. \qedhere
\end{proof}
As $\Re( \langle \rho,  \Li \rho \rangle)  = G(a-1) = \Re( \langle \rho,  \Li_0 \rho \rangle) $ 
 the states that survive longest are very diagonal. That is, dephasing suppresses coherence. This general principle was also noted for a simpler model in \cite[Chapter 8]{gerry2005introductory}, in \Cref{sec:exact_solv} and proven in larger generality in \cite{klausen2025decoherence}.
 
\tableofcontents

\vspace{-1cm}

\end{appendix}

\bibliographystyle{unsrt_new}
\bibliography{Lindblad}

\begin{thebibliography}{10}

\bibitem{ruelle1969remark}
D.~Ruelle.
\newblock A remark on bound states in potential-scattering theory.
\newblock {\em Il Nuovo Cimento A (1965-1970)}, 61(4):655--662, 1969.

\bibitem{amrein1973characterization}
W.~Amrein and V.~Georgescu.
\newblock Characterization of bound states and scattering states in quantum mechanics.
\newblock Technical report, Univ., Geneva, 1973.

\bibitem{enss1978asymptotic}
V.~Enss.
\newblock Asymptotic completeness for quantum mechanical potential scattering: I. short range potentials.
\newblock {\em Communications in Mathematical Physics}, 61(3):285--291, 1978.

\bibitem{Esposito2005EmergenceOD}
M.~Esposito and P.~Gaspard.
\newblock {Emergence of diffusion in finite quantum systems}.
\newblock {\em Physical Review B}, 71:1--12, 2005.

\bibitem{Znidaric2015RelaxationTO}
M.~Znidaric.
\newblock {Relaxation times of dissipative many-body quantum systems}.
\newblock {\em Physical review. E, Statistical, nonlinear, and soft matter physics}, 92 4:042143, 2015.

\bibitem{flynn2021topology}
V.~P. Flynn, E.~Cobanera, and L.~Viola.
\newblock Topology by dissipation: Majorana bosons in metastable quadratic markovian dynamics.
\newblock {\em Physical Review Letters}, 127(24):245701, 2021.

\bibitem{plenio2008dephasing}
M.~B. Plenio and S.~F. Huelga.
\newblock Dephasing-assisted transport: quantum networks and biomolecules.
\newblock {\em New Journal of Physics}, 10(11):113019, 2008.

\bibitem{clark2011diffusive}
J.~Clark, W.~De~Roeck, and C.~Maes.
\newblock Diffusive behavior from a quantum master equation.
\newblock {\em Journal of mathematical physics}, 52(8):083303, 2011.

\bibitem{xu2018interplay}
X.~Xu, C.~Guo, and D.~Poletti.
\newblock Interplay of interaction and disorder in the steady state of an open quantum system.
\newblock {\em Physical Review B}, 97(14):140201, 2018.

\bibitem{frohlich2016quantum}
J.~Fr{\"o}hlich and J.~Schenker.
\newblock Quantum brownian motion induced by thermal noise in the presence of disorder.
\newblock {\em Journal of Mathematical Physics}, 57(2):023305, 2016.

\bibitem{avron2012adiabatic}
J.~Avron, M.~Fraas, and G.~Graf.
\newblock {Adiabatic response for Lindblad dynamics}.
\newblock {\em Journal of Statistical Physics}, 148(5):800--823, 2012.

\bibitem{holevo}
A.~Holevo.
\newblock {\em {Statistical Structure of Quantum Theory}}, volume~67.
\newblock Springer, 01 2001.

\bibitem{tamura2016dynamical}
H.~Tamura and V.~A. Zagrebnov.
\newblock {Dynamical semigroup for unbounded repeated perturbation of an open system}.
\newblock {\em Journal of Mathematical Physics}, 57(2):023519, 2016.

\bibitem{PhysRevLett.123.234103}
T.~Can, V.~Oganesyan, D.~Orgad, and S.~Gopalakrishnan.
\newblock Spectral gaps and midgap states in random quantum master equations.
\newblock {\em Phys. Rev. Lett.}, 123:234103, Dec 2019.

\bibitem{CanRMT}
T.~Can.
\newblock {Random Lindblad Dynamics}.
\newblock {\em Journal of Physics A: Mathematical and Theoretical}, 52, 10 2019.

\bibitem{tarnowski2021random}
W.~Tarnowski, I.~Yusipov, T.~Laptyeva, S.~Denisov, D.~Chru{\'s}ci{\'n}ski, and K.~{\.Z}yczkowski.
\newblock {Random generators of Markovian evolution: A quantum-classical transition by superdecoherence}.
\newblock {\em Physical Review E}, 104(3):034118, 2021.

\bibitem{lange2021random}
S.~Lange and C.~Timm.
\newblock {Random-matrix theory for the Lindblad master equation}.
\newblock {\em Chaos: An Interdisciplinary Journal of Nonlinear Science}, 31(2):023101, 2021.

\bibitem{Gorini:1975nb}
V.~Gorini, A.~Kossakowski, and E.~C.~G. Sudarshan.
\newblock {Completely Positive Dynamical Semigroups of N Level Systems}.
\newblock {\em J. Math. Phys.}, 17:821, 1976.

\bibitem{lindblad1976generators}
G.~Lindblad.
\newblock {On the generators of quantum dynamical semigroups}.
\newblock {\em Communications in Mathematical Physics}, 48(2):119--130, 1976.

\bibitem{473322}
J.~P.~P. (https://physics.stackexchange.com/users/183551/jean-pierre polnareff).
\newblock Negativity of the real part of eigenvalues of lindblad operators.
\newblock Physics Stack Exchange.
\newblock URL:https://physics.stackexchange.com/q/473322 (version: 2019-04-18).

\bibitem{baumgartner2008analysis}
B.~Baumgartner and H.~Narnhofer.
\newblock {Analysis of quantum semigroups with GKS--Lindblad generators: II. General}.
\newblock {\em Journal of Physics A: Mathematical and Theoretical}, 41(39):395303, 2008.

\bibitem{buvca2012note}
B.~Bu{\v{c}}a and T.~Prosen.
\newblock A note on symmetry reductions of the lindblad equation: transport in constrained open spin chains.
\newblock {\em New Journal of Physics}, 14(7):073007, 2012.

\bibitem{Nigro}
D.~Nigro.
\newblock {On the uniqueness of the steady-state solution of the Lindblad-Gorini-Kossakowski-Sudarshan equation}.
\newblock {\em Journal of Statistical Mechanics: Theory and Experiment}, 2019, 03 2018.

\bibitem{symmetries}
V.~Albert and L.~Jiang.
\newblock {Symmetries and conserved quantities in Lindblad master equations}.
\newblock {\em Physical Review A}, 89, 10 2013.

\bibitem{Georgios}
G.~Styliaris and P.~Zanardi.
\newblock {Symmetries and monotones in Markovian quantum dynamics}.
\newblock {\em Quantum}, 4:261, 04 2020.

\bibitem{Holevo1993ANO}
A.~Holevo.
\newblock A note on covariant dynamical semigroups.
\newblock {\em Reports on Mathematical Physics}, 32:211--216, 1993.

\bibitem{reed1978iv}
M.~Reed and B.~Simon.
\newblock {\em {IV: Analysis of Operators}}, volume~4.
\newblock Elsevier, 1978.

\bibitem{halmos2012hilbert}
P.~R. Halmos.
\newblock {\em A Hilbert space problem book}, volume~19.
\newblock Springer Science \& Business Media, 2012.

\bibitem{azoff1974spectrum}
E.~A. Azoff.
\newblock Spectrum and direct integral.
\newblock {\em Transactions of the American Mathematical Society}, 197:211--223, 1974.

\bibitem{chowspectrum}
T.~R. Chow.
\newblock A spectral theory for direct integrals of operators.
\newblock {\em Mathematische Annalen}, 188, 1970.

\bibitem{ismailovsome}
Z.~I. Ismailov and E.~O. Cevik.
\newblock Some spectral properties of direct integral of operators.
\newblock {\em Journal of Analysis and Number Theory}, 3(1):1--7, 2015.

\bibitem{ismailov2020spectra}
Z.~I. Ismailov and I.~A. Pembe.
\newblock Spectra and pseudospectra of the direct sum operators.
\newblock {\em Sigma Journal of Engineering and Natural Sciences}, 38(3):1251--1259, 2020.

\bibitem{fialovatwo}
M.~Fialov{\'a}.
\newblock {Two-dimensional Dirac operator with translationally invariant electromagnetic field}.
\newblock {\em Master's Thesis}, 2018.

\bibitem{ng2020direct}
A.~C.~S. Ng.
\newblock Direct integrals of strongly continuous operator semigroups.
\newblock {\em Journal of Mathematical Analysis and Applications}, 489(2):124176, 2020.

\bibitem{durhuus2014mathematical}
B.~Durhuus and J.~P. Solovej.
\newblock {\em Mathematical Physics [Lecture Notes]}.
\newblock Department of Mathematical Sciences, University of Copenhagen, 2014.

\bibitem{vectorization}
T.~Havel.
\newblock {Procedures for Converting among Lindblad, Kraus and Matrix Representations of Quantum Dynamical Semigroups}.
\newblock {\em Journal of Mathematical Physics}, 44, 02 2002.

\bibitem{lecturenotes}
G.~T. Landi.
\newblock {Lecture Notes on Lindblad Master Equations}.
\newblock \url{http://www.fmt.if.usp.br/~gtlandi/11---lindblad-equation-2.pdf}.
\newblock Accessed: 2025-09-01.

\bibitem{355246}
N.~W. (https://mathoverflow.net/users/23141/nik weaver).
\newblock Fourier transform of a translation invariant operator on $l^2(\mathbb{Z}) \otimes l^2(\mathbb{Z})$.
\newblock MathOverflow.
\newblock URL:https://mathoverflow.net/q/355246 (version: 2020-03-19).

\bibitem{kato2013perturbation}
T.~Kato.
\newblock {\em Perturbation theory for linear operators}, volume 132.
\newblock Springer Science \& Business Media, 2013.

\bibitem{einsiedler2017functional}
M.~Einsiedler, T.~Ward, et~al.
\newblock {\em {Functional analysis, spectral theory, and applications}}, volume 276.
\newblock Springer, 2017.

\bibitem{LTTM}
A.~B{\"o}ttcher and B.~Silbermann.
\newblock {\em Introduction to large truncated Toeplitz matrices}.
\newblock Springer Science \& Business Media, 2012.

\bibitem{bottcher2002spectral}
A.~B{\"o}ttcher, M.~Embree, and M.~Lindner.
\newblock {Spectral approximation of banded Laurent matrices with localized random perturbations}.
\newblock {\em Integral Equations and Operator Theory}, 42(2):142--165, 2002.

\bibitem{hagen2000c}
R.~Hagen, S.~Roch, and B.~Silbermann.
\newblock {\em C*-algebras and numerical analysis}.
\newblock CRC Press, 2000.

\bibitem{hausdorff1937set}
F.~Hausdorff.
\newblock {\em Set Theory}.
\newblock Chelsea Publishing Company, New York, 1957.
\newblock English translation of \emph{Mengenlehre} (1935/1937).

\bibitem{Esposito2005ExactlySM}
M.~Esposito and P.~Gaspard.
\newblock {Exactly Solvable Model of Quantum Diffusion}.
\newblock {\em Journal of Statistical Physics}, 121:463--496, 2005.

\bibitem{Diehl}
S.~Diehl, E.~Rico~Ortega, M.~Baranov, and P.~Zoller.
\newblock Topology by dissipation in atomic quantum wires.
\newblock {\em Nature Physics}, 7, 05 2011.

\bibitem{LocalizationinOpenQuantumSystems}
I.~Yusipov, T.~Laptyeva, S.~Denisov, and M.~Ivanchenko.
\newblock {Localization in Open Quantum Systems}.
\newblock {\em Phys. Rev. Lett.}, 118:070402, Feb 2017.

\bibitem{vershinina2017control}
O.~Vershinina, I.~Yusipov, S.~Denisov, M.~V. Ivanchenko, and T.~Laptyeva.
\newblock Control of a single-particle localization in open quantum systems.
\newblock {\em EPL (Europhysics Letters)}, 119(5):56001, 2017.

\bibitem{klausen2025decoherence}
F.~R. Klausen and S.~Warzel.
\newblock {Decoherence is an echo of Anderson localization in open quantum systems}.
\newblock In {\em Annales Henri Poincar{\'e}}, pages 1--29. Springer, 2025.

\bibitem{lee2016anomalous}
T.~E. Lee.
\newblock {Anomalous edge state in a non-Hermitian lattice}.
\newblock {\em Physical Review Letters}, 116(13):133903, 2016.

\bibitem{bergholtz2021exceptional}
E.~J. Bergholtz, J.~C. Budich, and F.~K. Kunst.
\newblock {Exceptional topology of non-Hermitian systems}.
\newblock {\em Reviews of Modern Physics}, 93(1):015005, 2021.

\bibitem{song2019non}
F.~Song, S.~Yao, and Z.~Wang.
\newblock {Non-Hermitian skin effect and chiral damping in open quantum systems}.
\newblock {\em Physical Review Letters}, 123(17):170401, 2019.

\bibitem{okuma2020topological}
N.~Okuma, K.~Kawabata, K.~Shiozaki, and M.~Sato.
\newblock {Topological origin of non-Hermitian skin effects}.
\newblock {\em Physical Review Letters}, 124(8):086801, 2020.

\bibitem{Eisler_2011}
V.~Eisler.
\newblock {Crossover between ballistic and diffusive transport: the quantum exclusion process}.
\newblock {\em Journal of Statistical Mechanics: Theory and Experiment}, 2011(06):P06007, jun 2011.

\bibitem{PhysRevLett.123.140403}
S.~Denisov, T.~Laptyeva, W.~Tarnowski, D.~Chru\ifmmode \acute{s}\else \'{s}\fi{}ci\ifmmode~\acute{n}\else \'{n}\fi{}ski, and K.~\ifmmode~\dot{Z}\else \.{Z}\fi{}yczkowski.
\newblock {Universal Spectra of Random Lindblad Operators}.
\newblock {\em Phys. Rev. Lett.}, 123:140403, Oct 2019.

\bibitem{kunz1980spectre}
H.~Kunz and B.~Souillard.
\newblock {Sur le spectre des op{\'e}rateurs aux diff{\'e}rences finies al{\'e}atoires}.
\newblock {\em Communications in Mathematical Physics}, 78(2):201--246, 1980.

\bibitem{kirsch1982spectrum}
W.~Kirsch and F.~Martinelli.
\newblock {On the spectrum of Schr{\"o}dinger operators with a random potential}.
\newblock {\em Communications in Mathematical Physics}, 85(3):329--350, 1982.

\bibitem{toeplitz1918algebraische}
O.~Toeplitz.
\newblock {Das algebraische Analogon zu einem Satze von Fej{\'e}r}.
\newblock {\em Mathematische Zeitschrift}, 2(1):187--197, 1918.

\bibitem{hausdorff1919wertvorrat}
F.~Hausdorff.
\newblock {Der wertvorrat einer bilinearform}.
\newblock {\em Mathematische Zeitschrift}, 3(1):314--316, 1919.

\bibitem{colbrook2020pseudoergodic}
M.~Colbrook.
\newblock Pseudoergodic operators and periodic boundary conditions.
\newblock {\em Mathematics of Computation}, 89(322):737--766, 2020.

\bibitem{koekenbier2024transfer}
L.~Koekenbier and H.~Schulz-Baldes.
\newblock Transfer matrix analysis of non-hermitian hamiltonians: asymptotic spectra and topological eigenvalues.
\newblock {\em Journal of Spectral Theory}, 14(4):1563--1622, 2024.

\bibitem{vakulchyk2018signatures}
I.~Vakulchyk, I.~Yusipov, M.~Ivanchenko, S.~Flach, and S.~Denisov.
\newblock Signatures of many-body localization in steady states of open quantum systems.
\newblock {\em Physical Review B}, 98(2):020202, 2018.

\bibitem{406705}
N.~W. (https://mathoverflow.net/users/23141/nik weaver).
\newblock {If I multiply the coefficients of a trace-class operator with bounded complex numbers is it still trace class?}
\newblock MathOverflow.
\newblock URL:https://mathoverflow.net/q/406706 (version: 2021-10-22).

\bibitem{klausen2023random}
F.~R. Klausen.
\newblock {\em Random Problems in Mathematical Physics}.
\newblock PhD thesis, University of Copenhagen, Denmark, 2023.

\bibitem{Aizenman2015RandomOD}
M.~Aizenman and S.~Warzel.
\newblock {\em Random operators}, volume 168.
\newblock American Mathematical Soc., 2015.

\bibitem{simon2005trace}
B.~Simon.
\newblock {\em Trace Ideals and Their Applications}, volume 120 of {\em Mathematical Surveys and Monographs}.
\newblock American Mathematical Society, 2005.

\bibitem{Davies2007LinearOA}
E.~B. Davies.
\newblock {\em Linear operators and their spectra}, volume 106.
\newblock Cambridge University Press, 2007.

\bibitem{Falconi2016ScatteringTF}
M.~Falconi, J.~Faupin, J.~Fr{\"o}hlich, and B.~Schubnel.
\newblock {Scattering Theory for Lindblad Master Equations}.
\newblock {\em Communications in Mathematical Physics}, 350:1185--1218, 2016.

\bibitem{Attal}
S.~Attal.
\newblock {Quantum Channels}.
\newblock {\em Online course}, 2014.

\bibitem{olkiewicz1999environment}
R.~Olkiewicz.
\newblock {Environment-induced superselection rules in Markovian regime }.
\newblock {\em Communications in mathematical physics}, 208(1):245--265, 1999.

\bibitem{Perez}
D.~Pérez-García, M.~M. Wolf, D.~Petz, and M.~B. Ruskai.
\newblock {Contractivity of positive and trace-preserving maps under $L^p$ norms}.
\newblock {\em Journal of Mathematical Physics}, 47(8):083506, 2006.

\bibitem{reed1975ii}
M.~Reed and B.~Simon.
\newblock {\em II: Fourier Analysis, Self-Adjointness}, volume~2.
\newblock Elsevier, 1975.

\bibitem{zhu2018introduction}
K.~Zhu.
\newblock {\em An introduction to operator algebras}.
\newblock CRC press, 2018.

\bibitem{mori2020resolving}
T.~Mori and T.~Shirai.
\newblock {Resolving a Discrepancy between Liouvillian Gap and Relaxation Time in Boundary-Dissipated Quantum Many-Body Systems}.
\newblock {\em Physical Review Letters}, 125(23):230604, 2020.

\bibitem{szehr2015spectral}
O.~Szehr, D.~Reeb, and M.~M. Wolf.
\newblock Spectral convergence bounds for classical and quantum markov processes.
\newblock {\em Communications in Mathematical Physics}, 333:565--595, 2015.

\bibitem{khandelwal2021signatures}
S.~Khandelwal, N.~Brunner, and G.~Haack.
\newblock {Signatures of Liouvillian Exceptional Points in a Quantum Thermal Machine}.
\newblock {\em PRX Quantum}, 2(4):040346, 2021.

\bibitem{engel2000one}
K.-J. Engel, R.~Nagel, and S.~Brendle.
\newblock {\em One-parameter semigroups for linear evolution equations}, volume 194.
\newblock Springer, 2000.

\bibitem{measurability_reference}
B.~(https://mathoverflow.net/users/142961/bremen000).
\newblock Measurable selection involving measure valued random variable.
\newblock MathOverflow.
\newblock URL:https://mathoverflow.net/q/389578 (version: 2021-04-07).

\bibitem{hunter2020quantum}
M.~Hunter-Gordon, Z.~Szab{\'o}, R.~A. Nyman, and F.~Mintert.
\newblock {Quantum simulation of the dephasing Anderson model}.
\newblock {\em Physical Review A}, 102(2):022407, 2020.

\bibitem{rath2020prominent}
Y.~Rath and F.~Mintert.
\newblock {Prominent interference peaks in the dephasing Anderson model}.
\newblock {\em Physical Review Research}, 2(2):023161, 2020.

\bibitem{gerry2005introductory}
C.~Gerry and P.~Knight.
\newblock {\em Introductory quantum optics}.
\newblock Cambridge University Press, 2005.

\end{thebibliography}

\end{document}